\documentclass[a4paper,reqno]{amsart}
\usepackage[T1]{fontenc}
\usepackage[english]{babel}
\usepackage{amssymb,bbm,enumerate}
\usepackage{bm}
\usepackage{xcolor}
\usepackage[linktocpage=true,colorlinks=true, linkcolor=blue, citecolor=red, urlcolor=green]{hyperref}
\usepackage{multirow} 
\usepackage{graphicx}

\usepackage{float}
\restylefloat{table}
\usepackage{tabu}
\usepackage{ytableau}
\usepackage[makeroom]{cancel}

\renewcommand{\phi}{\varphi}
\renewcommand{\ker}{\Ker}
\renewcommand{\Re}{\operatorname{Re}}

\DeclareMathOperator{\Span}{Span}

\DeclareMathOperator{\im}{Im}
\DeclareMathOperator{\Ker}{Ker}

\newcommand{\Bna}{\overline{\mc{B}}_{N,\alpha}}
\newcommand{\bna}{\mc{B}_{N,\alpha}}
\newcommand{\Pjm}{\overline{\mc{P}}_{J,M}}
\newcommand{\pjm}{\mc{P}_{J,M}}
\newcommand{\Tna}{\overline{\mc{T}}_{N,\alpha}}
\newcommand{\tna}{\mc{T}_{N,\alpha}}
\newcommand{\vecu}{\underline{u}}
\newcommand{\vecv}{\underline{v}}
\newcommand{\vecx}{\underline{x}}
\newcommand{\vecX}{\underline{X}}
\newcommand{\vecY}{\underline{Y}}
\newcommand{\vectt}{\underline{t}}

\newcommand{\vecA}{\underline{A}}
\newcommand{\vecB}{\underline{B}}
\newcommand{\vecf}{\underline{f}}
\newcommand{\vecM}{\underline{M}}

\newcommand{\snu}{\scriptscriptstyle{[\nu]}}
\newcommand{\snus}{\scriptscriptstyle{[\nu^*]}}
\newcommand{\sn}{\scriptscriptstyle{(n)}}

\renewcommand{\phi}{\varphi}
\renewcommand{\ker}{\Ker}
\renewcommand{\Re}{\operatorname{Re}}

\newcommand{\bb}[1]{\mathbb{#1}}
\newcommand{\mc}[1]{\mathcal{#1}}
\newcommand{\mf}[1]{\mathfrak{#1}}

\newcommand{\e}{\varepsilon}
\newcommand{\C}{\mathbb{C}}

\newcommand{\Ca}{\widehat{\mathbb{C}}_{\alpha}}

\theoremstyle{plain}
\newtheorem{theorem}{Theorem}[section]
\newtheorem{lemma}[theorem]{Lemma}
\newtheorem{conjecture}[theorem]{Conjecture}
\newtheorem{proposition}[theorem]{Proposition}
\newtheorem{corollary}[theorem]{Corollary}

\theoremstyle{definition}
\newtheorem{definition}[theorem]{Definition}

\theoremstyle{remark}
\newtheorem{remark}[theorem]{Remark}

\numberwithin{equation}{section}

\definecolor{light}{gray}{.9}

\title{Counting monster potentials}

\author{R. Conti, D. Masoero}

\address{Grupo de F\'isica Matem\'atica da Universidade de Lisboa, Edif\'icio C6
Campo Grande, Lisboa, Portugal.}
\address{Faculdade de Ci\^encias da Universidade de Lisboa,
Campo Grande, Lisboa, Portugal.}
\email{dmasoero@gmail.com}
\email{riccardo.conti92@gmail.com }

\begin{document}

\pagestyle{plain}

\begin{abstract}
We study the large momentum limit of the monster potentials of Bazhanov-Lukyanov-Zamolodchikov,
which -- according to the ODE/IM correspondence -- should
correspond to excited states of the Quantum KdV model.

We prove that the poles of these potentials asymptotically condensate about the complex equilibria of the ground state potential, and 
we express the leading correction to such asymptotics in terms of the roots of Wronskians of Hermite polynomials.

This allows us to associate to each partition of $N$ a unique monster potential with $N$ roots, of which we compute the spectrum.
As a consequence, we prove -- up to a few mathematical
technicalities -- that, fixed an integer $N$, the
number of monster potentials with $N$ roots coincides with the number of integer partitions of $N$, which is the dimension
of the level $N$ subspace of the quantum KdV model. In striking accordance with the ODE/IM correspondence.
\end{abstract}

\maketitle

\tableofcontents

\section{Introduction}
For each triple $(N,\alpha,L)$, where $N$ is a positive integer, $\alpha$ a positive real and L an arbitrary complex number, one defines the Bazhanov-Lukyanov-Zamolodchikov (BLZ) system \cite{BLZ04}. This is the following system of $N$ algebraic equations for $N$ complex unknowns $z_1,\dots,z_N$, which are assumed to be pairwise distinct and non-zero
\begin{equation}\label{eq:algblz}
\sum_{j \neq k} \frac{z_k\big( z_k^2 +(3+\alpha)(1+2\alpha)z_k z_j+
\alpha (1+2 \alpha)z_j^2 \big)}{ (z_k-z_j)^3}- \frac{\alpha z_k}{ 4 (1+\alpha)}+\Delta=0 \;,\quad k=1,\dots,N \;,
\end{equation}
with
\begin{equation}
\label{eq:cdelta}
% c = 1 -{6\, \alpha^2\over \alpha + 1}\ , 
%\qquad
\Delta=\frac{{4L}+1-4\alpha^2}{ 16\,(\alpha + 1)} \;.
\end{equation}
The BLZ system \eqref{eq:algblz}, which is symmetric under any permutation of the roots, is the necessary and sufficient condition for the Sch\"odinger equation
\begin{equation}\label{eq:schrintro}
\psi''(x)=\big(V(x)-E\big)\psi(x) \;,\quad E \in \C \;,
\end{equation}
with the \textit{monster potential}
\begin{equation}\label{eq:V}
V(x)=\frac{L}{x^2}+x^{2\alpha}-2 \frac{d^2}{d x^2} \sum_{k=1}^N\,  \log(x^{2\alpha+2}-z_k) \;,
\end{equation}
to have trivial monodromy at any $x$ such that $x^{2\alpha+2}=z_k$ for all $k \in\lbrace1,\dots,N\rbrace$, for every value of $E \in \C$.

The relevance of the monster potentials, which may look strange at a first sight, is that spectral data of the Schr\"odinger equation \eqref{eq:schrintro}, namely the Stokes multiplier $T(E)$ and the spectral determinant of the radial problem $Q(E)$, provide solutions of the TQ relations of the Quantum KdV model \cite{BLZ04}. More precisely, according to the BLZ conjecture \cite{BLZ04}, the monster potentials with $N$ roots are in bijection with the level $N$ states of the Quantum KdV model, provided the central charge $c$ of the model is 
\begin{equation}\label{eq:c}
c = 1 - \frac{6 \alpha^2}{\alpha + 1} \;,
\end{equation}
and $\Delta$, as defined in \eqref{eq:cdelta}, is the highest weight of the corresponding Virasoro module. This is an instance of the celebrated ODE/IM correspondence, firstly discovered by Dorey and Tateo \cite{doreytateo98}, and later generalised by many more authors, see e.g. \cite{suzuki00,bazhanov01,dorey07,lukyanov10,FF11,Sun12,marava15,marava17,fh16,mara18,frenkel20}.

Provided the BLZ conjecture holds, it follows that the number of solutions of the BLZ system \eqref{eq:algblz} -- modulo the action of the symmetric group -- coincides with the dimension of the subspace of level $N$ of the irreducible highest weight Virasoro module with central charge $c$ and highest weight $\Delta$  as per \eqref{eq:c} and \eqref{eq:cdelta}, respectively. Such dimension, for generic values of $c$ and $\Delta$, is $p(N)$,
the number of integer partitions of the integer $N$, see e.g. \cite{difrancesco12}. Therefore we define the \textit{Weak BLZ conjecture}.
\begin{conjecture}[Weak BLZ conjecture]\label{conj:weakblz}
Fix $N \geq 1$. For every $\alpha$ and $L$, the number of solutions of the BLZ system \eqref{eq:algblz} is less or equal than $p(N) \times N!$. For generic values of $\alpha$ and $L$ the number of solutions is $p(N) \times N!$.
\end{conjecture}
In this paper we analyse the BLZ system in the large $L$ limit and we prove -- up to a few mathematical technicalities which we will discuss
later -- that the number of solutions of the BLZ system is indeed $N! \times p(N)$ for any $\alpha>0$, provided $L$ is large. We remark that in this paper we focus on the ODE side of the ODE/IM correspondence. The large momentum limit of the ODE/IM correspondence will be however addressed in a forthcoming paper.

\subsection{Notation}
Throughout the paper the following notation is used:
\begin{itemize}
 \item $\alpha>0$ is an arbitrary positive real number, $\Ca$ is the Riemann surface of the function $x^{2\alpha+2}$ 
 (i.e. the function $x^{2\alpha+2}$ is single-valued on $\Ca$ \footnote{
If $2\alpha \in \bb{N}$ then $\Ca=\C$; if $2\alpha \in \bb{Q} \setminus \bb{N}$ then $\Ca$ is a finite cover of $\C$ branched at $0$;
if $2\alpha \in \bb{R}\setminus \bb{Q}$ then $\Ca$ is the universal cover of $\C^*$.}) and
$\gamma_{\alpha} \in \C^*$ is defined as
\begin{equation}\label{eq:gamma}
       \gamma_{\alpha}=e^{\frac{i\pi}{\alpha+1}} \;.
\end{equation}
\item $L$ is an arbitrary complex number. One often writes $L=\ell(\ell+1)$ and name $\ell$ the angular momentum. Therefore, we refer to the limit $L \to \infty$ as large momentum limit.
\item For any monic polynomial $P$, we define the potential
\begin{equation}\label{eq:Vgen}
  V_P(x)=V_G(x)-2 \frac{d^2}{dx^2}\log P(x^{2\alpha+2}) \;,\quad x \in \Ca \;,
\end{equation}
where $V_G(x)$ is the \textit{ground-state potential}
      \begin{equation}\label{eq:ground}
      V_G(x)= \frac{L}{x^2}+x^{2\alpha} \;,\quad x \in \Ca \;.
\end{equation}
The potential $V_P$ satisfies the Dorey-Tateo symmetry
\begin{equation}\label{eq:DT}
V_P(\gamma_{\alpha} x)=\gamma_{\alpha}^{-2} V_P(x) \;,\quad \forall x \in \Ca \;,
\end{equation}
and it admits a $\tau$ function, namely
\begin{equation}\label{eq:tau}
 V(x)=-2\frac{d^2}{dx^2} \log \tau_V(x) \;,\quad
\tau_V(x)= x^{\frac{L}2}e^{\left(\frac{-x^{2 a+2}}{4 (a+1) (2 a+1)}\right)}P(x^{2\alpha+2}) \;.
\end{equation}
\item For any positive integer $N$, $p(N)$ is the number of distinct partitions of $N$. Each partition is uniquely described by a $j$-tuple of positive natural numbers $\nu$, such that
\begin{equation*}
\nu=(\nu_1,\dots,\nu_j) \;,\quad \sum_{l=1}^j \nu_l=N \;,\quad \nu_{l+1}\leq\nu_l \;.
\end{equation*}
\item
Given a partition $\nu$, we say that $\nu^*$ is the conjugate (or transpose) partition with respect to $\nu$ if its associated Young diagram is the transpose (i.e. obtained exchanging rows with columns) of the diagram associated to $\nu$. 
\item To any partition $\nu$ of $N$, we associate the following polynomial of degree $N$
\begin{equation}\label{eq:Pnu}
P^{\snu}(t)=c_{\nu}\mbox{Wr}[H_{\nu_j}(t),H_{\nu_{j-1}+1}(t),\dots,H_{\nu_1+j-1}(t)] \;,
\end{equation}
where $\displaystyle{H_{n}(t)=(-1)^ne^{\frac{t^2}{2}}\frac{d^n}{dt^n}e^{\frac{t^2}{2}}}$ is the $n-$th Hermite polynomial, $\mbox{Wr}$ is the Wronskian of $j$ functions and $c_\nu \neq 0$ a constant chosen in such a way that $P^{\snu}(t)$ is monic (see \cite{felder12}).
\item For every polynomial $P^{\snu}(t)$, we denote by
\begin{equation}\label{eq:vnu}
 \underline{v}^{\snu}=(v_1^{\snu},\dots,v_N^{\snu}) \in \C^N \;,
\end{equation}
the, not necessarily pairwise distinct, roots of $P^{\snu}(t)$. These are assumed to be ordered in such a way
that $ i \underline{v}^{\snu}= \underline{v}^{\snus}$ (see Lemma \eqref{thm:felder} below).
\end{itemize}

\begin{definition}\label{defi:monster} Fix $N$ and $\alpha$.
Identifying the space of monic polynomial of degree $N$ with $\C^N$,
we let $\mc{B}_{N,\alpha} \subset \C^{N+1}$ consists of the ordered pairs $(P,L)$
where $L$ is a complex number and $P$ is a monic polynomial of degree $N$,
with distinct and non-zero roots which satisfy the BLZ system \eqref{eq:algblz}. $\bna$ thus classifies monster potentials \eqref{eq:V}.

We also define $\Bna$ as the closure (with respect to the standard norm) of $\mc{B}_{N,\alpha}$ in $\C^{N+1}$.
Given a point in $\Bna$, we call the corresponding potential \eqref{eq:Vgen} a higher-state-potential.
\end{definition}

\subsection{Main results and organisation of the paper}
In Section 2 we review the derivation of the BLZ system and in Section 3, following Oblomkov \cite{oblomkov99}, we review the theory of the rational extensions of the harmonic oscillator -- hence of Wronskians of Hermite polynomials -- which plays a major role in the rest of the paper.

In Section 4 and 5, we prove our main results: A) and B).

A) Assume that $(P^{\sn},L^{\sn}) \in \Bna$ is a sequence (of higher-state-potentials) such that  $L^{\sn} \to \infty$, and denote by $z_k^{\sn}$
the corresponding roots of $P^{\sn}$. The sequence can be split into $J$ subsequences, where $1\leq J \leq p(N)$, with the following properties:
\begin{itemize}
 \item On each subsequence, there exists a partition $\nu$ of $N$ such that the roots $z_1^{\sn}$, $\dots$, $z_N^{\sn}$ admit the following expansion 
\begin{equation}\label{eq:zkexpansion}
z_k^{\sn}=\frac{L^{\sn}}{\alpha}+ \frac{(2\alpha+2)^{\frac34}}{\alpha}  v_k^{\snu} (L^{\sn})^{-\frac34}+ o\left((L^{\sn})^{-\frac34}\right) \;,\quad k=1,\dots,N \;,
\end{equation}
where $v_1^{\snu}$, $\dots$, $v_N^{\snu}$ are the roots of the polynomial $P^{\snu}(t)$.
\item Along such a subsequence,
the higher-state-potentials $V_{P^{(n)}}$ converge -- when the variables are rescaled conveniently -- to the
potential
\begin{equation*}
 U^{\snu}(t)=t^2-\frac{d^2}{dt^2} \log P^{\snu}(t) \;,
\end{equation*}
which is the rational extension of the harmonic oscillator associated to the partition $\nu$.
\end{itemize}

B) After the change of variable $z_k=\frac{L}{\alpha}\big(1+(2\alpha+2)^{-\frac{1}{4}}\e\,t_k \big)^{2\alpha+2}$, with $\e=L^{-\frac14}$,
we reduce the study of the monster potentials in the large momentum limit to a small perturbation of the rational extensions of the harmonic oscillator.
\begin{itemize}
\item The BLZ system \eqref{eq:algblz} reduces to
\begin{align}\nonumber
 & 2t_k+ 3 \,\e \, \frac{(5-2\alpha) }{3(2\alpha+2)^{\frac14}} t_{k}^2 - \sum_{j\neq k}\frac{4}{(t_k-t_j)^3}
  + \mathcal{O}\big(\e^2\big) = 0  \;,\quad \e=L^{-\frac14} \;, \\ \label{eq:pertintro}
&  \lim_{\e \to 0} t_k(\e)=v^{\snu}_k \;,\quad k=1,\dots,N \;.
\end{align}
This is a perturbation of the system of equations satisfied by the roots of the polynomial $P^{\snu}(t)$, and its solution yields
a one-parameter family of monster potentials satisfying the asymptotics \eqref{eq:zkexpansion}. \item We show that in order to prove the Weak BLZ conjecture \ref{conj:weakblz}, it suffices to show that the system \eqref{eq:pertintro} has a unique algebraic solution for every $\nu$, see Corollary \ref{cor:blzconj}. 
\item We compute the asymptotics of the (bottom of the) radial spectrum of a family of monster potentials satisfying the asymptotics
\eqref{eq:zkexpansion}, when $L$ is large and positive:
\begin{align}\nonumber
& E_n^{\snu}= (1+\alpha) \left(\frac{L}{\alpha} \right)^{\frac{\alpha}{\alpha+1}} + \big(2\alpha+2\big)^{\frac12} \alpha^{\frac{1}{\alpha+1}}
\big( 2(n-j)+1 \big) L^{\frac{\alpha-1}{2\alpha+2}} +  \\ \label{eq:spectrumepsintro}
& \qquad \qquad \qquad \qquad \qquad \qquad \qquad \qquad  + O\left( |L|^{-\frac{1}{\alpha+1}}\right) \;,\quad  n \in \bb{N}^{\snu} \;,
\end{align}
where $\bb{N}^{\snu}$ is the sequence of integer numbers which is obtained from $\bb{N}$ deleting the numbers $(\nu_j,\nu_{j-1}+1,\dots, \nu_1+j-1)$. This is in perfect accordance with the prediction from the IM side of the correspondence \cite[Appendix A]{BLZ04}:
in fact it shows that the spectrum is real and positive when the momentum is positive and sufficiently large, and it reproduces the fact that excitations are produced by inserting holes in the Fermi-sea.
\end{itemize}

In Sections 6, 7, and 8 we study the well-posedness (the existence and uniqueness of solutions) of the perturbation series \eqref{eq:pertintro}. In Section 6, we deal with the case that the roots $v^{\snu}_1$, $\dots$, $v^{\snu}_N$ of the polynomial $P^{\snu}(t)$ are distinct and, as a byproduct, we obtain a (conjectural) closed combinatorial formula (see equation \eqref{eq:spectrumJ}) for the spectrum of the matrix
$$J^{\snu}_{ij}= 
2\delta_{ij}\left( 1+\sum_{l\neq j} \frac{6}{(v^{\snu}_j - v^{\snu}_l)^4}\right) - (1-\delta_{ij})\frac{12}{(v^{\snu}_i- v^{\snu}_j)^4} \;,\quad
i,j=1,\dots, N \;,$$
which is an important object in the theory of Wronskians of Hermite polynomials, see e.g. \cite{ahmed79}. In Section 7, we deal with the case where $P^{\snu}$ has a unique root, namely $0$, of multiplicity greater than one. Finally, in Section 8, we deal with the mixed case, when $0$ is a root of multiplicity greater than one but $P^{\snu}$ has other, all simple, roots.\footnote{There could be in principle other possibilities but these are ruled-out by the well-tested and well-known conjecture by Felder et al. \cite{felder12}. More about this below.}
Unfortunately, the system \eqref{eq:pertintro} is extremely intricate already for a moderate multiplicity (of the root $0$ of $P^{\snu}$) and
we have to restrict here to the case that the multiplicity is not greater than $6$. \textit{This is the mathematical technicality which prevents us from obtaining a rigorous proof of the Weak BLZ conjecture for $N\geq10$}.

Section 9 is devoted to extending the above results to study the large $L$ limit of potentials of the form
$$
x^{M-2}+\frac{L}{x^2}- 2 \sum_{k=1}^J \frac{1}{(x-x_k)^2} \;,\quad \bb{N} \ni M \geq3 \;,\quad J \in \bb{N} \;,
$$
with trivial monodromy about $x_1$, $\dots$, $x_J$. The latter potentials are of interest to us because they coincide with the monster potentials,
if $2\alpha+2=M$ and the set of poles $x_1$, $\dots$, $x_J$ is symmetric under rotation by $e^{\frac{2\pi i}{M}}$.

In Section 10, we  test numerically our results and, in Section 11, we wrap up the paper by laying out future research directions.

The paper is essentially self-contained and it was designed to be readable by theoretical physicists and mathematicians alike. We hope to have succeeded in this.

\begin{remark}
We recall that the BLZ conjecture was extended to the generalised Quantum $\mathfrak{g}$-KdV model, where $\mathfrak{g}$ is an affine Kac-Moody algebra, in the seminal work by Feigin and Frenkel \cite{FF11} (the Quantum KdV model coincides with the case $\mathfrak{g}=\mathfrak{sl}_2^{(1)}$).
Building on this work and using heavily the theory of opers, the analogue of the monster potentials and of the BLZ system \eqref{eq:algblz} are explicitly described in \cite{mara18} in the case $\mathfrak{g}=g^{(1)}$, with $g$ a simply-laced simple Lie algebra. The sub-case $\mathfrak{g}=\mf{sl}_3^{(1)}$ is treated with more elementary techniques in \cite{mara19}. In the present work, we do not study such generalisations.
\end{remark}

\subsection*{Acknowledgements}
The research was initially inspired
by the paper \cite{langlands95}, which eventually played a minor role in our analysis.

Davide Masoero gratefully acknowledges support from the Simons Center for Geometry and Physics, Stony Brook University
at which the research for this paper began in November 2018. 

We are very grateful to Sergei Lukyanov who read an early draft of the paper and showed us
his support and appreciation. The authors also thanks Vladimir Bazhanov, Clare Dunning, Davide Fioravanti, Stefano Negro, Andrea
Raimondo, and Roberto Tateo for useful discussions and feedback.

The authors are partially supported by the FCT Project PTDC/MAT-PUR/ 30234/2017 `Irregular
connections on algebraic curves and Quantum Field Theory', and by the
FCT Investigator grant IF/00069/2015 `A mathematical framework for the ODE/IM correspondence'.

\section{First Considerations}
Here we review the derivation of the BLZ system, and we briefly consider the case of roots coalescing to a value different from zero. Finally, we discuss
the case of one or more roots coalescing to zero and we show that this cannot happen in the large $L$ limit. 

To start with, we review the concept of trivial monodromy about a Fuchsian singularity. Assume that we are given a function $V$, analytic in the punctured neighbourhood $D_{x_0}$ of a point $x_0 \in \C$. We choose a small loop $c:[0,1] \to D_{x_0}$ with winding number $1$ about $x_0$ and, for any $E \in \C$, we let $X_E\cong\C^2$ be the space of solutions of the differential equation
\begin{equation}\label{eq:schr}
\psi''(x)=\big(V(x)-E)\psi(x) \;,
\end{equation}
whose domain is a simply-connected neighbourhood of $c(0)$. The monodromy about $x_0$ of equation \eqref{eq:schr} is the linear operator $M_E:X_E\to X_E$ that associates to each solution its analytic continuation along $c$. 
\begin{definition}
Let $V$ be an analytic function with a double pole at $x_0 \in \C$. We say that
$V$ has trivial monodromy at $x_0$ if the monodromy $M_E$ of \eqref{eq:schr} is the identity operator for all $E \in \bb{C}$.
\end{definition}
We have the following well-known local characterisation of potentials with trivial monodromy at a point, in terms of the coefficients of the Laurent expansion.
\begin{theorem}[\cite{duistermaat86}]\label{thm:dui}
Let $V(x)$ be a meromorphic function with a double pole at $x_0$. The monodromy of $V$ at $x_0$ is trivial if and only if 
\begin{equation}\label{eq:cs}
 V(x)=\frac{d(d+1)}{(x-x_0)^2}+\sum_{k\geq-1}c_k (x-x_0)^k \;,
\end{equation}
where $d \in \bb{N}^*$ and the coefficients $c_k \in \bb{C} $ satisfy the following
identities
$$c_{2k-1}=0 \;,\quad \forall k=0,\dots,d \;.$$
\begin{proof}
See \cite{duistermaat86}.
\end{proof}
\end{theorem}
The BLZ system is a corollary of the above theorem. 
\begin{proposition}[\cite{BLZ04}]\label{pro:BLZ}
The monster potential \eqref{eq:V}, where  $z_1$, $\dots$, $z_N$ are pairwise distinct and non-zero, has, for all $k \in 1,\dots,N$, trivial monodromy at any $x \in \Ca$ such that $x^{2\alpha+2}=z_k$, if and only if $z_1$, $\dots$, $z_N$ satisfy the BLZ system \eqref{eq:algblz}. 

\begin{proof}
For every $k$, the points $x \in \Ca$ such that $x^{2\alpha+2}=z_k$ are of the form  $\gamma_{\alpha}^l x_k$, with $l \in \mathbb{Z}$, for some $x_k$ such that $x_k^{2\alpha+2}=z_k$.
From \eqref{eq:tau} it follows that,
at the points $\gamma_{\alpha}^l x_k$, $V_P$ admits the Taylor expansion \eqref{eq:cs} with $d=1$ and $c_{-1}=0$.
According to Theorem \ref{thm:dui}, the monodromy is trivial at $\gamma_{\alpha}^l x_k$ if and only if
the coefficient $c_1=0$ of the same Laurent expansion vanishes.
In order to compute the coefficient $c_1$, we split the potential $V$ into $3$ terms:
\begin{enumerate}
\item  `Self-interaction': $\displaystyle{-2 \frac{d^2}{d x^2}  \log(x^{2\alpha+2}-z_k)\;;}$
\item   `Not-Self-Interaction': $\displaystyle{-2 \frac{d^2}{d x^2} \sum_{j\neq k} \log(x^{2\alpha+2}-z_j) \;;}$
\item  `External field': $\displaystyle{\frac{L}{x^2}+x^{2a} \;.}$
\end{enumerate}
Each of the above term yields the following contribution to the coefficient $c_1$:
\begin{align*}
& \gamma_{\alpha}^{-3 l}x_k^{-3} \times  \frac{4 \alpha^2-1}{2} \;, \\
& \gamma_{\alpha}^{-3 l}x_k^{-3} \times \sum_{j\neq k} \frac{-8 (1 + \alpha ) z_k \big(z_k^2 + (3 + 7 \alpha + 2 \alpha^2) z_k z_l + 
  \alpha (1 + 2 \alpha) z_l^2\big)}{(z_k-z_l)^3} \;, \\ 
& \gamma_{\alpha}^{-3 l} x_k^{-3} \times  \big( -2 L + 2 \alpha z_k \big) \;.
\end{align*}
Collecting the three terms we obtain that $c_1$ vanishes if and only if
$$
\sum_{j \neq k}
\frac{z_k\big( z_k^2 +(3+\alpha)(1+2\alpha)z_k z_j+
\alpha (1+2 \alpha)
z_j^2 \big)}{ (z_k-z_j)^3}- \frac{\alpha z_k}{ 4 (1+\alpha)}+
\Delta=0 \;,
$$
where $\Delta$ is as per \eqref{eq:cdelta}. Notice in particular that the above expression vanishes at $x_k$ if and only if it vanishes at $\gamma_{\alpha}^lx_k$ for every $l\in \mathbb{Z}$ (a fact in accordance with the Dorey-Tateo symmetry). Therefore the thesis is proven.
\end{proof}

\end{proposition}

\subsection*{Coalescing roots}
The space of monic polynomials of degree $N$, such that $0$ is not a root, is divided in $p(N)$ strata:
for every partition $\nu=(\nu_1,\dots,\nu_j)$ of $N$, the corresponding stratum $\mc{S}^*_{\snu}$ consists of those polynomials $P$ with $j\leq N$ distinct (and all non-zero) roots $z_1$, $\dots$, $z_j$ with multiplicity $\nu_1$, $\dots$, $\nu_j$.

By definition, the intersection of $\overline{\mc{B}}_{N,\alpha}$ with the generic stratum, the one where all roots are simple, is $\bna$. The problem of roots-coalescing to a non-zero value is the problem of understanding $\Bna \cap \mc{S}^*_{\snu}$ for all deeper strata. A full discussion of this interesting problem is outside the scope of this paper, however a simple counting argument, which follows from Theorem \ref{thm:dui}, gives us some hint of what we should expect.

We reason as follows.
\begin{itemize}
 \item Since the monodromy at any root $z_1$, $\dots$, $z_j$ is trivial, it follows from equation \eqref{eq:cs} of Theorem \ref{thm:dui}
 that each multiplicity must be a triangular number,
i.e. $\nu_k=\frac{d_k(d_k+1)}{2}$ for some integers $d_k \geq 1$ with $k=1,\dots,j$.
\item After Theorem \ref{thm:dui}, at each point $x_k$ such that $x_k^{2\alpha+2}=z_k$, the coefficients $c_{2l-1}$, $l=0,\dots,d_k$ of the Laurent expansion of $V_P$ at $x_k$ must vanish. Since by construction of $V_P$, the coefficient $c_{-1}$ always vanishes, it is straightforward to see that the roots of the polynomial $P$ must satisfy $\sum_{k=1}^N d_k$ algebraic relations. This is an over-determined algebraic system if $j<N$.
\item The intersection of $\overline{\mc{B}}_{N,\alpha}$ with the `second stratum', consisting of polynomials with $N-3$ simple roots and $1$ triple root, should be a finite non-empty set. In fact, such a stratum has dimension $N-2$ and the trivial monodromy conditions are a system of $N-1$ equations: the trivial monodromy conditions on the second stratum are a complete system of algebraic equations for the roots of the polynomials and the parameter $L$ of potential. 
\item From the same counting argument, we expect that, for generic values of $\alpha$,
the intersection of $\overline{\mc{B}}_{N,\alpha}$ with any deeper strata is empty.
\end{itemize}

\subsection*{Zero roots}
We now turn our attention to the case of $m \geq 1$ roots coalescing to zero.
We show that this cannot happen in the large momentum limit, if $N$ is fixed. In fact we have the following Lemma.
\begin{lemma}\label{lemma:smallz}
Fix $N \in \mathbb{N}$ and $\alpha \in \bb{R}^+$. The subset of those $L \in \C$ such that there exists a $(P,L) \in \Bna$ with $P(0)=0$, is finite.
\begin{proof}
We assume that there is a sequence of points $(P^{\sn},L^{\sn}) \in \bna$ converging to $(P,L) \in \C^{N+1}$, where $P$ is a monic polynomial
of degree $N$ such that $0$ is a root with multiplicity $m\geq 1$.
We denote by $z_{m+1}$, $\dots$, $z_N$ the non trivial roots of $P$ and write
\begin{align*}
&  V^{\sn}_P(x)=x^{2\alpha}+ \frac{L^{\sn}}{x^2}-2 \frac{d^2}{dx^2} \log P^{\sn}(x^{2\alpha+2}) \;, \\
& V_P(x)=x^{2\alpha}+ \frac{L+ (2\alpha+2) m}{x^2}-2\sum_{k=m+1}^N \frac{d^2}{dx^2} \log \big( x^{2\alpha+2}-z_k\big) \;.
\end{align*}
Let us now study the local monodromy of $V^{\sn}_P$ and $V_P$ about $0$. To this aim we write,
\begin{equation}\label{eq:00}
L^{\sn}=\ell^{\sn}(\ell^{\sn}+1) \;,\quad L= \ell(\ell+1) \;,
\end{equation}
for unique $\ell^{\sn},\ell$ such that $\Re \ell^{\sn}, \Re \ell \geq-\frac12$.

Since the point $x=0$ is a ramification point of the potential, one needs to introduce the monodromy in the infinite-dimensional space of solutions of the equation \eqref{eq:schr} which depend analytically on the energy $E$. The monodromy is defined via the following formula $M \psi(x,E)=\psi(\gamma_{\alpha} x, \gamma_{\alpha}^{-2} E)$, where $\gamma_{\alpha}$ is as per \eqref{eq:gamma}. Summarising what is known from the ODE/IM literature (see \cite[Section 5]{mara18} for a rigorous discussion), the monodromy operator $M$ is known to have two (possibly coinciding) eigenvalues, which have the same dependence on the residue of the potential at $0$ as the eigenvalues of the monodromy about a Fuchsian singularity.
In fact, we have that the monodromy operator $M$ for the potential $V^{\sn}_P$ has eigenvalues $\lambda^{\sn}_{\pm}=e^{\pm 2\pi i \ell^{\sn}}$, where $\ell^{\sn}$ is as per \eqref{eq:00}, while the monodromy operator for the potential $V_P$ has eigenvalues
$\lambda_{m,\pm}=e^{\pm 2\pi i \ell_m}$, where 
\begin{equation}\label{eq:01}
 \ell_m(\ell_m+1)= L +(2\alpha+2) m \;,\quad \Re \ell_m \geq - \frac12 \;.
\end{equation}

Since the potentials $V_P^{\sn}$ have trivial monodromy at all roots which are coalescing to $0$, the eigenvalues of the monodromy of $V_P$ must coincide with the limit of the eigenvalues of the monodromy for the potentials $V_P^{(n)}$. This means that $\lambda_{m,\pm}=e^{\pm i 2\pi \ell}$ or $\lambda_{m,\pm}=e^{\mp i 2 \pi \ell}$. It follows that
 \begin{equation}\label{eq:02}
  \ell_m =\ell + \frac{d}{2} \;,\quad \mbox{ for some } d \in \bb{N}^* \;.
 \end{equation}
Combining (\ref{eq:00},\ref{eq:01},\ref{eq:02}), we obtain
\begin{equation}
 \ell=-\frac12-\frac{d}{2} + \frac{(2\alpha+2)m}{d} \;,\quad \Re \ell\geq \frac12 \;,\quad d \in \bb{N}^* \;,
\end{equation}
For any given $m$, there are a finite number of $\ell$ which satisfy the above system of one equation and one inequality. It follows that the subset of those $L \in \C$ such that there exists a $(P,L) \in \Bna$ for which $0$ is a root of order $m$, is finite. Since $1\leq m \leq N$, the thesis is proven.
\end{proof}

\end{lemma}

\begin{remark}
The phenomenon of roots coalescing to $0$ is widely believed to correspond, on the IM side of the correspondence, to
those values of the highest weight such that the Verma module is reducible (or degenerate).
We agree with this interpretation. Let us in fact assume, as in Lemma \ref{lemma:smallz} above, that
there is a sequence of points $(P^{\sn},L^{\sn}) \in \bna$ converging to $(P,L) \in \C^{N+1}$, where $P$ is a monic polynomial
of degree $N$ such that $0$ is a root with multiplicity $m\geq 1$.
The limit potential is
\begin{align*}
 V_P(x)=x^{2\alpha}+ \frac{L+ (2\alpha+2) m}{x^2}-2\sum_{k=m+1}^N \frac{d^2}{dx^2} \log \big( x^{2\alpha+2}-z_k\big) ,
\end{align*}
which is, by definition, a monster potential with $N-m$ roots and momentum $L \to L+ (2\alpha+2) m$. Since the parameter $L$ and
the highest weight $\Delta$ are related by equation \eqref{eq:cdelta}, this corresponds to a shift in the highest weight
$\Delta \to \Delta + m$.
In the particular case that $m=N$, $V_P(x)$ is the ground state potential with $\Delta = \Delta+ N$.
Therefore $V_P$ corresponds to a null vector, i.e. a primary field of level $N>0$.

On the contrary, we do not expect that the coalescence of roots to a value different from zero has any representation-theoretical interpretation.

Finally, we remark that one can \textit{blow-up} $\Bna$ in order to obtain a new closed set $\Bna'$,
which does not contain the potentials obtained by the coalescence of roots
to $0$, but does contain the potentials obtained by the coalescence of roots to a value different from $0$:
If $\mc{B}'_{N,\alpha} \subset \C^{N+2}$ consists of the ordered triples $(P,\frac{1}{P(0)},L)$
where $L$ is a complex number and $P$ is a monster potential,
$\Bna'$ is the closure of $\mc{B}_{N,\alpha}'$ in $\C^{N+2}$.
\end{remark}

\section{Rational extensions of the harmonic oscillator and related problems}\label{section:harmonic}
In this Section, we briefly \footnote{This subject is a very active topic of investigation, see e.g. \cite{clarkson20,dunning20}.}
discuss the problem of the rational extensions of the harmonic oscillator, a problem which plays a major role in the large momentum limit of the BLZ equation. It indeed provides the next-to-leading term in the large $L$ expansion of its solutions, as per \eqref{eq:zkexpansion}. We also touch upon two related problems, the rational extensions of the perturbed harmonic oscillator and the Airault-McKean-Moser locus, which govern the second-order corrections, whenever the polynomial $P^{\snu}(t)$ has multiple roots.

Let $P$ a monic polynomial of degree $N$. We say that the potential
\begin{equation}\label{eq:extendedharmonic}
U_P(t)=t^2-2 \frac{ d^2}{dt^2} \log P(t) \;,
\end{equation}
is a \textit{rational extensions of the harmonic oscillator} of level $N$, if it has trivial monodromy at all zeroes of $P$.

The characterisation of all rational extensions of the harmonic oscillator was obtained by Oblomkov \cite{oblomkov99} (after a conjecture by Veselov).
\begin{theorem}[\cite{oblomkov99}]\label{thm:oblomokov}
The potential \eqref{eq:extendedharmonic} is a rational extension of the harmonic oscillator if and only if there exists a partition $\nu$ of $N$ such that $P= P^{\snu}$.

Therefore, for each $N$, there are exactly $p(N)$ rational extensions of the harmonic oscillator, which we denote
by $U^{\snu}(t)$.
\end{theorem}

\subsection*{Spectrum of a rational extension}
Let $\nu=(\nu_1,\dots,\nu_j)$ be a partition and $U^{\snu}(t)$ the corresponding rational extension of the harmonic oscillator. We say that $\lambda$ belongs to its $\bb{R}$-spectrum if there exists a \textbf{meromorphic} solution of the following boundary value problem
\begin{equation*}
 \psi_{\lambda}''(t)=\big(U^{\snu}(t)-\lambda\big)\psi_{\lambda} \;,\quad \lim_{t\to \pm \infty} \psi_{\lambda}(t)=0 \;.
\end{equation*}
The $\bb{R}$-spectrum can be computed explicitly since $U^{\snu}(t)$ is obtained by a sequence of Darboux transformations from the harmonic oscillator (whose $\bb{R}$-spectrum  is $\lambda_n=2n+1$, $n \in \bb{N}$). In fact, following \cite{oblomkov99}, we obtain that the $\bb{R}$-spectrum of $U^{\snu}(t)$ is discrete and simple, and its points
are
\begin{equation}\label{eq:spectrumnu}
 \lambda^{\snu}_n= 1-2j+ 2 n \;,\quad n \in \bb{N}^{\snu} \;,
\end{equation}
where $\bb{N}^{\snu}$ is the sequence obtained from $\bb{N}$ by deleting the numbers $(\nu_j,\nu_{j-1}+1,\dots,\nu_1+j-1)$.
These are the degrees of the Hermite polynomials entering into the definition of $P^{\snu}(t)$ according to formula \eqref{eq:Pnu}.

\subsection*{Four-fold symmetry}
The harmonic oscillator has a four-fold symmetry $t \to i t$ which has a non-trivial action on the set of rational extensions.
\begin{lemma}[\cite{felder12}]
\label{thm:felder}
For partition $\nu$ of $N$, we have that
\begin{itemize}
\item $P^{\snus}(t) = (-i)^N\,P^{\snu}(it) \;;$
\item $P^{\snu}(-t) = (-1)^N\,P^{\snu}(t) \;.$
\end{itemize}

Therefore
\begin{itemize}
\item $t$ is a root of $P^{\snu}(t)$ if and only if $-t$ is a root of $P^{\snu}(t) \;;$
\item $t$ is a root of $P^{\snu}(t)$ if and only if $i  t$ is a root of $P^{\snus}(t) \;.$
\end{itemize}
\end{lemma}

\subsection*{Multiplicity of poles}
We begin with a definition.
\begin{definition}\label{def:fproperty}
We say that a partition $\nu$ of $N$ (or a polynomial $P^{\snu}(t)$) is non-degenerate if $P^{\snu}(t)$ has $N$ distinct zeroes.

Otherwise, we say that the partition and its polynomial $P^{\snu}(t)$ are degenerate.

We say that a partition admits the F-property if the following holds: $t$ is root of $P^{\snu}$ and $t \neq0$ then $t$ is a simple root. 
\end{definition}
The authors of \cite{felder12} noted the following property: the multiplicity of the root $t=0$ of $P^{\snu}(t)$ is $\frac{d(d+1)}{2}$,
where $d$ is the difference between the number of odd and even elements of the sequence $(\nu_{j},\nu_{j-1}+1\dots,\nu_{1}+j-1)$. They moreover made the following conjecture, which has the strongest numerical evidence \cite{felder12}.\footnote{The conjecture is proven in the case of rectangular partitions, namely when $N=n \times m$ and $\nu=(n,\dots,n)$, in which case $P^{\snu}(t)$ coincides with the generalised Hermite polynomial $H_{m,n}(t)$, see \cite{maro18}.}
\begin{conjecture}[\cite{felder12}]\label{conj:felder}
Any partition $\nu$ admits the F-property: there cannot be a non-zero root of $P^{\snu}$ with multiplicity greater than one.
\end{conjecture}
Now, assume that $\nu$ admits the F-property, $P^{\snu}(t)$ has $\frac{d(d+1)}{2}$ trivial roots and $N-\frac{d(d+1)}{2}$ non-trivial pairwise distinct root, $t_1,\dots,t_{N-\frac{d(d+1)}{2}}$. After Theorem \ref{thm:dui}, these solve the following system of algebraic equations
\begin{eqnarray}\label{eq:algrational}
& \displaystyle{2t_k - \sum_{j\neq k}\frac{4}{(t_k-t_j)^3} - \frac{2 d(d+1)}{t_k^3}=0 \;,\quad k=1,\dots,N-\frac{d(d+1)}{2} \;,} \\
& \sum_{k} t_k^{-(2l+1)}=0 \;,\quad l=1,\dots, d \;.
\end{eqnarray}
Conversely every solution of the above system yields a degree $N$ rational extension of the harmonic oscillator satisfying the F-property. 

\subsection{Perturbed harmonic oscillator}\label{sub:perturbed}
It is a beautiful problem the one of studying rational extensions of the harmonic oscillator perturbed by a small cubic term. Here we assume that we are given a series $\sum_{n\geq 4} c_n t^{n}$, $c_n \in \C$ with a positive radius of convergence, say $R$, and
we define the perturbed potential to be
\begin{equation}\label{eq:perturbedh}
 U(t;\e)= t^2+\e \kappa t^3 + \sum_{n\geq 4} \e^{n-2} c_n t^{n} \;,\quad  \kappa \in \C \;,
\end{equation}
which has radius of convergence $\e^{-1} R$. We want to find all monic polynomials $P(t;\e)$ of degree $N$ such that the potential
\begin{equation}\label{eq:rationale}
 U_P(t;\e)=U(t;\e)-2\frac{d^2}{dt^2}\log P(t;\e) \;,
\end{equation}
has trivial monodromy about all roots of $P(t;\e)$.
According to Theorem \ref{thm:dui}, the roots $t_k$ must satisfy the following system provided they are all distinct
\begin{equation}\label{eq:tperturbed}
 2 t_k + 3 \e \kappa\, t_k^2+ \sum_{n \geq 4}  n c_n \e^{n-2} t_k^{n-1} - \sum_{j\neq k}\frac{4}{(t_k-t_j)^3}=0 \;,\quad k=1,\dots,N \;.
\end{equation}
In fact, according to our numerical and analytical experiments, all roots
are distinct if $\e \neq 0$ is small enough and the coefficient $\kappa$ does not vanish.
We omit any further discussion of this problem, however we will see below that the equation \eqref{eq:tperturbed} coincides up to order $O(\e^2)$ with the equation describing the roots of the BLZ equation in the large $L$ limit, with $\e=L^{-\frac14}$ and $\kappa=\frac{(5-2\alpha) }{3(2\alpha+2)^{\frac14}}$. We think of it as a toy model of the BLZ system in the large $L$ limit.

\subsection{The Airault-McKean-Moser locus} \label{sec:AirMcMos} 
A rational extension of the trivial potential $U(t)=0$ is a potential of the form
\begin{equation}\label{eq:extensionvoid}
 W_P(t)=-2\frac{d^2}{dt^2} \log P(t) \;,\quad P \mbox{ monic polynomial} \;,
\end{equation}
which has trivial monodromy at all roots of $P$.

It is well-known that rational extensions exist if and only if the degree $N$ of $P$ is a triangular number, namely $N=\frac{d(d+1)}{2}$ for some $d\geq1$. In this case the set of rational extensions is naturally an algebraic variety isomorphic to $\C^{d+1}$, the Airault-McKean-Moser locus, which is invariant under the KdV flows; see \cite{airault77}.

According to Theorem \ref{thm:dui}, the roots of the polynomial $P$ satisfy the following system of algebraic equations
\begin{equation}\label{eq:moser}
 \sum_{j\neq k}\frac{1}{(t_k-t_j)^3}=0 \;,\quad k=1,\dots,N=\frac{d(d+1)}{2} \;,
\end{equation}
provided they are all distinct. In the case $d=2$, the set of solutions of the above equation is particularly simple to describe
\begin{equation}\label{eq:locusd2}
\lbrace (t_1,t_2,t_3) =  (a+b,a+b\omega,a+b\omega^2) \mbox{ with } a,b\in\C, b\neq 0 \mbox{ and } \omega=e^{2\pi i/3} \rbrace \;,
\end{equation}
as it can be checked by hand.

\section{On large solutions}
Here we prove the first important results of the paper. We prove that the solutions of the BLZ system \eqref{eq:algblz} are bounded if $L$ is bounded, while
\begin{equation}\label{eq:zkla}
 z_k = \frac{L}{\alpha} + o\big(L\big) \;,\quad \forall k\in\lbrace1,\dots,N\rbrace \mbox{ as } L \to \infty \;.
\end{equation}
From the latter limit, it follows that in the large momentum regime the poles of the higher state potentials condensate about the \textit{minima} of $V_G$.
In fact, $V'_G(x)=0$ if and only if $x^{2\alpha+2}=\frac{L}{\alpha}$. The rest of this paper emanates from this crucial point.

We have the following
\begin{proposition}\label{pro:largez}
Fix $\alpha \in \bb{R}^+$ and $N \in \mathbb{N}$.

\begin{enumerate}
 \item For every $L \in \mathbb{C}$, there exists a $C_{L}>0$ such that if $(P,L) \in \Bna$ and $z_1$, $\dots$, $z_N$ are its roots, then
 $|z|<C_{L}$.
\item  Let $L^{\sn}$ be a diverging sequence of complex numbers  $L^{\sn} \to \infty$ and $(P^{\sn},L^{\sn})$ a sequence
of points in $\Bna$. Let moreover $z_1^{\sn}$, $\dots$, $z_N^{\sn}$ denote the corresponding sequence of roots of $P^{\sn}$. We have that
\begin{equation*}
      \lim_{n \to \infty} \frac{\alpha z_k^{\sn}}{L^{\sn}} \to 1 \;,\quad \forall  k\in\lbrace1,\dots,N\rbrace  \;.
     \end{equation*}
\end{enumerate}
\end{proposition}
\paragraph*{Proof}
To start with, we notice that it is sufficient to prove the theorem for $\bna$ since by definition $\Bna$ is the closure of $\bna$.
I.e., we can assume that $z_1$, $\dots$, $z_N$ are solutions of the BLZ system \eqref{eq:algblz}.

We divide the $k$-th equation of the BLZ system \eqref{eq:algblz} by $-\frac{\alpha}{4(1+\alpha)}z_k^{\frac{3}{2+2\alpha}}$ to obtain
\begin{align}\nonumber
& \sum_{j\neq k}I(z_k,z_j) + U(z_k)=0 \;,\quad k=1,\dots,N \;, \\ \nonumber
& I(z,w)=\frac{ -4 (1+\alpha)}{\alpha}\frac{z\big(z^2 + (3+\alpha)(1+2\alpha)z w+\alpha(1+2\alpha)w^2\big)}{z^{\frac{3}{2+2\alpha}}(z-w)^3} \;,
\\ \label{eq:BLZlarge}
& U(z)= z^{-\frac{3}{2+2\alpha}}\left( z -\frac{L}{ \alpha} -\frac{1-4\alpha^2}{4 \alpha} \right) \;.
\end{align}
Next we state and prove two preparatory lemmas about \eqref{eq:BLZlarge}.

\begin{lemma}\label{lem:boundedinteratcion}
 Let $\left(z^{\sn},w^{\sn}\right)$ be a sequence of a pair of complex numbers such that 
$\inf_{n \to \infty}\big|\frac{w^{\sn}}{z^{\sn}}-1\big|>0 $.

There exists a $C>0$ such that
\begin{equation}\label{eq:boundedint}
\left| I(z^{\sn},w^{\sn}) \right| \leq C |z^{\sn}|^{-\frac{3}{2+2\alpha}} \;,\quad \forall n\in\mathbb{N} \;.
\end{equation}
\begin{proof}
A straightforward computation leads to the following identity
 \begin{equation*}
%  \label{eq:Izwfracted}
 I(z,w)=\frac{ -4 (1+\alpha)}{\alpha z^{\frac{3}{2+2\alpha}}} 
 \frac{ 4 (1 +\alpha)^2-3 (1 + 3 \alpha + 2 \alpha^2) (1-\frac{z}{w}) + \alpha(1+2\alpha) (1-\frac{z}{w})^2}{(1-\frac{z}{w})^3} \;,
 \end{equation*}
 from which the thesis follows.
\end{proof}
\end{lemma}

\begin{lemma}\label{lem:boundedinteratcionsymm}
Let $(z^{\sn},w^{\sn})$ be a sequence of a pair of non-zero complex numbers such that
 \begin{itemize}
 \item  $z^{\sn}\neq w^{\sn}$ for all $n\in\mathbb{N}\;;$
  \item  $\displaystyle{\lim_{n \to \infty}\frac{w^{\sn}}{z^{\sn}}=1 \;.}$
 \end{itemize}
 Then there exists a $C>0$ such that 
 \begin{equation}\label{eq:boundedsymm}
 \left| I(z^{\sn},w^{\sn})+I(w^{\sn},z^{\sn}) \right| \leq C |z^{\sn}|^{-\frac{3}{2+2\alpha}} \;,\quad \forall n\in\mathbb{N} \;.
 \end{equation}
\begin{proof}
We have
$$J(z,w):=I(z,w)+I(w,z)=\frac{w^{\frac{3}{2+2\alpha}} z (z^2+ A z w+ b w^2) + z^{\frac{3}{2+2\alpha}} w( w^2+ A w z + B w^2)}{(x-w)^3 
z^{\frac{3}{2+2\alpha}} w^{\frac{3}{2+2\alpha}}},$$
with  $ A =(3+\alpha)(1+2\alpha)$, $B=\alpha(1+2\alpha)$. We expand the numerator of $J(z,w)$ at $w=z$ to obtain
$$
J(z,w)=z^{-\frac{3}{2+2\alpha}}  \left(\sum_{m\geq0} C_m w^{-m} (w-z)^{m}\right)
=z^{-\frac{3}{2+2\alpha}}  \left(\sum_{m\geq0} C_m \left(1-\frac{z}{w}\right)^{m}\right) \;,
$$
for some $C_m$ such that $\limsup_{m \to \infty}\frac{|C_{m+1}|}{|C_m|}=1$.  The thesis follows immediately from the above expansion.

\end{proof}

\end{lemma}

Now we suppose to have a sequence of complex numbers $L^{\sn}$ as well as a sequence of corresponding solutions of the BLZ system, $z_1^{\sn}$, $\dots$, $z_N^{\sn}$. Since $\C$ is sequentially compact, we can assume that the sequence $z_k^{\sn}$ as well as the sequence $\frac{z_k^{\sn}}{z_j^{\sn}}$ have a well-defined limit in $\C$ for all $k,j\in\lbrace1,\dots,N\rbrace$.
  
Fix $l \in\lbrace1,\dots,N\rbrace$. We define $J_l$ to be the subset of indices such that $\frac{z_l^{\sn}}{z_j^{\sn}} \to 1$. Summing the equations \eqref{eq:BLZlarge} for all the indices $k \in J_l$, we obtain
 \begin{equation*}
 \sum_{k\neq j \in  J_l}\frac{I(z_k,z_j)+I(z_j,z_k)}{2}+ \sum_{k \in J_l, j \notin J_l} I(z_k,z_j) + \sum_{k \in  J_l}U(z^{\sn}_k)=0 \;.
 \end{equation*}
Inserting estimates (\ref{eq:boundedint},\ref{eq:boundedsymm}) into the latter equation, we deduce that there exists a $C$ such that
 \begin{equation}\label{eq:estimatelarge}
 \left| \sum_{k \in  J_l}U(z^{(n)}_k) \right| \leq C |z_l^{\sn}|^{\frac{-3}{2+2\alpha}}\;,\quad \forall n\in\mathbb{N} \;.
 \end{equation}

To prove (1), we impose that $L^{\sn}=L$ is fixed and argue by contradiction, i.e. we assume that there exists a sequence of solutions such that
$|z_l^{\sn}|\to \infty$ for some $l \in\lbrace1, \dots,N\rbrace$. Then, after equation \eqref{eq:BLZlarge}, we have that $$\sum_{k \in  J_l} U(z_k^{\sn})= |J_l| \big(z^{\sn}_l\big)^{1-\frac{3}{2\alpha+2}} +o\big( |z^{\sn}_l|^{1-\frac{3}{2\alpha+2}}\big) \;,$$
where $|J_l|\geq 1$ is the cardinality of $J_l$. The latter asymptotics contradicts \eqref{eq:estimatelarge}.

To prove (2), we multiply \eqref{eq:estimatelarge} by $|z_l^{\sn}|^{\frac{3}{2+2\alpha}}$ and we obtain 
$$ \left| \sum_{k \in J_l} \left(\frac{z_l^{\sn}}{z_k^{\sn}}\right)^{\frac{-3}{2+2\alpha}} \left(z^{\sn}_k
-\frac{L^{\sn}}{ \alpha} -\frac{1-4\alpha^2}{4 \alpha}\right) \right| \leq C \;,\quad \forall n\in\mathbb{N} \;.$$
Since $\frac{z_k^{\sn}}{z_l^{\sn}} \to 1$, from the latter inequality it follows that there exits a $C'>0$ such that
$$ \left|  z^{\sn}_l -\frac{L^{\sn}}{ \alpha} -\frac{1-4\alpha^2}{4 \alpha} \right| \leq C' \;,\quad \forall k \in J_l \;,\quad \forall n\in\mathbb{N}\;.$$ 
Since, by hypothesis $L^{\sn}\to \infty$, this implies $\lim_{n \to \infty}\frac{\alpha z^{\sn}_k }{L^{\sn}}\to 1$ for all $k \in J_l$.
Since $l$ is arbitrary, the thesis is proven.
\begin{flushright}
 $\square$
\end{flushright}

\begin{remark}
 The reader may wonder why we have chosen to divide the $k$-th equation of the BLZ system by the term $z_k^{\frac{3}{2+2\alpha}}$ to prove
 Proposition \ref{pro:largez}. This was done in order to obtain  the estimate \eqref{eq:boundedsymm}. In fact, the reader may check that for any other choice of a factor (but for a constant multiple  of the one we chose) the analogous of the estimate \eqref{eq:boundedsymm} would contain a pre-factor $(z-w)^{-1}$ which cannot be a-priori
 bounded.
\end{remark}

\section{The higher-state-potentials in the large momentum limit}
In the previous section, see Proposition \ref{pro:largez}(2),  we have shown that roots of the higher-state-potentials, $z_1$, $\dots$, $z_N$, satisfy
the asymptotics $z_k=\frac{L}{\alpha} +o\big(L\big)$ as $L \to \infty$ (see equation \eqref{eq:zkla}). This implies that the poles of the higher-state-potentials
condensates about the points $x \in \Ca$ such that $x^{2\alpha+2}=\frac{L}{\alpha}$,
i.e. the roots of $V'_G(x)=0$. 
These have the expression
\begin{equation}\label{eq:minimum}
m_l=\gamma_{\alpha}^l \left(\frac{L}{\alpha}\right)^{\frac{1}{2\alpha+2}} \in \Ca \;,\quad l \in \mathbb{Z} \;,
\end{equation}
where $\gamma_{\alpha}$ is as per \eqref{eq:gamma} and $m_0$ is a given fixed root of
$\big(\frac{L}{\alpha}\big)^{\frac{1}{2\alpha+2}}$.

In this Section we expand the higher-state-potential $V_P$ about one of these minima, say $m_0$ (the choice is immaterial since the potential enjoys the Dorey-Tateo symmetry \eqref{eq:DT}) and show that, in some sense to be made precise below, the higher-state-potentials converge to rational extensions of the harmonic oscillators; see Theorem \ref{thm:toharmonic}. In particular, we show that the asymptotics \eqref{eq:zkexpansion} holds.

\subsection{Expanding higher-state-potentials in the large $L$ limit}
In this subsection, we change the variable of the Schr\"odinger equation \eqref{eq:schr} with a higher state potential $V_P$, as per \eqref{eq:Vgen},
to obtain a finite expression in the large $L$ limit. Before we enter into the details of the computations, we remark that in order to find a meaningful expansion, we need to change both the variable $x$ in the Schr\"odinger equation and the unknowns $z_1$, $\dots$, $z_N$ which are the roots of the polynomial $P$ of the potential  \footnote{This is an example of a multiple scale limit.}. For sake of clarity, let us write
\begin{equation}\label{eq:Vzk}
V(x;z_1,\dots,z_N):=V_P(x)\;,\mbox{  with} \quad P=\prod_{k=1}^N (x-z_k) \;,
\end{equation}
If we introduce the linear change of variable in the Schroedinger equation
\begin{equation}\label{eq:Axt}
x(t)=A t+m_0 \;,
%\mbox{ or } \quad t(x)=A^{-1}(x-m_0),
\end{equation}
with $m_0$ as per \eqref{eq:minimum}, and for some $A \neq 0$ depending on $L$, for consistency we also define $z_1$, $\dots$, $z_N$ in terms of a new set of unknowns $t_1$, $\dots$, $t_N$ according to the formula
\begin{equation}\label{eq:Azk}
 z_k:=\big(x(t_k)\big)^{2\alpha+2}=\big(At_k+m_0\big)^{2\alpha+2}=
 m_0^{2\alpha+2}\left(1+\frac{ A t_k}{m_0}\right)^{2\alpha+2}=\frac{L}{\alpha}\left(1+\frac{A t_k}{ m_0}\right)^{2\alpha+2} \;,
\end{equation}
where we have used \eqref{eq:Axt} and \eqref{eq:minimum}.
Therefore, taking into consideration the law of a potential under an affine change of co-ordinates
\footnote{$-\psi''(x)+( V(x)-E )\psi(x)=0$ is mapped into $-\psi''(t)+( A^{2}V(At+m_0) -A^{2} E) \psi(t)=0$.}
the resulting Schr\"odinger equation is
\begin{equation}\label{eq:AVx}
-\psi''(t)+ \left( A^{2}V\left(At+m_0;\frac{L}{\alpha}\biggl(1+\frac{A t_1}{ m_0}\biggr)^{2\alpha+2},\dots,\frac{L}{\alpha}\biggl(1+\frac{A t_N}{ m_0}\biggr)^{2\alpha+2}
%,\frac{L}{\alpha}(1+\frac{A t_N}{ m_0})^{2\alpha+2}
\right)-A^{2} E\right)\psi(t)=0 \;.
\end{equation}
As it will turn out, $\frac{A}{m_0} \sim L^{-\frac14}$, and the potential \eqref{eq:AVx} will be indeed a convergent
power series in $L^{-\frac14}$. 

Having clarified this point, we begin by computing the Taylor series of the ground-state potential $V_G$ at $m_0$. This reads
\begin{align}\label{eq:VTaylor}
& V_G(x) = \sum_{n\geq 0}  v^{\sn} L^{\frac{2 a - n}{2\alpha+2}}(x-m_0)^n \;,
\\ \nonumber
& v^{\scriptscriptstyle{(0)}}=(1+\alpha) \alpha^{\frac{-\alpha}{\alpha+1}} \;,\quad v^{\scriptscriptstyle{(1)}}=0 \;,\quad v^{\scriptscriptstyle{(2)}} =(2\alpha+2)\alpha^\frac{2}{\alpha+1} \;,
  \\ \label{eq:vns}
& v^{\sn} =   \alpha^\frac{n-2\alpha}{2\alpha+2}\left((-1)^{n+2}(n+1)\alpha + \binom{2\alpha}{n}\right) \;,\quad  n\geq 3 \;,
\end{align}
In the above formula $\binom{2\alpha}{n}=\frac{(2\alpha (2\alpha-1)\dots (2\alpha-k+1)}{k!}$ and $\alpha^\frac{n-2\alpha}{2\alpha+2}$ is assumed to be real and positive for all $n$. We notice that the Taylor series \eqref{eq:VTaylor} has radius of convergence  $\big|\frac{L}{\alpha}\big|^{\frac{1}{2\alpha+2}}$, since the only singularity of $V$ is at $x=0$.

Now we specify the change of variables (\ref{eq:Axt},\ref{eq:Azk},\ref{eq:AVx}) by setting the coefficient $A=\e^{\frac{\alpha-1}{\alpha+1}}(v^{\scriptscriptstyle{(2)}})^{-\frac14}$, so that the quadratic term in the expansion \eqref{eq:VTaylor} in the new variable $t$ is $1$.

Explicitly, we have
\begin{equation}\label{eq:xtot}
 x=(v^{\scriptscriptstyle{(2)}})^{-\frac14}\e^{\frac{\alpha-1}{\alpha+1}} t+m_0 \;,
\mbox{ with  } \e=L^{-\frac14} \;,
\end{equation}
where $v^{\scriptscriptstyle{(2)}}=(2\alpha+2)\alpha^\frac{2}{\alpha+1}$ as per \eqref{eq:vns}.
Moreover,
$z_k=\varphi(t_k;\e)$ where
\begin{align} \label{eq:varphi}
& \varphi(t;\e): \mathcal{D}_{\e} \to \mc{E}_{\e} : t
\mapsto \frac{\e^{-4}}{\alpha}\left( 1+(2\alpha+2)^{\frac{3}{4}} \e \big( t+ \e \sigma (t;\e)\big)  \right) \;, \\
& \sigma(t,\e)=\sum_{l\geq2}\binom{2\alpha+2}{l}(2\alpha+2)^{-\frac{l+3}{4}}\, \e^{l-2} t^{l} \;.\label{eq:sigma}
\end{align}
The domain $\mc{D}_{\e}$ and the co-domain $\mc{E}_{\e}$ are chosen to be the maximal sets such that
the map $\varphi(t;\e)$ is holomorphic and bijective:
\begin{align}\label{eq:Dom} \nonumber
& \mathcal{D}_{\e} = \left\lbrace t \in\C \,:\,
\left| \frac{\e\,t}{(2\alpha+2)^{\frac14}} \right|<1 \mbox{ et } -\frac{\pi}{2\alpha+2}<\arg \left(1+\frac{\e\,t}{(2\alpha+2)^{\frac14}}\right)
\leq \frac{\pi}{2\alpha+2}
\right\rbrace \;, \\
& \mc{E}_{\e}= \left\lbrace z \in \C \,:\,  0< |z| < \frac{2}{\alpha\e^4} \right\rbrace \;.
\end{align}

Finally, the Schr\"odinger equation with a higher-state-potential reads
 \begin{align}\label{eq:psieps}
&\frac{d^2}{dt^2}\psi(t) = \big( V(t;\e) - \widetilde E_\varepsilon \big) \psi(t) \;, \\
\label{eq:thigherstates}
 & V(t;\e)=U(t;\e)-2\frac{d^2}{dt^2}\sum_{k=1}^N \log\left( t - t_k + \e \left( \sigma(t;\e)-\sigma(t_k;\e)\right)\right) \;, \\
 \label{eq:Ue}
& U(t;\e)  = t^2 + \sum_{n\geq 1}u^{\sn}\,\varepsilon^n\,t^{n+2}\;,\quad u^{\sn}
=  (v^{\scriptscriptstyle{(2)}})^{-\frac{n+4}{4}} v^{\scriptscriptstyle{(n+2)}} ,\\ \label{eq:Eeps}
& \widetilde{E}_\varepsilon = (v^{\scriptscriptstyle{(2)}})^{-1/2}\left(\e^\frac{2\alpha-2}{\alpha+1}E - \e^{-2}v^{\scriptscriptstyle{(0)}}\right) \;,
\end{align}
where the coefficient $v^{\sn}$'s are as per (\ref{eq:vns}), and
the series $U(t;\e)$ has radius of convergence
$|(2\alpha+2)^{\frac14}\,\e^{-1} \big| $. In fact, the Taylor series \eqref{eq:VTaylor} has radius of convergence
$\big|\frac{L}{\alpha}\big|^{\frac{1}{2\alpha+2}}$ and $x,t$ are related by \eqref{eq:xtot}.

We naturally have the following definition
\begin{definition}\label{defi:tna}
 $\tna \subset \C^{N}\times \C^*$ consists of the ordered pairs $(P,\e)$
 where $\e \neq 0$ is a complex number, and $P$ is a monic polynomial of degree $N$ with distinct and non-zero roots $t_1$, $\dots$, $t_N$ such that
 \begin{enumerate}
  \item $t_k \in \mc{D}_{\e}$ for all $k\in\lbrace1,\dots,N\rbrace$, where $\mc{D}_{\e}$ is as per \eqref{eq:Dom};
  \item The potential \eqref{eq:thigherstates} has trivial monodromy at all $t_1$, $\dots$, $t_N$.
 \end{enumerate}
$\Tna$ is the closure of $\tna$ in $\C^{N}\times\C^*$.

$\bna^*$ (resp. $\Bna^*$) is the set of those $(P,L) \in \bna$ (resp.  $ \in \Bna$) such that the
roots of $P$ have all modulus strictly less that $\big|\frac{L}{\alpha}\big|$.
\end{definition}

We can summarise and formalise our computations in the following Lemma.
\begin{lemma}\label{lem:ztot}
The map  $\big(z_k=\varphi(t_k;\e) \;,\; k=1\dots N \;,\; L=\e^{-4}\big)$, induces a surjective analytic map $ \Phi: \Tna \to \Bna^*$.

Moreover, if we introduce a cut $C$ -- connecting $0$ with $\infty$ -- in the complex $L$ plane and we choose a branch of $L^{\frac14}$,
the restriction of the map $\Phi$ from $\Tna \setminus \lbrace (P,\e) \in \Tna\;,\; \e^4 \in C \rbrace$ 
to $\Bna^* \setminus \lbrace (P,L) \in \Bna^*\;,\; L \in C \rbrace $ is a bijection.
\begin{proof}
It follows from the computations leading to formula \eqref{eq:psieps} that
\begin{itemize}
 \item If $(P,L)\in \Tna$, $\left(\prod_{k=1}^N(z-\varphi(t_k)),\e^{-4}\right) \in \Bna^*$, where $t_1$, $\dots$, $t_N$ are the roots of $P$;
\item Conversely if $(P,L) \in \Bna^*$  and we choose a quartic root of $L$,
then $\left(\prod_{k=1}^N(t-\varphi^{-1}(z_k)), L^{-\frac14}\right) \in \Tna$,  where $z_1$, $\dots$, $z_N$ are the roots of $P$.
\end{itemize}
This proves the thesis.
\end{proof}
\end{lemma}

\begin{remark}
We have seen that for $L$ large the poles of the potentials $V_P$ are condensating about the zeroes of $V'_G$. Now it is a well-known fact, which was discovered at the very beginning of quantum mechanics, that in the large momentum limit, a (convex) potential looks like a harmonic oscillator localised at its unique real minimum. However the rational extensions of the perturbed harmonic oscillator, as defined in formula \eqref{eq:rationale}, do not yield
solutions of the BLZ system because they do not implement the Dorey-Tateo symmetry. The term  $\e \big( \sigma(t;\e)-\sigma(t_k;\e)\big)$ in \eqref{eq:thigherstates} takes care of this.
\end{remark}

\subsection{Convergence to rational extensions of the harmonic oscillator}
Suppose that we have a sequence of $(P^{\sn},\e^{\sn}) \in \Tna$ such that $\e^{\sn} \to 0$. Let $t_1^{\sn}$, $\dots$, $t_N^{\sn}$ be the (possibly not pairwise distinct) roots of $P^{\sn}$. The sequence of potentials $V^{\sn}(t):=V(t;\e^{\sn})$, where $V(t;\e^{\sn})$ is as per \eqref{eq:thigherstates} with $t_k:=t_k^{\sn}$, formally converges to a rational extension of the harmonic oscillator with $N$ poles. Here we prove that the sequence does convergence. More precisely, we have the following
\begin{proposition}\label{pro:ztoe}
Let $(P^{\sn},\e^{\sn}) \in \Tna$ such that $\e^{\sn} \to 0$.
Assume moreover that the roots $t_1^{\sn}$, $\dots$, $t_N^{\sn}$ of $P^{\sn}$ satisfy the limit
\begin{equation}\label{eq:te0}
 \lim_{n\to \infty}t_k^{\sn} \, \e^{\sn} \to 0 \;,\quad \forall k\in\lbrace1,\dots,N\rbrace \;.
\end{equation}
The sequence $P^{\sn}$ can be split into $J$ subsequences, with $1\leq J\leq p(N) $, each subsequence associated to a unique partition $\nu$ of $N$ in such a way that the following $3$ equivalent properties hold:
\begin{enumerate}
 \item $\lim_{n \to \infty}P^{\sn}(t)=P^{\snu}(t)$,  uniformly on compact subsets, where $P^{[\nu]}(t)$, as per \eqref{eq:Pnu}, is the Wronskian of Hermite polynomials associated to $\nu$.
\item $\lim_{n \to \infty}t_k^{\sn}=v_k^{\snu}$, $k=1,\dots,N$, where $v_1^{\snu}$, $\dots$, $v_N^{\snu}$ are the roots of the polynomial $P^{\snu}(t)$. \item Denote by $V^{\sn}(t)$ the potential $V(t;\e^{\sn})$, as per \eqref{eq:thigherstates} with $t_k:=t_k^{\sn}$. We have that
 $$\lim_{n \to \infty}V^{\sn}(t)=t^2-2\frac{d^2}{dt^2} \log P^{\snu}(t) \;,$$ uniformly on compact subsets which do not contain any $v^{\snu}_k$.
 This is the rational extension of the harmonic oscillator, associated to the partition $\nu$ according to
formula \eqref{eq:extendedharmonic}.
\end{enumerate}
\begin{proof}
We notice that the $3$ properties above are indeed equivalent, and that
it suffices to prove the thesis for sequences $(P^{\sn},\e^{\sn}) \in \tna$.

First we prove the thesis assuming that,
$\limsup_{n \to \infty}|t_k^{\sn}|<\infty$ for all $k\in\lbrace 1,\dots,N\rbrace$. Take any converging subsequence, and define $v_k^{\snu}=\lim_{n\to \infty}
t_k^{\sn}$. The potential $V^{\sn}(t)$ converges -- uniformly on compact subsets which do not contain any $v_k^{\snu}$ -- to
 $$
 t^2-2\frac{d^2}{dt^2} \log \big(t-v_k^{\snu}\big) \;.
 $$
 Since, by continuity, the monodromy is trivial about all $v_1^{\snu}$, $\dots$, $v_N^{\snu}$, the above potential is a rational extension of the harmonic oscillator.

 Now we prove that the condition \eqref{eq:te0} implies that  $\limsup_{n \to \infty}|t_k^{\sn}|<\infty$ for all $k\in\lbrace 1,\dots,N\rbrace$.
 To this aim we suppose that $\limsup_{n \to \infty}|t_l^{\sn}|=\infty$ for some $l$. Since $\C$ is sequentially compact,
 there is a subsequence such that $\frac{t_k^{\sn}}{t_j^{\sn}}$ as well as $t_k^{\sn}$ converge in $\C$ for all indices $j,k\in\lbrace 1,\dots,N\rbrace$.
 We name $J_l$ the set of indices $j \in 1 \dots N$ such that $ \frac{t_j^{\sn}}{t_l^{\sn}} \to 1$ on this subsequence.
 
 Due to Theorem \ref{thm:dui} and formula \eqref{eq:thigherstates}, the quantities $t_1^{\sn}$, $\dots$, $t_N^{\sn}$ satisfy the system of equations
 \begin{equation}\label{eq:perturbedextensions}
  U'(t_k^{\sn};\e^{\sn})+ f(t_k^{\sn}; \e^{\sn})+\sum_{j \neq k} \,I(t_k^{\sn},t_j^{\sn}; \e^{\sn}) =0\;,\quad k=1,\dots,N \;,
 \end{equation}
where $f(t_k;\e)$ is the coefficient of the term $(t-t_k)$ in the expansion of 
$$-2\frac{d^2}{dt^2}\log\left\{ t - t_k + \e \big[ \sigma(t;\e)-\sigma(t_k;\e)\big]\right\} \;,$$
 and
 $$I(t_j,t_k;\e)=-2\frac{d^3}{dt^3}\log\left\{ t - t_j + \e \big[ \sigma(t;\e)-\sigma(t_j;\e)\big]\right\}_{t=t_k} \;.$$
 
 After the hypothesis $\e^{\sn}t_k^{\sn} \to 0$, $\forall k \in\lbrace1,\dots,N\rbrace$, the following asymptotics follow
 \begin{align*}
&  U'(t_k^{\sn};\e^{\sn})=2 t_l^{\sn} +o(t_l^{\sn}) \;,\quad  \forall k \in J_l \;, \\
& f(t_k^{\sn})=o(t_l^{\sn}) \;,\quad \forall k \in J_l  \;, \\
& I(t_k^{\sn},t_j^{\sn};\e^{\sn})=o(t_l^{\sn}) \;,\quad \forall k \in J_l \mbox{ and } j \notin J_l  \;,\\
& I(t_k^{\sn},t_j^{\sn};\e)+I(t_j^{\sn},t_l^{\sn};\e)=o (t_l^{\sn}) \;,\quad \forall k\neq j \in J_l \;.
 \end{align*}
The first two rows are straightforward to verify. The third and fourth row follows from the very same reasoning that led to
 the proof of Lemmas \ref{lem:boundedinteratcion} and \ref{lem:boundedinteratcionsymm}, hence we omit
 to write the explicit computations. It suffices to say that, to analyse the third row
 we write
 $$
 I(u,v;\e)=\frac{A + B \big(u-v+\e(\sigma(u,\e)-\sigma(v,\e)\big)+ C \big(u-v+\e(\sigma(u,\e)-
 \sigma(v,\e)\big)^2}{\left(u-v+\e\big(\sigma(u,\e)-\sigma(v,\e)\big)\right)^3} \;,
 $$
 for some $A,B,C$ with a finite limit as $\e^{\sn} \to 0$, and to analyse the fourth row, we expand $I(u,v)+I(v,u)$ at $u=v$,
 and use the identity $\e \frac{d^l}{dt^l}\sigma(t_k^{\sn};\e^{\sn})\to 0$ as $n \to \infty$, for all $l \geq 1$.
 
 Summing the equation \eqref{eq:perturbedextensions} for all $k \in J_l$, and using the four estimates above, we obtain that
 \begin{equation*}
 \lim_{n \to \infty}2 \, |J_l| \,  t^{\sn}_l + o (t_l^{\sn}) = 0 \;,
 \end{equation*}
 where $|J_l|$ is the cardinality of the set $J_l$. This contradicts the assumption that $\lim_{n \to \infty}|t_l^{\sn}|= \infty$.
 \end{proof}

\end{proposition}
Due to Lemma \ref{lem:ztot} and Proposition \ref{pro:ztoe}, we can say the higher-state-potentials in the large $L$ limit converge to rational extensions of the harmonic oscillator, provided the variables $x$ and $z_k$ are scaled according to (\ref{eq:xtot},\ref{eq:varphi}).
More precisely, we have the following theorem.
\begin{theorem}
\label{thm:toharmonic}
Fix $\alpha \in \bb{R}^+$ and $N \in \mathbb{N}$.
Let $(P^{\sn},L^{\sn}) \in \Bna$ be a sequence such that $L^{\sn} \to \infty$.
Furthermore assume that
$L^{\sn}$ belongs, for $n$ large, to the complex plane minus a cut connecting $0$ with $\infty$, so that
$(L^{\sn})^{\frac14}$ can be unambiguously defined.

The sequence $P^{\sn}$ can be split into $J$ subsequences, with $1\leq J\leq p(N) $, each subsequence associated to a unique partition $\nu$ of $N$ in such a way that the following properties holds:
\begin{itemize}
 \item The, appropriately ordered, roots $z_1$, $\dots$, $z_N$ of $P^{\sn}$ satisfy the asymptotics
\begin{equation}\label{eq:zkexpthm}
z_k^{\sn}=\frac{L^{\sn}}{\alpha}+ \frac{(2\alpha+2)^{\frac34}}{\alpha} v_k^{\snu} \big(L^{\sn}\big)^{-\frac34}+ o\left(\big(L^{\sn}\big)^{-\frac34}\right) \;,\quad k=1,\dots,N \;,
\end{equation}
where $v_1^{\snu}$, $\dots$, $v_N^{\snu}$ are the roots of the polynomial $P^{\snu}(t)$, associated to the partition $\nu$ of $N$ as per
\eqref{eq:Pnu}.
\item After the change of variables,
$x=x(t;\e)$, and $E \to \widetilde{E}_\e$ as per formulae (\ref{eq:xtot},\ref{eq:Eeps}),
the Schr\"odinger equation with the higher-state-potential
\begin{equation*}
 \psi''(x)=\big(V^{\sn}(x)-E\big)\psi(x) \;,
\end{equation*}
converges to the rational extension of the harmonic potential $U^{\snu}(t)$,
\begin{equation}\label{eq:toharmonicthm}
 \psi''(t)=\left(t^2-2\frac{d^2}{dt^2}\log P^{\snu}(t) - \widetilde{E}_{\e} \right)\psi(t) \;.
\end{equation}
The convergence of the potential of the above equation is uniform on all compact subsets of $ \C \setminus
\lbrace v_1^{\snu},\dots,v_N^{\snu}\rbrace $.
\end{itemize}

\begin{proof}
 After Proposition \ref{pro:largez}(2), we have that $\frac{z_k^{\sn}\alpha}{L^{\sn}} \to 1$ for all $k\in\lbrace 1,\dots,N\rbrace$.
This implies that $\left|\frac{z_k^{\sn}\alpha}{L^{\sn}}\right|< 2$ for $n$ large enough.  
Therefore Lemma \ref{lem:ztot} applies:
Choosing a fourth root of $L$, the quantities $t_k^{\sn}=\varphi^{-1}(z^{\sn}_k, \e^{\sn})$ satisfy the system \eqref{eq:BLZt} with $\e^{\sn}
=(L^{\sn})^{-\frac14}$. Moreover, the limit $\frac{z^{\sn}_k\alpha}{L^{(n)}} \to 1$ implies the limit $t^{\sn}_k \e^{\sn} \to 0$.
Therefore Proposition \ref{pro:ztoe} applies, from which the thesis follows.
\end{proof}

\end{theorem}

\subsection{Spectrum of the monster potentials in the large momentum limit}
Here we briefly derive an asymptotic formula \eqref{eq:spectrumeps} for the eigenvalues of the radial spectrum of monster potentials, in the regime $L$ large and positive. A lengthier and more complete study, with proofs, will be addressed in a forthcoming publication. 

Assume that for a given $\nu=(\nu_1,\dots,\nu_j)$ there exists an algebraic family of monster potentials $V^{\snu}_P(x;L)$, where $P:=P^{\snu}(z;L)$ satisfies the expansion \eqref{eq:zkexpthm}. We say that $E^{\snu} \in \bb{N}$ belongs to the radial spectrum (which is obviously discrete) of the monster potential if there exists a meromorphic solution 
$\psi:\C \setminus \bb{R}^- \to \overline{\C}$ of the following boundary value problem
\begin{align}\nonumber
 & \psi''(x)=\left( x^{2\alpha}+\frac{L}{x^2}-2 \frac{d^2}{dx^2} \log P^{\snu}(x^{2\alpha};L)- E^{\snu} \right) \psi(x) \;,\\
 & \lim_{x \to 0} \psi(x)=\lim_{x\to + \infty}\psi(x)=0 \;. \label{eq:boundaryradial}
\end{align}

After Theorem \ref{thm:toharmonic}, in a large disc with centre $x=m_0=\big(\frac{L}{\alpha}\big)^{\frac{1}{2\alpha+2}}>0$ and radius
$(\frac{L}{\alpha})^{\frac{1}{2\alpha+2}}$, 
the monster potential $V_P^{\snu}(x;L)$ looks like
the rational extension of the harmonic oscillator $U^{\snu}(t)$, provided we make the change of variables $x \to t$ and
$E^{\snu} \to \widetilde{E}^{\snu}$ according to formulas
(\ref{eq:xtot}) and (\ref{eq:Eeps}). It follows from this that the boundary value problem \eqref{eq:boundaryradial}
in the new variables $t$, $\widetilde{E}_n$, approximately reads
\footnote{The fact that the limits $ \lim_{x \to 0} \psi(x)=\lim_{x\to + \infty}\psi(x)=0$ read
$\lim_{t\to\pm \infty}\psi(t)=0$, follows from
a rather standard matching procedure, see e.g. \cite[A.2]{cinese},
which will be studied more carefully in the forthcoming publication to which we referred above.}
\begin{equation}\label{eq:pertubedRspectrum}
 \psi''(t)=\left( t^2-2\frac{d^2}{dt^2}\log P^{\snu}(t) +O(\e^2)-\widetilde{E}^{\snu}\right)\psi(t) \;,\quad
 \lim_{t\to\pm \infty}\psi(t)=0 \;,
\end{equation}
where $\e=L^{-\frac14}$. The above boundary value problem describes a small perturbation of the boundary
value problem which defines the $\bb{R}$-spectrum of
the rational extension of the harmonic oscillator $U^{\snu}(t)$, which was introduced in
Section \ref{section:harmonic}. Therefore, applying the standard perturbation theory to the exact formula
for the eigenvalues of the $\bb{R}$-spectrum
\eqref{eq:spectrumnu}, we obtain that the spectrum of the problem \eqref{eq:pertubedRspectrum} is
\begin{equation}\label{eq:pertubedEn}
 \widetilde{E}^{\snu}_n= 2(n-j) + 1 + O(\e^2) \;,\quad n \in \bb{N}^{\snu} \;,
\end{equation}
where $\bb{N}^{\snu}$ is the sequence of integer numbers which is obtained from $\bb{N}$ deleting
the numbers $(\nu_j,\nu_{j-1}+1,\dots, \nu_1+j-1)$.

From equation \eqref{eq:pertubedEn} and 
the transformation law \eqref{eq:Eeps},
we obtain the following asymptotics formula for the eigenvalues of the radial spectrum of the monster potentials
\begin{multline}\label{eq:spectrumeps}
 E^{\snu}_n=  (1+\alpha) \biggl(\frac{L}{\alpha} \biggr)^{\frac{\alpha}{\alpha+1}} + \big(2\alpha+2\big)^{\frac12} \alpha^{\frac{1}{\alpha+1}}
 L^{\frac{\alpha-1}{2\alpha+2}} 
\big( 2(n-j)+1 \big)+ O\big( |L|^{-\frac{1}{\alpha+1}}\big) \\ ,\quad n \in \bb{N}^{\snu} \;.
\end{multline}
We notice that formula \eqref{eq:spectrumeps} is valid for any fixed eigenvalue level, or equivalently, for any fixed subset
of eigenvalues, and hence it describes the bottom of the radial spectrum. In fact, if we let $n \to \infty$,
the standard perturbation theory may not correctly compute the spectrum of (\ref{eq:boundaryradial},\ref{eq:pertubedRspectrum})
and other approaches should be used.

We remark that the radial spectrum is the fundamental object of the ODE/IM correspondence. In fact,
the points of the radial spectrum are the (exponential of two times the) Bethe Roots of a solution of the Bethe Ansatz equations
of the quantum KdV model \cite{BLZ04}. Formula \eqref{eq:spectrumeps} is
in striking accordance with the prediction of the NLIE integral equation on the IM side of the correspondence,
see \cite[Appendix A]{BLZ04}: in fact it shows that the spectrum is real and positive when the momentum is
positive and sufficiently large, and it reproduces the Fermi-sea structure of the excitations.

\begin{remark}
We can check the formula \eqref{eq:spectrumeps} in the case $\alpha=1$, the harmonic oscillator, against known formulas.
Writing $L=\ell(\ell+1)$,
\eqref{eq:spectrumeps} reads
\begin{equation}\label{eq:specharmonic}
 E^{\snu}_n= 2 l + 3 + 4 n- 2j + O \big(l^{-1}\big) \;,\quad n \in  \bb{N}^{\snu} \;.
\end{equation}
Now, it is well known that eigenvalues of the radial spectrum of the ground state potential $x^2+\frac{\ell(\ell+1)}{x^2}$ are
$E_n=2 l + 3 + 4 n$, $ n \in \bb{N}$. Equation \eqref{eq:specharmonic} correctly agrees with the latter formula,
when one chooses the empty partition:
$j=0$, $\bb{N}^{\snu}=\bb{N}$. Moreover, one can actually show that the
truncation of  \eqref{eq:specharmonic} yields the exact result for all higher-state-potentials. 
One should however be aware that, with regard to the ODE/IM correspondence, for some particular values of $\ell$
some of the higher-state-potentials correspond to null-vectors of the Quantum KdV model
and have to be excluded \cite{dunningpers}.
\end{remark}

\subsection{The BLZ system for large $L$}

We write the BLZ system with respect of the variable $t$, using the change of variable
$z_k=\varphi(t_k;\e)$ as per \eqref{eq:varphi}, where $L=\e^{-4}$. Namely,
\begin{align*}
& z_k= \frac{\e^{-4}}{\alpha}\left( 1+ (2\alpha+2)^{\frac{3}{4}} \e \big( t+ \e \sigma (t;\e)\big)  \right) \;, \\
& \sigma(t,\e)=\sum_{l\geq2}\binom{2\alpha+2}{l}(2\alpha+2)^{-\frac{l+3}{4}}\, \e^{l-2} t^{l} \;.
\end{align*}
Inserting the above change of variable as well as the relation $L=\e^{-4}$, the BLZ system reads
\begin{equation}
\label{eq:BLZt}
\e^{-3} F_k(\underline{t};\e) = 0 \;,\quad k=1,\dots,N \;,
\end{equation}
where the functions $F_k:\mathcal{D}^N \to \C$ admit the following explicit expression
\begin{align} \label{eq:Fk}
F_k(\underline{t};\e) =  \frac{\e^{3} (1-4 \alpha^2)}{8 M} - 
\frac{t_k+ \e \sigma(t_k;\e)}{2M^{\frac14}}+
\sum_{j \neq k}\frac{M^{-\frac94}\mathcal{F}(t_k,t_j;\e)}{\big(t_k-t_j+ \e (\sigma(t_k;\e)-\sigma(t_j;\e))\big)^3} \;,
\end{align}
where $\mathcal{F}(t_k,t_j;\e)$ is given by
\begin{align*}
& \mathcal{F}(t_k,t_j;\e) =\big(1+ M^{-\frac14} \e (t_k+ \e \sigma(t_k;\e))\big)^3 +  \\
&+(3+\alpha)(1+2\alpha)\big(1+ M^{-\frac14} \e (t_k+ \e \sigma(t_k;\e))\big)^2 \big(1+ M^{-\frac14} \e (t_j+ \e \sigma(t_j;\e))\big) \\
&+\alpha(1+2\alpha)\big(1+ M^{-\frac14} \e (t_k+ \e \sigma(t_k;\e))\big) \big(1+ M^{-\frac14} \e (t_j+ \e \sigma(t_j;\e))\big)^2 \;,
\end{align*}
with $M=2\alpha+2$. Moreover,
\begin{align}\label{eq:Dk}
\mathcal{D}^N  = \left\lbrace (t_1,\dots,t_N;\e) \in \C^{N+1} \,:\, t_k \in  \mc{D}_{\e} \, , \, \forall k \mbox{ and }
 t_k\neq t_j \, ,\, \forall j \neq k \right\rbrace  \subset \C^{N+1} \;.
\end{align}

Notice that the expansion of $F_k(\vectt,\e)=0$ around $\e=0$ truncated at $\mathcal{O}(\e)$ yields the system
\begin{equation}
\label{eq:approximateBLZ}
G_k(\vectt;\e) = 2t_k + 3 \e \kappa\, t_{k}^2 - \sum_{j\neq k}\frac{4}{(t_k-t_j)^3} = 0 \;,\quad k=1,\dots,N \;,
\end{equation}
where $\kappa=\frac{(5-2\alpha) }{3(2\alpha+2)^{\frac14}}$. Formula \eqref{eq:approximateBLZ} is the equation
governing the rational extensions of the harmonic oscillator perturbed by the cubic term $ \e \kappa\, t^3$;
see (\ref{eq:perturbedh},\ref{eq:tperturbed}).

\subsection{Conditional proof of the Weak BLZ conjecture}
\begin{definition}
 Wee call a convergent Puiseux series with rational exponents
 an algebraic function.
 
 We thus write $f=\sum_{q \in \bb{Q}}c_q \e^q$, $c_q \in \C^N$, and we define $\lim_{\e \to 0}f(\e)=c_0$, provided $c_q=0$ if $q<0$.
\end{definition}

\begin{conjecture}\label{conj:Pte}
Fix $N\geq1$ and $\alpha>0$. Identify the space of monic polynomials of degree $N$ with $\C^N$.

For every partition $\nu$ of $N$ there exists a neighbourhood $D$ of $0$, and a unique algebraic function
$$
P:D \to \C^N: \e \mapsto P^{\snu}(t;\e),
$$
such that
\begin{enumerate}
\item $\lim_{\e \to 0}P^{\snu}(t;\e)= P^{\snu}(t)$, where $P^{\snu}(t)$ is as per \eqref{eq:Pnu};
 \item The roots $t_1^{\snu}(\e)$, $\dots$, $t_N^{\snu}(\e)$ of $P^{\snu}(t;\e)$ satisfy the system \eqref{eq:BLZt} for all
 $\e \in D \setminus \lbrace 0 \rbrace$.
\end{enumerate}
\end{conjecture}

The above conjecture is equivalent to the following one, which is better suited for explicit verification.
\begin{conjecture}\label{conj:te}
Fix $N\geq1$ and $\alpha>0$. Given a partition $\nu$ of $N$,
let $v_1^{\snu}$, $\dots$, $v_N^{\snu}$ be the, possibly not pairwise distinct, roots of the polynomial
$P^{\snu}$. Denote by $\mc{G}$ the subgroup of permutations $\mc{S}_N$, $\mc{G}=\left\lbrace \sigma \in \mc{S}_N,
\; v_{\sigma(k)}^{\snu}=v_k^{\snu} \,,\, k=1 \dots, N \right\rbrace$.

For every partition $\nu$ of $N$ there exists a neighbourhood $D$ of $0$, and an algebraic function
$$
\underline{t}:D \to \C^N:\e \mapsto \big(t_1(\e), \dots, t_N(\e)\big) \;,
$$
such that
\begin{enumerate}
 \item The components $t_1(\e),\dots,t_N(\e)$ of $\underline{t}(\e)$ satisfy the system \eqref{eq:BLZt}
 for all $\e \in D \setminus \lbrace 0 \rbrace$;
 \item $\lim_{\e \to 0}t_k(\e)=v_k^{\snu}$, for all $k\in\lbrace 1,\dots,N\rbrace$;
 \item The function $\underline{t}$ is unique up to the action of $\mc{G}$: If $\underline{u}$ is another solution satisfying (1,2) then
 $u_k(\e)=t_{g(k)}(\e)$, $k=1 \dots N$ for some $g \in \mc{G}$.
\end{enumerate}
\end{conjecture}
We have the following Lemma
\begin{lemma}\label{lem:psnute}
With the same notation of Conjecture \ref{conj:Pte}.
If the Conjecture \ref{conj:Pte} holds, then
\begin{itemize}
  \item[(i)] $P^{\snu}(-it;i\e)=(i)^N P^{\snus}(t;\e)$;
  \item[(ii)] $P$ is an analytic function.
 \end{itemize}
 \begin{proof}
(i)  We have $\lim_{\e \to 0} P^{\snu}(-i t;i\e)=(i)^N P^{\snus}(t)$, since $P^{\snu}(-i t)=(i)^N P^{\snus}(t) $, see Lemma
  \ref{thm:felder}. Since $(\vectt;\e)\to(-i\vectt; i\e)$ is a symmetry of \eqref{eq:BLZt}, it follows that
  $P^{\snu}(-i t; i \e)=(i)^N P^{\snus}(t;\e)$.
  
(ii) Applying four times (i), we obtain $P^{\snu}(t;e^{2 \pi i} \e)=P^{\snu}(t;\e)$, where $P^{\snu}(t;e^{2 \pi i} \e)$ is the analytic
continuation of $P^{\snu}(t;\e)$ along a small loop about $\e=0$. This means that $P^{\snu}(t,\e)$ is single valued. Moreover, the
limit $\lim_{\e \to 0}P(t;\e)$ is well-defined, by hypothesis. Hence $\e=0$ is a removable singularity.
 \end{proof}

\end{lemma}

\begin{proposition}\label{pro:znue}
Fix $N \geq 1$ and $\alpha>0$. Assume that Conjecture \ref{conj:Pte} holds. Let $v_1^{\snu},\dots,v_N^{\snu}$ be the roots of $P^{\snu}(t)$,
as defined in \eqref{eq:Pnu}.

For every partition $\nu$ of $N$, there is a unique algebraic function defined in a neighbourhood $D$ of $L=\infty$, with values
in the space of monic polynomials of degree $N$,
 $$P^{\snu}:D \to \C^N: L \mapsto P^{\snu}(z;L) \;,$$
 such that
 \begin{itemize}
  \item[(i)] The roots $z_1(L),\dots,z_N(L)$ of $P^{\snu}(z;L)$ satisfy the BLZ system \eqref{eq:algblz};
  \item[(ii)] The roots $z_1(L),\dots,z_N(L)$ satisfy the asymptotics 
 \begin{equation*}
z_k^{\snu}(L) = \frac{L}{\alpha}+
 \frac{(2\alpha+2)^{\frac34}\, v_k^{\snu}}{\alpha} L^{-\frac{3}{4}}+ o(L^{-\frac34}) \;,\quad k=1 \dots N \;.
\end{equation*}
\end{itemize}

Moreover, the identity $P^{\snu}(z;e^{2\pi i} L)=P^{\snus}(z;L)$ holds, where $P^{\snu}(z;e^{2\pi i} L)$ denotes
the analytic continuation
of $P^{\snu}(z;L)$ along a small loop about $L=\infty$.
\begin{proof}
  After Lemma \ref{lem:ztot} -- provided $z_k=\varphi(t_k;\e)$ with $\varphi$ as per \eqref{eq:varphi} and $\e=L^{-\frac14}$ is small enough --
  the BLZ system \eqref{eq:algblz} and the system \eqref{eq:BLZt}. Therefore (i,ii) follow directly from Conjecture \ref{conj:Pte} (1,2).
  
  The identity $P^{\snu}(z;e^{2\pi i} L)=P^{\snus}(z;L)$ follows from Lemma \ref{lem:psnute}(i) since
  $\varphi(-it_k;i \e)=\varphi(t_k;\e)$, as it can be directly checked.
 \end{proof}

\end{proposition}

The Weak BLZ conjecture is a corollary of the above Proposition \ref{conj:te}. More precisely we have,
\begin{corollary}\label{cor:blzconj}
Fix $N \in \bb{N}$ and $\alpha >0$. Assume that Conjecture \ref{conj:Pte} holds.
\begin{enumerate}
 \item If $L$ is large enough, for each $L$ there are exactly $p(N)$ monster potentials.
 \item If $L$ is large enough, $(P,L) \in \Bna \Longrightarrow (P,L) \in \bna$.
 \item The set $\lbrace L \in \C, \exists P \mbox{ s.t. } (P,L) \in \Bna \setminus \bna\rbrace$  is discrete.
 \item For all $L \in \C$, the number of higher-state-potentials, and hence the number of monster potentials, is less than
 or equal to $p(N)$. Moreover, for a generic $L$, the number of higher state potentials and the number of monster potentials is
 exactly $p(N)$.
\end{enumerate}
\begin{proof}
 $\bna$ is an algebraic variety, since it is defined by algebraic equations, which have a natural projection over the complex $L$ plane.
 Unless it is empty, we can parametrise such a variety in a neighbourhood of $L=\infty$ by a family of algebraic functions. Due to
 Theorem \ref{thm:toharmonic} any of these functions must satisfy the asymptotics \eqref{eq:zkexpthm}, for some partition $\nu$ of $N$.
 After Proposition \ref{pro:znue}, there exists a unique algebraic function with the given asymptotic. Therefore
 $\bna$ is an algebraic variety of dimension $1$, i.e. a non-compact, possibly singular, possibly disconnected Riemann surface, which
 is parametrised, in a punctured neighbourhood of $\infty$, by $p(N)$ algebraic functions $P^{\snu}(z;L)$: $\bna$ is locally
 $p(N)$-sheeted smooth Riemann surface. In other words, in a neighbourhood of $L=\infty$
 there are $p(N)$ solutions of the BLZ system. This proves (1).
 
 Since, in a punctured neighbourhood of infinity, $\bna$ is the disjoint union of a finite number of sets, in the same neighbourhood
 it coincides with its closure, $\Bna$. Therefore (2) holds.
 
 It follows that $\Bna$ is also a 1 dimensional algebraic variety. The set $\Bna \setminus \bna$ is
 the sub-variety characterised by the vanishing of the discriminant.
 This function cannot have an accumulation point of zeroes since it is non-zero
 for $L$ large enough.
 
 Finally, since $\Bna$ is a Riemann surface, and the number of sheets is constantly $p(N)$ above an open subset (the punctured neighbourhood of
 $L=\infty$), the number of sheets  above any point $L$ is always less than or equal to $p(N)$. Therefore (4) holds.
\end{proof}

\end{corollary}

\begin{remark}
 Since $\Bna$ is a Riemann surface whose projection over $L$ has a finite number of sheets, it is natural to consider its
 normalisation (loosely speaking, the desingularised compactification). If Conjecture \ref{conj:Pte} holds,
 it follows from Proposition \ref{pro:znue} that in a neighbourhood of $L=\infty$ the surface can be parametrised by the
 $p(N)$ functions $P^{\snu}(z;L)$, one for each partition $\nu$ of $N$. We can define a partial compactification of $\Bna$
 by adding points over $L=\infty$
 in such a way to obtain a smooth surface in the neighbourhood of $L=\infty$, on which the map 
 $(P,L) \mapsto L$ is holomorphic and of degree $p(N)$.
 In fact, according to the general theory of Riemann surfaces (see e.g. \cite[Section 4.2]{donaldson11}), we must add one point for each orbit of the monodromy about $\infty$.
 Since $P^{\snu}(z;e^{2\pi i} L)=P^{\snus}(z;L)$, we need to add exactly one point for each
 unordered pairs of partitions $(\nu,\nu^*)$ of $N$.
 
 Such is the local structure of $\Bna$. What is however its global structure? Is $\Bna$ irreducible \footnote{Conversely,
 it could be just a disjoint union of
 $\widetilde{N}$ copies of $\bb{P}^1$, where $\widetilde{N}$ is the number of unordered pairs $(\nu,\nu^*)$
 of partitions $(\nu,\nu^*)$ of $N$}?
 What are the other branch-points of $L$, and what is their monodromy?
 Is the partial compactification that we have defined a compact smooth Riemann surface (i.e. is $\Bna$ smooth and $L$ proper)?
\end{remark}
In the rest of the paper we address the proof of the Conjecture \ref{conj:Pte}.
In Section \ref{sec:nondeg}, we address the case where $\nu$ is a non degenerate partition of $N \geq 1$ such that the polynomial
$P^{\snu}(t)$, as defined by formula \eqref{eq:Pnu}, has simple roots only (i.e.
a non-degenerate partition satisfying the $F$-property of Definition \ref{def:fproperty}). This is the simplest case, since
the proof of the Conjecture follows from the implicit function theorem, provided the Jacobian of the equation at $\e=0$ is
invertible.

In Section \ref{sec:completely}, we address the case where $P^{\snu}(t)$ has a single root of multiplicity greater than $1$,
namely $t^{\frac{d(d+1)}{2}}$, for some $d \geq2$ and $\nu=(1,2,\dots,d)$. We call such a case the completely degenerate case.

In Section \ref{sec:partially}, we address the partially degenerate case.
That is when $0$ is a root of multiplicity greater than $1$, and $P^{\snu}$ has other roots all of multiplicity one.

There are two considerations to be done. First, provided the Conjecture \ref{conj:felder} by Felder et al. holds, any partition
$\nu$ belongs to one of the above $3$ cases.
We assume that it indeed holds, even though our methods of analysis could be in principle used
if it does not hold.
Second, in the case of completely or partially degenerate partitions, the perturbation series which the function
$t_k(\e)$'s must satisfy are already very intricate for moderate $d$. In fact, we
were not able to analyse the case $d\geq 4$, which imply that we are not able to prove our Conjecture \ref{conj:Pte}
for $N\geq 10$. We do think that this is not a major obstacle but simply
a technical difficulty, which will be overcome when we (or other researchers) will
be able to find an effective functional-analytic setting to analyse these series: A
not-too-mathematically-minded reader can safely assume that the Conjecture \ref{conj:felder} by Felder et al.
and our Conjecture \ref{conj:Pte} hold.

\section{Perturbation Series. The non-degenerate case}\label{sec:nondeg}
Let $\nu$ be a non degenerate partition of $N \geq 1$ such that the polynomial
$P^{\snu}(t)$, as defined by formula \eqref{eq:Pnu}, has simple roots only (i.e.
a non-degenerate partition satisfying the $F$-property of Definition \ref{def:fproperty}).

Recall that the system \eqref{eq:BLZt} is of the form $\underline{F}(\underline{t};\e)=0$, $\underline{F}=(F_1,\dots,F_N)$ and
$\underline{t}=(t_1,\dots ,t_N)$, where the functions $F_1,\dots,F_N$ are as per \eqref{eq:Fk}.
We denote by $J$ the Jacobian of the map $\underline{F}(\cdot;0):  \C^N \to \C^N$.
In components, we have
\begin{equation}\label{eq:jacobian}
J_{ij}(\underline{t}):=\frac{\partial F_i(\underline{t};0)}{\partial t_j}= 
2\delta_{ij}\left( 1+\sum_{l\neq j} \frac{6}{(t_j - t_l)^4}\right) - (1-\delta_{ij})\frac{12}{(t_i- t_j)^4} \;,\quad i,j=1,\dots,N \;.
\end{equation}
In particular, if $\underline{v}^{\snu}=(v_1^{\snu},\dots,v_N^{\snu})$ are the distinct roots of $P^{\snu}(t)$, we define
\begin{equation}\label{eq:Jnu}
 J^{\snu}:=J(\underline{v}^{\snu}) \;.
\end{equation}

We have the following Proposition.
\begin{proposition}\label{thm:notdegenerate}
 Let $\nu$ be a non-degenerate partition of $N$ such that all zeros the polynomial $P^{\snu}(t)$ are simple, and such that
 the Jacobian $J^{\snu}$ is invertible.
 
 The Conjecture \ref{conj:te} holds for $\nu$ and $\nu^*$.
 \begin{proof}
  We notice that
$J^{\snu}=J^{\snus}$, since 
$v_k^{\snus}= i v_k^{\snu}$ by equation \eqref{eq:vnu}.
A direct application of the inverse function theorem furnishes us with the unique solutions
$t_k^{\snu}:D \to \bb{C}$ and $t_k^{\snus}:D \to \bb{C}$, $k=1\dots,N$ of \eqref{eq:BLZt} such that
$t_k^{\snu}(0)=v_k^{\snu}$ and $t_k^{\snus}(0)=v_k^{\snus}$. This proves the thesis.
\end{proof}
\end{proposition}

\subsection{Analysis of the Jacobian $J^{\snu}$}
Here we complete the analysis of the non-degenerate case, by studying the spectrum of the
Jacobian $J^{\snu}$, in order to show that is invertible. We present here
a (conjectural) closed combinatorial formula, equation \eqref{eq:spectrumJ},
for the spectrum of $J^{\snu}$. After this formula all the eigenvalues of $J^{\snu}$ are integer and strictly positive numbers, hence
$J^{\snu}$ is invertible.

We start with the following definition.
\begin{definition}\label{def:tableau}
To each partition $\nu$ of $N\geq 1$, we associate a sequence $\rho^{\snu}$ of, not-necessarily-distinct, $N$ integer and positive numbers,
according to the following rules.

First of all to the Young diagram of the partition we associate
\begin{enumerate}
 \item The unique non-standard Young tableau, whose entries are integer numbers. They start with one and increase
by $1$ in each row from right to left and down each column;
 \item The sequence $\ell^{\snu}$ of $M$ positive integers, where $M$ is the number of columns of the diagram, obtained by
 collecting -from right to left - the entries at the bottom of each column;
 \item A partition $\nu'$ of $N-M$ (if $N=M$, this is the empty diagram), whose Young diagram is obtained  
 by erasing the boxes at the end of each column (of the Young diagram of $\nu$).
\end{enumerate}

Then we iterate the procedure until we reach the empty diagram. The sequence $\rho^{\snu}$ is defined by
reordering, so that it is monotone non-decreasing, the sequence obtained by joining the sequences $\ell^{\snu},\ell^{\scriptscriptstyle{[\nu']}}, \dots $ 
\end{definition}
\noindent Let us illustrate our definition through the example of the partition $\nu=(3,2,2,1,1)$.\\
At the first step, we fill the Young tableau associated to
$\nu$ with integer numbers (see figure below in the middle),
collect the sequence $\ell^{\snu}=\lbrace 1,2,4,5,7\rbrace$ (marked in red), and
we obtain the Young tableau associated to the partition $\nu'=(2,1,1)$ (see figure below on the right).
\vspace{0.5cm}
\begin{center}
Step $1:$ \hspace{0.2cm} \begin{ytableau}
$$&&&& \cr && \cr \cr
\end{ytableau} 
\hspace{0.3cm} $\longrightarrow$ \hspace{0.3cm}
\begin{ytableau}
5&4&3&{\color{red}2}&{\color{red}1} \cr 6&{\color{red}5}&{\color{red}4} \cr {\color{red}7} \cr
\end{ytableau}
\hspace{0.3cm} $\longrightarrow$ \hspace{0.3cm}
\begin{ytableau}
$$&& \cr \cr
\end{ytableau} 
\end{center}
\vspace{0.5cm}
\noindent
At the second step, one finds $\ell^{[\nu']} = \lbrace 1,2,4 \rbrace$ and $\nu''=(1)$.
\vspace{0.5cm}
\begin{center}
Step $2:$ \hspace{0.2cm}\begin{ytableau}
$$&& \cr \cr
\end{ytableau} 
\hspace{0.3cm} ${\longrightarrow}$ \hspace{0.3cm}
\begin{ytableau}
3&{\color{red}2}&{\color{red}1} \cr {\color{red}4} \cr
\end{ytableau}
\hspace{0.3cm} ${\longrightarrow}$ \hspace{0.3cm}
\begin{ytableau}
$$ \cr
\end{ytableau}
\end{center}
\vspace{0.5cm}

\vspace{0.5cm}
\noindent
At the third step, we obtain $\ell^{\scriptscriptstyle{[\nu'']}} = \lbrace 1 \rbrace$ and $\nu'''$ is the empty partition; therefore we stop.
\begin{center}
Step $3:$ \hspace{0.2cm}
\begin{ytableau}
$$ \cr
\end{ytableau} 
\hspace{0.3cm} ${\longrightarrow}$ \hspace{0.3cm}
\begin{ytableau}
{\color{red}1} $$ \cr
\end{ytableau}
\hspace{0.3cm} ${\longrightarrow}$ \hspace{0.3cm} $\emptyset$
\end{center}
\vspace{0.5cm}
\noindent
The final outcome is the sequence 
\begin{equation*}
\rho^{[(3,2,2,1,1)]} = \lbrace 1,1,1,2,2,4,4,5,7 \rbrace .
\end{equation*}
Given the above definition, we have found the following closed formula for the spectrum of $J^{\snu}$ \footnote{We have verified it numerically
for all non-degenerate partitions $\nu$ for $N\leq 10$}.
\begin{conjecture}
\label{conj:eigenconj}
Let $\nu$ be a partition of $N$ such that the corresponding polynomial $P^{\snu}$ has simple roots only, and
$J^{\snu}$ be the $N \times N$ matrix defined by formula \eqref{eq:Jnu}.

The eigenvalues of $J^{\snu}$ are described by the following sequence $\lambda_1,\dots,\lambda_N$ of $N$, not necessarily distinct,
rational numbers
\begin{equation}\label{eq:spectrumJ}
 \lambda_k= 2(\rho^{\snu}_k)^2, \; k=1 \dots N,
\end{equation}
where $\rho^{\snu}_k$ is the $k-th$ entry of the sequence $\rho^{\snu}$, as per Definition \ref{def:tableau}.
It follows that the determinant of $J^{\snu}$ is the following strictly positive integer number
\begin{equation}
\label{eq:detJ}
\det{(J^{\snu})} =  2^N\prod_{k=1}^N \big(\rho^{\snu}_k\big)^2 .
\end{equation}
\end{conjecture}
\begin{remark}
 Notice that $J^{\snu}=J^{\snus}$, since by Lemma \ref{thm:felder},
 the roots of $P^{\snu}$ and $P^{\snus} $ satisfy the relation $v_k^{\snu}=i v_k^{\snus}$ (if appropriately ordered).
 It is amusing, but not straightforward, to show that $\rho^{\snu}=\rho^{\snus}$, as required.
\end{remark}

\begin{remark}
In the case of the partition $\nu_0=(N)$, $P^{\scriptscriptstyle{[\nu_0]}}(t)$ is the Hermite polynomial $H_N(t)$ and
$\rho^{\scriptscriptstyle{[\nu_0]}}=(1,2,\dots,N)$. The conjecture therefore states that
the eigenvalues of $J^{\scriptscriptstyle{[\nu_0]}}$ are $2(1,4,\dots,N^2)$.
Such a result is known in the literature about Hermite polynomials, see \cite{ahmed79}. 
After \eqref{eq:detJ}, we have the beautiful formula
\begin{equation}\label{eq:Jhermite}
\det{(J^{\scriptscriptstyle{[\nu_0]}})} =   2^N N!^2  .
\end{equation}

\end{remark}

\section{Perturbation Series. Completely degenerate case}\label{sec:completely}
In this Section we study Conjecture \ref{conj:te}, in the case $\nu$ is a completely degenerate partition,
that is $\nu=(d,d-1,\dots,1)$, $d \geq 2$ and $P^{\snu}(t)=t^N$ with $N=\frac{d(d+1)}{2}$. Notice that 
such partitions are self-conjugate.

Recall that our aim is to find a (unique up to the action of $\mc{S}_N$) algebraic solution $t_1(\e),\dots,t_N(\e)$ of \eqref{eq:BLZt},
by means of a convergent Puiseux series, such that
$\lim_{\e \to 0}t_k(\e)=0$ for all $k\in\lbrace 1,\dots,N\rbrace$.
Since the point $(\underline{t},\e)=(\underline{0},0)$ is a singularity of \eqref{eq:BLZt}, the perturbation series that we obtain will be much more intricate than
in the case of a non-degenerate partition. In fact, we were able to fully solve the case $d=2,3$ (i.e. $N=3,6$) only. We present our results below omitting the details of the case $d=3$ because they are too long to be transcribed here.

\subsection{The case $d=2$} We have the following Proposition.

\begin{proposition}\label{prop:d2}
There exists a neighbourhood $D$ of $\e=0$ and a unique -- up to permutation of the index $k=1,2,3$ --
algebraic solution
$t_k(\e)$, $k=1,2,3$ of the system \eqref{eq:BLZt} with the following Puiseux expansion
\begin{equation}  
\label{eq:ansatzvec}
\vectt(\e) = \e^{\frac13}\vectt^{\scriptscriptstyle{(0)}} +
  \sum_{l\geq 1} \e^{\frac{4l+1}{3}}\vectt^{\scriptscriptstyle{(l)}} \;,
\end{equation}
where $\vectt^{\scriptscriptstyle{(l)}}=(t_1^{\scriptscriptstyle{(l)}},t_2^{\scriptscriptstyle{(l)}},t_3^{\scriptscriptstyle{(l)}})$.
The above solution has dominant term
\begin{equation}\label{eq:ansatzt0}
  t_k^{\scriptscriptstyle{(0)}}=\omega^{k-1} \kappa^{\frac13} \;,\quad k=1,2,3 \;,
 \end{equation}
 with $\kappa=\frac{5-2\alpha}{3 (2\alpha+2)^{\frac14}}$ and $\omega=e^{\frac{2\pi i}{3}}$.

Consequently, the corresponding solution $z_k(L)=\varphi(t_k(\e),\e)) \;,\; k=1,2,3$, where 
  $\varphi$ is as per \eqref{eq:varphi} and $\e=L^{-\frac{1}{4}}$, of the BLZ system \eqref{eq:algblz} has expansion 
  \begin{equation}\label{eq:zkd2}
   z_k(L)= \frac{L}{\alpha}+\frac{ \omega^{k-1} \kappa^{\frac13} (2\alpha+2)^{\frac34}}{\alpha} L^{\frac23} + \mathcal{O}(L^{\frac13}) \;,\quad k=1,2,3 \;.
   \end{equation}
 Moreover, letting $P(z;L)=\prod_{k=1}^3 \big(z-z_k(L)\big)$, we have that
  \begin{equation} \label{eq:monodromyd2}
P(z;e^{2\pi i }L) = P(z;L) \;,
\end{equation}
where $P(z;e^{2\pi i }L)$ denotes the analytic continuation of $P(z;e^{2\pi i }L)$ along a small loop about $L=\infty$.

\end{proposition}

\paragraph*{\it Proof}

We begin with three remarks:
\begin{enumerate}
 \item Due to M. Artin's approximation theorem, a unique formal solution of an algebraic equation is a convergent algebraic
solution \cite{artin69,hauser16p}. Hence we just need to prove the existence and uniqueness of a formal solution with the expansion
\eqref{eq:ansatzvec}.
\item As we proved in greater generality in Lemma \ref{lem:psnute} and Proposition \ref{pro:znue},
if $P^{\snu}(t;\e)=\prod_{k=1}^3 \big(t-t_k(\e)\big)$, then the uniqueness of the solution implies
that $$P^{\snu}(it;-i \e)=i P^{\snus}(t;\e) =i P^{\snu}(t;\e) ,$$
where the latter identity derives from the fact that in the case at hand $\nu=(2,1)=\nu^*$. From the above
equality, \eqref{eq:monodromyd2} follows.
\item In the case $\alpha=\frac52$, the dominant term in the asymptotics \eqref{eq:ansatzt0} vanishes. In fact,
 one can show that, for $\alpha=\frac52$,
%$\alpha=\frac52$ is not singularity of the system and the corresponding solution is obtained either by suitably taking the limit $\alpha\to\frac52$ of the solution constructed below for all $\alpha\neq\frac52$ or repeating the computation starting from the Ansatz
the solution to \eqref{eq:BLZt} admits the following expansion
$$\vectt(\e) = \e^3\vectt^{\scriptscriptstyle{(0)}} +\sum_{l\geq 1} \e^{4l+3}\vectt^{\scriptscriptstyle{(l)}} \;,$$
with dominant term 
$$t_k^{\scriptscriptstyle{(0)}}=-\frac{45}{2\times 7^{\frac34}} \;,\quad k=1,2,3 .$$
For sake of brevity we omit the proof of this fact, and below we simply
assume that $\alpha \neq \frac52$.
\end{enumerate}
Given the above remarks, we prove below the existence and uniqueness of a formal solution satisfying the Ansatz \eqref{eq:ansatzvec}.

We start with the following definitions.
\begin{definition}\label{def:hadamard}
 Let $\mathbbm{N} \ni p >0 $. Given two generic vectors $\vecv=(v_1,\dots,v_p)\in\C^p$ and $\underline{w}=(w_1,\dots,w_p)\in\C^p$, we denote by $\langle\underline{v},\underline{w}\rangle = \sum_{i=1}^p v_i w_i^*$
 %$$\langle,\rangle: \C^3\times\C^3 \to \C : (\vecv,\underline{w})\to\langle\underline{v},\underline{w}\rangle = \sum_{i=1}^3 v_i w_i^* \;,$$
 the standard Hermitian product on $\C^p$. %and by $\underline{v}\circ\underline{w} = (v_1w_1,\dots,v_pw_p)$ the standard Hadamard product on $\C^p$. Conventionally, $\vecv^{\circ m}=(v_1^m,\dots,v_p^m)  \;,\; m\in\mathbbm{Z}\backslash\lbrace 0 \rbrace$, denotes the $m-$th Hadamard power of $\vecv$.
\end{definition}

Thereafter we shall use the following notation. 
\begin{definition}
\label{def:basis}
We denote by $\mathcal{B}$ the basis of $\C^3$ generated by the vectors $\vecX_1=(1,1,1)$, $\vecX_2=(1,\omega,\omega^2)$, $\vecX_3=\vecX_2^*=(1,\omega^2,\omega)$ and by $\mathcal{B}'$ the basis of $\C^3$ generated by the vectors $\vecX_1'=\vecX_1$, $\vecX_2'=(1,0,-1)$, $\vecX_3'=(0,1,-1)$.
\end{definition}
\begin{definition} \label{def:multimaps}
Let $\vecx=(x_1,x_2,x_3)$ and $\bb{N} \ni n >0$. We define the following symmetric $n$-linear maps
\begin{equation} 
\vecM_n^{\vecx}=(M_{n,1}^{\vecx},M_{n,2}^{\vecx},M_{n,3}^{\vecx})\,:\,\mathbbm{C}^3\underset{n-\text{times}}{\times\dots\times}\mathbbm{C}^3\to\mathbbm{C}^3 : (\vecv_1,\dots,\vecv_n)\to \vecM_{n}^{\vecx}(\vecv_1,\dots,\vecv_n) \;,
\end{equation}
where 
\begin{equation}
M_{n,k}^{\vecx}(\vecv_1,\dots,\vecv_n) = \sum_{j\neq k} \frac{\left(v_{1,k}-v_{1,j}\right)\dots\left(v_{n,k}-v_{n,j}\right)}{\left(x_k-x_j\right)^{n+3}} \;,\quad k=1,2,3 \;,
\end{equation}
and $\vecv_m=(v_{m,1},v_{m,2},v_{m,3})\in\C^3 \;,\; m=1,\dots,n$. If $\vecx=a\vecX_1+b\vecX_2 \;,\; a,b\in\C \;,\; b\neq 0$ , we set $\vecM_n\equiv\vecM_n^{\vecx}$.
\end{definition}
From the latter Definitions, we have the following Lemmas.
\begin{lemma} 
\label{thm:rankker}
Let $\ker{(\vecM_1^{\vecx})}$ and $\im{(\vecM_1^{\vecx})}$ denote the kernel and the range of the linear map $\vecM_1^{\vecx}$, respectively. If $\vecx = a\vecX_1+b\vecX_2 \;,\; a,b\in\C \;,\; b\neq 0$, then
\begin{itemize}
\item $\ker{(\vecM_1)} = \Span\lbrace \vecX_1,\vecX_2 \rbrace$;
\item $\im{(\vecM_1)} = \Span\lbrace \vecX_2\rbrace$.
\end{itemize}
Otherwise,
\begin{itemize}
\item $\ker{(\vecM_1^{\vecx})} = \Span\lbrace \vecX_1'\rbrace$;
\item $\im{(\vecM_1^{\vecx})} = \Span\lbrace \vecX_2',\vecX_3'\rbrace$.
\end{itemize}
\end{lemma}
\begin{lemma}
\label{thm:Mndecomposition}
In the basis $\mathcal{B}$, the only non-vanishing components of $\lbrace \vecM_n \rbrace_{n=1}^4$ are
\begin{align}
&\vecM_1(\vecX_3)=-\frac{1}{3b^4}\,\vecX_2 \;, \notag \\
&\vecM_2(\vecX_i,\vecX_3) = -\frac{1}{3b^5}\,\vecX_i \;,\quad i=2,3 \;, \notag \\
&\vecM_3(\vecX_i,\vecX_2,\vecX_3) = -\frac{1}{3b^6}\,\vecX_i \;,\quad i=2,3 \;,\notag\\
&\vecM_4(\vecX_3,\vecX_3,\vecX_3,\vecX_3) = \frac{1}{3b^7}\,\vecX_2 \;,\quad \vecM_4(\vecX_i,\vecX_2,\vecX_2,\vecX_3) = -\frac{1}{3b^7}\,\vecX_i \;,\quad i=2,3 \;,
\end{align}
up to permutations of the arguments.
\end{lemma}
 
The idea of the proof is to construct a solution $\vectt(\e)$ of the form \eqref{eq:ansatzvec} to the system $F_k(\vectt,\e)=0 \;,\; k=1,2,3$, with $F_k(\vectt,\e)$ defined as per \eqref{eq:Fk}, and show that such solution is unique. Plugging the Ansatz \eqref{eq:ansatzvec} in $F_k(\vectt,\e)$, the first perturbative contribution to the expansion of $F_k(\vectt(\e),\e)$ about $\e=0$ is $\mathcal{O}(\e^{-1})$ and its cancellation corresponds to the following set of equations for $\vectt^{\scriptscriptstyle{(0)}}$
 \begin{equation}
 \label{eq:locust0}
 \sum_{j\neq k} (t_k^{\scriptscriptstyle{(0)}}-t_j^{\scriptscriptstyle{(0)}})^{-3} = 0 \;,\quad k=1,2,3 \;.
 \end{equation}
 The latter is the system of algebraic equations
 \eqref{eq:moser} for the distinct roots of the rational extensions of the
 trivial potential for $d=2$ (see Section \ref{sec:AirMcMos}).
 Choosing the parametrisation \eqref{eq:locusd2} and using Definition \ref{def:basis}, we write $\vectt^{\scriptscriptstyle{(0)}}$ as
 \begin{equation}
 \label{eq:t0}
 \vectt^{\scriptscriptstyle{(0)}} = a\vecX_1 + b \vecX_2 \;,\quad a,b\in\C \;,\quad b\neq 0 \;.
 \end{equation}
 To fix the coefficients $a$ and $b$ in \eqref{eq:t0}, we need to take into account higher perturbative contributions in the expansion of $F_k(\vectt(\e),\e)$. Using the Ansatz \eqref{eq:ansatzvec} with \eqref{eq:t0}, the expansion of $F_k(\vectt(\e),\e)$ about $\e=0$ yields
 
 \begin{equation}
 F_k(\vectt(\e),\e) = -\frac{1}{4M^\frac14}\sum_{l\geq 1}A_k^{\scriptscriptstyle{(l)}}\e^{\frac{4l}{3}-1} \;,\quad k=1,2,3 \;,
 \end{equation}
 where we recall $M=2\alpha+2$. Thus the system $F_k(\vectt(\e),\e)=0 \;,\; k=1,2,3$,  becomes in vector form
\begin{equation}
\label{eq:systemvec}
S:\vecA^{\scriptscriptstyle{(m)}} = \underline{0} \;,\quad m\geq 1 \;,
\end{equation}
with $\vecA^{\scriptscriptstyle{(m)}} = (A_1^{\scriptscriptstyle{(m)}},A_2^{\scriptscriptstyle{(m)}},A_3^{\scriptscriptstyle{(m)}})$ and
 \begin{align}
&A_k^{\scriptscriptstyle{(1)}} = 12M_{1,k}(\vectt^{\scriptscriptstyle{(1)}}) + 2t_k^{\scriptscriptstyle{(0)}} \;, \notag \\
&A_k^{\scriptscriptstyle{(2)}} = 12M_{1,k}(\vectt^{\scriptscriptstyle{(2)}}) + 2t_k^{\scriptscriptstyle{(1)}} + \frac{36\,t_k^{\scriptscriptstyle{(0)}}}{M^{\frac14}}M_{1,k}(\vectt^{\scriptscriptstyle{(1)}}) - 24M_{2,k}(\vectt^{\scriptscriptstyle{(1)}},\vectt^{\scriptscriptstyle{(1)}}) +\frac{M-1}{M^{\frac14}}(t_k^{\scriptscriptstyle{(0)}})^{2} \;, \notag \\
&A_k^{\scriptscriptstyle{(3)}} = 12M_{1,k}(\vectt^{\scriptscriptstyle{(3)}}) + 2t_k^{\scriptscriptstyle{(2)}} + \frac{36\,t_k^{\scriptscriptstyle{(0)}}}{M^{\frac14}}M_{1,k}(\vectt^{\scriptscriptstyle{(2)}}) - 48M_{2,k}(\vectt^{\scriptscriptstyle{(1)}},\vectt^{\scriptscriptstyle{(2)}}) + \frac{2(M-1)}{M^{\frac14}} t_k^{\scriptscriptstyle{(0)}}t_k^{\scriptscriptstyle{(1)}}  \notag \\
&- \frac{72\,t_k^{\scriptscriptstyle{(0)}}}{M^{\frac14}}M_{2,k}(\vectt^{\scriptscriptstyle{(1)}},\vectt^{\scriptscriptstyle{(1)}}) + 40M_{3,k}(\vectt^{\scriptscriptstyle{(1)}},\vectt^{\scriptscriptstyle{(1)}},\vectt^{\scriptscriptstyle{(1)}}) +\frac{36}{M^{\frac12}}\left((t_k^{\scriptscriptstyle{(0)}})^{2}+M^{\frac{1}{4}}t_k^{\scriptscriptstyle{(1)}}\right)M_{1,k}(\vectt^{\scriptscriptstyle{(1)}}) \notag \\
&+\frac{1}{3M^{\frac{1}{2}}}(M-1)(M-2)(t_k^{\scriptscriptstyle{(0)}})^{3} -\frac{3}{2M^{\frac{5}{4}}}(M-1)(M-3) \;, \notag
\end{align}
\begin{align}
&A_k^{\scriptscriptstyle{(4)}} = 12M_{1,k}(\vectt^{\scriptscriptstyle{(4)}}) + 2t_k^{\scriptscriptstyle{(3)}} + \frac{36\,t_k^{\scriptscriptstyle{(0)}}}{M^{\frac14}}M_{1,k}(\vectt^{\scriptscriptstyle{(3)}}) - 48M_{2,k}(\vectt^{\scriptscriptstyle{(1)}},\vectt^{\scriptscriptstyle{(3)}}) \notag \\
&+ \frac{2\,t_k^{\scriptscriptstyle{(2)}}}{M^{\frac14}}\left((M-1)\,t_k^{\scriptscriptstyle{(0)}}+18M_{1,k}(\vectt^{\scriptscriptstyle{(1)}})\right) - \frac{144\,t_k^{\scriptscriptstyle{(0)}}}{M^{\frac14}}M_{2,k}(\vectt^{\scriptscriptstyle{(1)}},\vectt^{\scriptscriptstyle{(2)}})  - 24M_{2,k}(\vectt^{\scriptscriptstyle{(2)}},\vectt^{\scriptscriptstyle{(2)}}) \notag \\
&+\frac{36}{M^{\frac12}}\left((t_k^{\scriptscriptstyle{(0)}})^{2}+M^{\frac{1}{4}}t_k^{\scriptscriptstyle{(1)}}\right)M_{1,k}(\vectt^{\scriptscriptstyle{(2)}})+ 120M_{3,k}(\vectt^{\scriptscriptstyle{(1)}},\vectt^{\scriptscriptstyle{(1)}},\vectt^{\scriptscriptstyle{(2)}}) +\frac{36\,t_k^{\scriptscriptstyle{(0)}}t_k^{\scriptscriptstyle{(1)}}}{M^{\frac12}}M_{1,k}(\vectt^{\scriptscriptstyle{(1)}}) \notag \\
&+ \frac{(M-1)\,t_k^{\scriptscriptstyle{(1)}}}{M^{\frac12}}\left((M-2)(t_k^{\scriptscriptstyle{(0)}})^{2} + M^{\frac{1}{4}}t_k^{\scriptscriptstyle{(1)}}\right) - \frac{72}{M^{\frac12}}\left((t_k^{\scriptscriptstyle{(0)}})^{2} + M^{\frac{1}{4}}t_k^{\scriptscriptstyle{(1)}}\right)M_{2,k}(\vectt^{\scriptscriptstyle{(1)}},\vectt^{\scriptscriptstyle{(1)}})
\notag \\
&+ \frac{12\,(t_k^{\scriptscriptstyle{(0)}})^3}{M^{\frac34}}M_{1,k}(\vectt^{\scriptscriptstyle{(1)}}) +\frac{120\,t_k^{\scriptscriptstyle{(0)}}}{M^{\frac14}}M_{3,k}(\vectt^{\scriptscriptstyle{(1)}},\vectt^{\scriptscriptstyle{(1)}},\vectt^{\scriptscriptstyle{(1)}}) - 60M_{4,k}(\vectt^{\scriptscriptstyle{(1)}},\vectt^{\scriptscriptstyle{(1)}},\vectt^{\scriptscriptstyle{(1)}},\vectt^{\scriptscriptstyle{(1)}}) \notag \\
&+\frac{(t_k^{\scriptscriptstyle{(0)}})^4}{12M^{\frac{3}{4}}}(M-1)(M-2)(M-3) + \frac{t_k^{\scriptscriptstyle{(0)}}}{20M}(M^4-19M^2+19)\;, \notag  \\ 
 &\vdots \notag \\
&A_k^{\scriptscriptstyle{(m)}} = 12M_{1,k}(\vectt^{\scriptscriptstyle{(m)}}) + 2t_k^{\scriptscriptstyle{(m-1)}} +  \frac{36\,t_k^{\scriptscriptstyle{(0)}}}{M^{\frac14}}M_{1,k}(\vectt^{\scriptscriptstyle{(m-1)}}) - 48M_{2,k}(\vectt^{\scriptscriptstyle{(1)}},\vectt^{\scriptscriptstyle{(m-1)}}) \notag \\
&+ \frac{2\,t_k^{\scriptscriptstyle{(m-2)}}}{M^{\frac14}}\left((M-1)\,t_k^{\scriptscriptstyle{(0)}}+18M_{1,k}(\vectt^{\scriptscriptstyle{(1)}})\right) - \frac{144\,t_k^{\scriptscriptstyle{(0)}}}{M^{\frac14}}M_{2,k}(\vectt^{\scriptscriptstyle{(1)}},\vectt^{\scriptscriptstyle{(m-2)}}) \notag \\
&- 48M_{2,k}(\vectt^{\scriptscriptstyle{(2)}},\vectt^{\scriptscriptstyle{(m-2)}}) +\frac{36}{M^{\frac12}}\left((t_k^{\scriptscriptstyle{(0)}})^{2}+M^{\frac14}t_k^{\scriptscriptstyle{(1)}}\right)M_{1,k}(\vectt^{\scriptscriptstyle{(m-2)}}) \notag \\
&+ 120M_{3,k}(\vectt^{\scriptscriptstyle{(1)}},\vectt^{\scriptscriptstyle{(1)}},\vectt^{\scriptscriptstyle{(m-2)}}) + f_k^{\scriptscriptstyle{(m-3)}}(\vectt^{\scriptscriptstyle{(0)}},\dots,\vectt^{\scriptscriptstyle{(m-3)}}) \;,\quad m\geq 5 \;,
 \label{eq:Amgeq2n}
 \end{align}
 where $\vecf^{\scriptscriptstyle{(m-3)}}=\left(f_1^{\scriptscriptstyle{(m-3)}},f_2^{\scriptscriptstyle{(m-3)}},f_3^{\scriptscriptstyle{(m-3)}}\right)$ is a vector function that collects all the terms depending on $\vectt^{\scriptscriptstyle{(0)}}$, $\dots$, $\vectt^{\scriptscriptstyle{(m-3)}}$. The vector equation $\underline{A}^{\scriptscriptstyle{(1)}}=\underline{0}$ immediately fixes the coefficient $a$. In fact, since $\ker{(\vecM_1)}$ is non-trivial, we need to impose that $\vectt^{\scriptscriptstyle{(0)}}\in\im{(\vecM_1)}$. Hence, by Lemma \ref{thm:rankker}, it follows that $a=0$ and
 \begin{equation}
 \vectt^{\scriptscriptstyle{(0)}} =b\vecX_2 \;.
 \label{eq:constr0}
 \end{equation}
 To fix $b$, we proceed as follows. We start by expressing the vector $\vectt^{\scriptscriptstyle{(l)}}$, $l\geq 1$ in the basis $\mathcal{B}$ with coefficients $ c_1^{\scriptscriptstyle{(l)}}$, $ c_2^{\scriptscriptstyle{(l)}}$, $ c_3^{\scriptscriptstyle{(l)}}$ as 
 \begin{equation}
 \label{eq:tldecomp}
 \vectt^{\scriptscriptstyle{(l)}} = \sum_{i=1}^3 c_i^{\scriptscriptstyle{(l)}} \vecX_i \;,\quad l\geq 1 \;.
 \end{equation}
 Then, we consider the vector equations $\vecA^{\scriptscriptstyle{(1)}}$, $\vecA^{\scriptscriptstyle{(2)}}=\underline{0}$ and project them along the basis $\mathcal{B}$ so as to arrive at the following system of equations
 \begin{align}
 \label{eq:systemt1}
 S^{\scriptscriptstyle{(1)}}&:\begin{cases}
 \displaystyle{\langle \underline{A}^{\scriptscriptstyle{(1)}},\vecX_2\rangle = 6b - \frac{12}{b^4}\,c_3^{\scriptscriptstyle{(1)}} = 0} \\
 \displaystyle{\langle \underline{A}^{\scriptscriptstyle{(2)}},\vecX_1\rangle = 6\,c_1^{\scriptscriptstyle{(1)}} = 0} \\
 \displaystyle{\langle \underline{A}^{\scriptscriptstyle{(2)}},\vecX_3\rangle = \frac{24}{b^5}\,(c_3^{\scriptscriptstyle{(1)}})^2 + \left(6-\frac{36}{M^{\frac14}b^3}\right)c_3^{\scriptscriptstyle{(1)}}+\frac{3(M-1)}{M^{\frac14}}\,b^2= 0}
 \end{cases}\notag \\
 &\longrightarrow\quad 
 \begin{cases}
 \displaystyle{c_3^{\scriptscriptstyle{(1)}} = \frac{1}{2}\left(\frac{7-M}{3M^{\frac{1}{4}}}\right)^{\frac{5}{3}}} \\
 \displaystyle{c_1^{\scriptscriptstyle{(1)}} = 0} \\
 \displaystyle{b=\left(\frac{7-M}{3M^{\frac{1}{4}}}\right)^{\frac{1}{3}}}
 \end{cases} \;,
 \end{align}
 where we used \eqref{eq:tldecomp} and Lemma \ref{thm:Mndecomposition} to evaluate the Hermitian products.
 
 In conclusion, $\vectt^{\scriptscriptstyle{(0)}}$ yields
 \begin{equation}
 \label{eq:systemt1sol}
 \vectt^{\scriptscriptstyle{(0)}} = \left(\frac{7-M}{3M^{\frac{1}{4}}}\right)^{\frac{1}{3}}\vecX_2 \;,
 \end{equation}
 which proves \eqref{eq:ansatzt0} using $M=2\alpha+2$. The strategy displayed above can be generalised to construct $\vectt^{\scriptscriptstyle{(m)}}$ in terms of $\vectt^{\scriptscriptstyle{(0)}}$, $\dots$, $\vectt^{\scriptscriptstyle{(m-1)}}$. The main idea is to project the vector equations $\vecA^{\scriptscriptstyle{(m)}}=\underline{0}$ that make up the system \eqref{eq:systemvec} along the basis $\mathcal{B}$ to form the subsystems
 \begin{equation}
 \label{eq:generalsystem}
 S^{\scriptscriptstyle{(m)}}:\begin{cases}
 \langle \underline{A}^{\scriptscriptstyle{(m)}},\vecX_2\rangle = 0 \\
 \langle \underline{A}^{\scriptscriptstyle{(m+1)}},\vecX_1\rangle = 0 \\
 \langle \underline{A}^{\scriptscriptstyle{(m+1)}},\vecX_3\rangle = 0
 \end{cases} \;,\quad m\geq 1 \;,
 \end{equation}
which are such that 
$$S=\bigcup_{m\geq 1} S^{\scriptscriptstyle{(m)}} \;,$$
since $\langle \underline{A}^{\scriptscriptstyle{(1)}},\vecX_1\rangle$ and $\langle \underline{A}^{\scriptscriptstyle{(1)}},\vecX_3\rangle$ are identically fulfilled upon setting $a=0$.

To prove in full generality that $\vectt^{\scriptscriptstyle{(m)}}$ can be expressed in terms of $\vectt^{\scriptscriptstyle{(0)}}$, $\dots$, $\vectt^{\scriptscriptstyle{(m-1)}}$ for any $m$, we shall make use of the expression of $\vecA^{\scriptscriptstyle{(m)}}$ valid for $m\geq 5$ as per \eqref{eq:Amgeq2n}. 
However, before dealing with the general case, we must first evaluate $\vectt^{\scriptscriptstyle{(1)}}$ and $\vectt^{\scriptscriptstyle{(2)}}$ which appear explicitly in $\vecA^{\scriptscriptstyle{(m)}}$, $m\geq 5$. From $S^{\scriptscriptstyle{(2)}}$ we get
 \begin{align}
 \label{eq:systemt2}
 S^{\scriptscriptstyle{(2)}}&:\begin{cases}
 \displaystyle{\langle \underline{A}^{\scriptscriptstyle{(2)}},\vecX_2\rangle = 30\,c_2^{\scriptscriptstyle{(1)}} - 36\left(\frac{3M}{(7-M)^4}\right)^{\frac{1}{3}}c_3^{\scriptscriptstyle{(2)}} = 0} \\
 %4\,\vecX_2.\vectt^{\scriptscriptstyle{(2)}} - 10\kappa^{4/3}\,\vecX_3.\vectt^{\scriptscriptstyle{(1)}} = 0 
 \displaystyle{\langle \underline{A}^{\scriptscriptstyle{(3)}},\vecX_1\rangle = 6\,c_1^{\scriptscriptstyle{(2)}} - \frac{(M-1)(53-23M)}{6M^{\frac{3}{4}}} = 0} \\
 %2\,\vecX_1.\vectt^{\scriptscriptstyle{(2)}} + 9\kappa^3 %- 12u_1u_2 
 \displaystyle{\langle \underline{A}^{\scriptscriptstyle{(3)}},\vecX_3\rangle = (8M-11)\left(\frac{8(7-M)}{3M}\right)^{\frac{1}{3}}c_2^{\scriptscriptstyle{(1)}} + \frac{6(17-5M)}{7-M}\,c_3^{\scriptscriptstyle{(2)}} = 0}
 %10\,\vecX_2.\vectt^{\scriptscriptstyle{(2)}} - 16\kappa^{4/3}\,\vecX_3.\vectt^{\scriptscriptstyle{(1)}} = 0
 \end{cases} \notag \\
 &\longrightarrow\quad 
 \begin{cases}
 \displaystyle{c_3^{\scriptscriptstyle{(2)}} = 0} \\
 \displaystyle{c_1^{\scriptscriptstyle{(2)}} = -\frac{(M-1)(23M-53)}{36M^{\frac{3}{4}}}}\\
 %\frac{3}{2}u_1\left(4u_2-3u_1^2\right) %-\frac{9}{2}\kappa^3 \\
 \displaystyle{c_2^{\scriptscriptstyle{(1)}} = 0}
 \end{cases} \;,
 \end{align}
 where we used \eqref{eq:systemt1}. The first two equations in (\ref{eq:systemt1}) together with the last equation in (\ref{eq:systemt2}) fix $\vectt^{\scriptscriptstyle{(1)}}$ as
 \begin{equation}
 \label{eq:systemt2sol}
 \vectt^{\scriptscriptstyle{(1)}} =  \frac{1}{2}\left(\frac{7-M}{3M^{\frac14}}\right)^{\frac{5}{3}}\vecX_3 \;.
 \end{equation}
 From $S^{\scriptscriptstyle{(3)}}$ we get
 \begin{align}
 \label{eq:systemt3}
 S^{\scriptscriptstyle{(3)}}&:\begin{cases}
 \displaystyle{\langle \underline{A}^{\scriptscriptstyle{(3)}},\vecX_2\rangle = 30\,c_2^{\scriptscriptstyle{(2)}} - 36\left(\frac{3M}{(7-M)^4}\right)^{\frac{1}{3}}c_3^{\scriptscriptstyle{(3)}} = 0} \\
 %4\,\vecX_2.\vectt^{\scriptscriptstyle{(3)}} - 10\kappa^{4/3}\,\vecX_3. \vectt^{\scriptscriptstyle{(2)}} = 0 \\
 \displaystyle{\langle \underline{A}^{\scriptscriptstyle{(4)}},\vecX_1\rangle = 6\,c_1^{\scriptscriptstyle{(3)}} = 0} \\
 %\vecX_1.\vectt^{\scriptscriptstyle{(3)}} = 0 \\
 \displaystyle{\langle \underline{A}^{\scriptscriptstyle{(4)}},\vecX_3\rangle = \left(\frac{8(7-M)}{3M}\right)^{\frac{1}{3}}(8M-11)\,c_2^{\scriptscriptstyle{(2)}} + \frac{6(17-5M)}{7-M}\,c_3^{\scriptscriptstyle{(3)}} = 0}
 %10\,\vecX_2.\vectt^{\scriptscriptstyle{(3)}}-16\kappa^{4/3}\,\vecX_3.\vectt^{\scriptscriptstyle{(2)}} = 0
 \end{cases} \notag \\
 &\longrightarrow\quad 
 \begin{cases}
 c_3^{\scriptscriptstyle{(3)}} = 0 \\
 c_1^{\scriptscriptstyle{(3)}} = 0 \\
 c_2^{\scriptscriptstyle{(2)}} = 0
 \end{cases} \;,
 \end{align}
 where we used \eqref{eq:systemt1} and \eqref{eq:systemt2}. The first two equations in (\ref{eq:systemt2}) together with the last equation in (\ref{eq:systemt3}) fix $\vectt^{\scriptscriptstyle{(2)}}$ as
 \begin{equation}
 \vectt^{\scriptscriptstyle{(2)}} = %\frac{u_1}{2}\left(4u_2 - 3u_1^2\right)
 -\frac{(M-1)(23M-53)}{36M^{\frac{3}{4}}}\,\vecX_1 \;.
 \end{equation}
 Now, we are finally ready to deal with the general case. Using the explicit expressions of $\vectt^{\scriptscriptstyle{(0)}}$, $\vectt^{\scriptscriptstyle{(1)}}$ and $\vectt^{\scriptscriptstyle{(2)}}$ together with $\vecA^{\scriptscriptstyle{(m)}}$, $m\geq 5$, the system $S^{\scriptscriptstyle{(m)}}$ for $m\geq 5$ yields
 \begin{align}
 S^{\scriptscriptstyle{(m)}}&:\begin{cases}
\displaystyle{\langle \underline{A}^{\scriptscriptstyle{(m)}},\vecX_2\rangle = - 36\left(\frac{3M}{(7-M)^4}\right)^{\frac{1}{3}}c_3^{\scriptscriptstyle{(m)}} +(M-4)\left(\frac{72(7-M)}{M}\right)^{\frac{1}{3}}c_1^{\scriptscriptstyle{(m-2)}}} \\
\displaystyle{+ 30\,c_2^{\scriptscriptstyle{(m-1)}} + \langle \vecf^{\scriptscriptstyle{(m-3)}},\vecX_2 \rangle = 0} \\
 %4\vecX_2.\vectt^{\scriptscriptstyle{(m)}} - 10\kappa^{4/3}\,\vecX_3.\vectt^{\scriptscriptstyle{(m-1)}} + 6\kappa^{8/3}\,\vecX_1.\vectt^{\scriptscriptstyle{(m-2)}} - \kappa^{4/3}\,\vecX_3.\vecF_2^{\scriptscriptstyle{(m-3)}}\bigl(\vectt^{\scriptscriptstyle{(0)}},\dots,\vectt^{\scriptscriptstyle{(m-3)}}\bigr) = 0 \\
 \displaystyle{\langle \underline{A}^{\scriptscriptstyle{(m+1)}},\vecX_1\rangle = 6\,c_1^{\scriptscriptstyle{(m)}} - (M^2-2M-17)\left(\frac{72}{M(7-M)^2}\right)^{\frac{1}{3}}c_3^{\scriptscriptstyle{(m-1)}}} \\
+\langle \vecf^{\scriptscriptstyle{(m-2)}},\vecX_1 \rangle = 0 \\
 %2\,\vecX_1.\vectt^{\scriptscriptstyle{(m)}} - 6\kappa^{4/3}\,\vecX_2.\vectt^{\scriptscriptstyle{(m-1)}} + \vecX_1.\vecF_2^{\scriptscriptstyle{(m-2)}}\bigl(\vectt^{\scriptscriptstyle{(0)}},\dots,\vectt^{\scriptscriptstyle{(m-2)}}\bigr) = 0 \\
 \displaystyle{\langle \underline{A}^{\scriptscriptstyle{(m+1)}},\vecX_3\rangle = \frac{6(17-5M)}{7-M}\,c_3^{\scriptscriptstyle{(m)}} + (8M-11)\left(\frac{8(7-M)}{3M}\right)^{\frac{1}{3}}c_2^{\scriptscriptstyle{(m-1)}}} \\
 + \langle \vecf^{\scriptscriptstyle{(m-2)}},\vecX_3 \rangle = 0
 %10\,\vecX_2.\vectt^{\scriptscriptstyle{(m)}} - 16\kappa^{4/3}\,\vecX_3.\vectt^{\scriptscriptstyle{(m-1)}} + \vecX_2.\vecF_2^{\scriptscriptstyle{(m-2)}}\bigl(\vectt^{\scriptscriptstyle{(0)}},\dots,\vectt^{\scriptscriptstyle{(m-2)}}\bigr) = 0
 \end{cases} \notag \\
 &\longrightarrow\quad
 \begin{cases} 
 c_3^{\scriptscriptstyle{(m)}} = \mathcal{H}^{\scriptscriptstyle{(m-1)}}\bigl(\vectt^{\scriptscriptstyle{(0)}},\dots,\vectt^{\scriptscriptstyle{(m-1)}}\bigr)  \\
 c_1^{\scriptscriptstyle{(m)}} = \mathcal{F}^{\scriptscriptstyle{(m-1)}}\bigl(\vectt^{\scriptscriptstyle{(0)}},\dots,\vectt^{\scriptscriptstyle{(m-1)}}\bigr)  \\
 c_2^{\scriptscriptstyle{(m-1)}} = \mathcal{G}^{\scriptscriptstyle{(m-2)}}\bigl(\vectt^{\scriptscriptstyle{(0)}},\dots,\vectt^{\scriptscriptstyle{(m-2)}}\bigr) 
 \end{cases} \;,
 \label{eq:generalsystemsol}
 \end{align}
 with 
 \begin{align}
 \mathcal{F}^{\scriptscriptstyle{(m-1)}} &= \left(\frac{1}{3M(7-M)^2}\right)^{\frac{1}{3}}(M^2-2M-17)\,c_3^{\scriptscriptstyle{(m-1)}} - \frac{1}{6}\langle \vecf^{\scriptscriptstyle{(m-2)}},\vecX_1 \rangle \;,\notag\\
 \mathcal{G}^{\scriptscriptstyle{(m-1)}} &= \frac{1}{18}\left(\frac{8}{3M(7-M)}\right)^{\frac{1}{3}}(M-4)(5M-17)\,c_1^{\scriptscriptstyle{(m-1)}} + \frac{5M-17}{54(7-M)}\langle \vecf^{\scriptscriptstyle{(m-2)}},\vecX_3 \rangle \notag \\
&-\frac{1}{3}\left(\frac{M}{9(7-M)^4}\right)^{\frac{1}{3}}\langle \vecf^{\scriptscriptstyle{(m-1)}},\vecX_2 \rangle \;,\notag\\
 \mathcal{H}^{\scriptscriptstyle{(m-1)}} &= \frac{(M-4)(8M-11)}{54}\left(\frac{8(7-M)^2}{9M^2}\right)^{\frac{1}{3}}c_1^{\scriptscriptstyle{(m-1)}}-\frac{5}{54}\langle \vecf^{\scriptscriptstyle{(m-2)}},\vecX_2 \rangle \notag \\
&+ \frac{8M-11}{162}\left(\frac{7-M}{3M}\right)^{\frac{1}{3}}\langle \vecf^{\scriptscriptstyle{(m-3)}},\vecX_3 \rangle \;.
 \end{align}
 Consider now $S^{\scriptscriptstyle{(m+1)}}$, which is obtained from \eqref{eq:generalsystemsol} replacing $m\to m+1$
 \begin{equation}
 \begin{cases} 
 c_3^{\scriptscriptstyle{(m+1)}} = \mathcal{H}^{\scriptscriptstyle{(m)}}\bigl(\vectt^{\scriptscriptstyle{(0)}},\dots,\vectt^{\scriptscriptstyle{(m)}}\bigr) \\
 c_1^{\scriptscriptstyle{(m+1)}} = \mathcal{F}^{\scriptscriptstyle{(m)}}\bigl(\vectt^{\scriptscriptstyle{(0)}},\dots,\vectt^{\scriptscriptstyle{(m)}}\bigr) \\
 c_2^{\scriptscriptstyle{(m)}} = \mathcal{G}^{\scriptscriptstyle{(m-1)}}\bigl(\vectt^{\scriptscriptstyle{(0)}},\dots,\vectt^{\scriptscriptstyle{(m-1)}}\bigr)
 \end{cases} \;.
 \label{eq:generalsystemsol1}
 \end{equation}
 The first two equations in (\ref{eq:generalsystemsol}) together with the last equation in (\ref{eq:generalsystemsol1}) fix uniquely $\vectt^{\scriptscriptstyle{(m)}}$ as a function of $(\vectt^{\scriptscriptstyle{(0)}},\dots,\vectt^{\scriptscriptstyle{(m-1)}})$.  In this way, we proved that the vector function $\vectt(\e)$ with Puiseux expansion \eqref{eq:ansatzvec} is the unique solution to $F_k(\vectt,\e)=0 \;,\; k=1,2,3$, about $\e=0$ with $\vectt(0)=\underline{0}$, up to permutation of the arguments.
 
 \begin{flushright}
 $\square$
\end{flushright} 

\subsection{The case d=3}

We state the analogous of Proposition \ref{prop:d2}, when $d=3$, namely when $\nu=(3,2,1)$ and $P^{\snu}(t)=t^6$.
We omit the proof because it is very long and involved. 

\begin{proposition}\label{prop:d3}
There exists a neighbourhood $D$ of $\e=0$ and a unique -- up to permutation of the index $k=1,\dots,6$ --
algebraic solution
$t_k(\e) \;,\; k=1,\dots,6$, of the system \eqref{eq:BLZt} with the following Puiseux expansion
\begin{equation}  
\label{eq:ansatzvecd3}
\vectt(\e) = \e^{\frac13}\vectt^{\scriptscriptstyle{(0)}} +
  \sum_{l\geq 1} \varepsilon^{\frac{4l+1}{3}}\vectt^{\scriptscriptstyle{(l)}} \;,
\end{equation}
where $\vectt^{\scriptscriptstyle{(l)}}=(t_1^{\scriptscriptstyle{(l)}},\dots,t_6^{\scriptscriptstyle{(l)}})$.
The above solution has dominant term
\begin{align}\label{eq:ansatzt0d3}
t_k^{\scriptscriptstyle{(0)}}=a^{\scriptscriptstyle{-}}\omega^{k-1} \kappa^{\frac13} \;,\quad k=1,2,3 \;,\\
t_k^{\scriptscriptstyle{(0)}}=a^{\scriptscriptstyle{+}}\omega^{k-1} \kappa^{\frac13} \;,\quad k=4,5,6 \;,
\end{align}
 with $\kappa=\frac{5-2\alpha}{3 (2\alpha+2)^{\frac14}}$ ,  
 $a^{\scriptscriptstyle{\pm}}=\left(\frac{5\pm3\sqrt{5}}{2}\right)^{\frac13}$ and $\omega=e^{\frac{2\pi i}{3}}$.

Consequently, the corresponding solution of the BLZ system \eqref{eq:algblz} $z_k(L)=\varphi(t_k(\e),\e))$, $k=1,\dots, 6$, where 
  $\varphi$ is as per \eqref{eq:varphi} and $\e=L^{-\frac{1}{4}}$, has expansion
  \begin{align}\label{eq:zkd3}
   z_k(L)= \frac{L}{\alpha}+
\frac{ a^{\scriptscriptstyle{-}}\omega^{k-1} \kappa^{\frac13} (2\alpha+2)^{\frac34}}{\alpha} L^{\frac23} + \mathcal{O}(L^{\frac13}) \;,\quad k=1,2,3 \;,\notag \\
z_k(L)= \frac{L}{\alpha}+
\frac{a^{\scriptscriptstyle{+}}\omega^{k-1} \kappa^{\frac13} (2\alpha+2)^{\frac34}}{\alpha} L^{\frac23} + \mathcal{O}(L^{\frac13}) \;,\quad k=4,5,6 \;,
   \end{align}
Moreover, letting $P(z;L)=\prod_{k=1}^6 \big(z-z_k(L)\big)$, we have that
  \begin{equation*} 
P(z;e^{2\pi i }L) = P(z;L) \;,
\end{equation*}
where $P(z;e^{2\pi i }L)$ denotes the analytic continuation of $P(z;e^{2\pi i }L)$ along a small loop about $L=\infty$.

\end{proposition}

\begin{remark}
 We remark that Proposition \ref{prop:d2} and Proposition \ref{prop:d3} are not exactly the proof of the Conjecture \ref{conj:Pte}
 for the two partitions considered. In fact, in the two propositions we prove uniqueness and existence of an algebraic solution
 with the Puiseux series \eqref{eq:ansatzvec} and \eqref{eq:ansatzvecd3}. We do not show that any algebraic solution
 must necessarily has such an expansion. This is outside the scope of the present paper.
\end{remark}

\section{Perturbation Series. Partially degenerate case}
\label{sec:partially}
\noindent
In this section we consider partitions $\nu$ of $N$ such that $P^{\snu}(t)$ has $N-3$ non-trivial distinct roots, with $N=2n+3 \;,\; n\in\mathbb{N}^*$, and a triple trivial root. We order the roots $v_1^{\snu},\dots,v_{2n+3}^{\snu}$ as follows:
$v_1^{\snu},v^{\snu}_2,v^{\snu}_3=0$, and  $v_k^{\snu} := u^{\snu}_{k-3} \neq 0 \;,\; k=4,\dots, 2n+3$. According to formula \eqref{eq:algrational}, the roots $u_1^{\snu},\dots,u_{2n}^{\snu}$ solve the system
\begin{gather}
\widetilde{F}_a^{\snu}:= 2u_a^{\snu} -\sum_{b\neq a} \frac{4}{(u_a^{\snu}-u_b^{\snu})^3} -\frac{12}{(u_a^{\snu})^3} =0\;,\quad a=1,\dots,2n \;, \notag \\
\label{eq:Fta}
\sum_{a=1}^{2n} (u_a^{\snu})^{-(2l+1)}  =0 \;,\quad l=1,2 \;. 
\end{gather}
We denote by $\widetilde{\mathbf{J}}^{\snu}$ the Jacobian of the vector function
$\underline{\widetilde{F}}^{\snu} = \bigl(\widetilde{F}_1^{\snu},\dots,\widetilde{F}_{2n}^{\snu}\bigr)$:
\begin{equation}
\label{eq:jtildesnu}
\widetilde{\mathbf{J}}_{ab}^{\snu} = 2\delta_{ab}\left(1+\frac{18}{(u_b^{\snu})^4} + 
\sum_{c\neq b}\frac{12}{(u_b^{\snu}-u_c^{\snu})^{4}}\right) -\frac{12(1-\delta_{ab})}{(u_a^{\snu}-u_b^{\snu})^{4}} \;,\quad a,b=1,\dots,2n\;.
\end{equation}
Observe that if $\nu$ is one of such partitions then $\nu^*$ is too,
since $v^{\snus}_k=i v_k^{\snu}$. Moreover $\widetilde{\mathbf{J}}^{\snus}=\widetilde{\mathbf{J}}^{\snu}$.

We have the following Proposition.
 \begin{proposition}
 \label{prop:d2mixed}
 Let $\nu$ be a partition as described above.
Assume that the matrix $\widetilde{\mathbf{J}}^{\snu}$ is invertible
as well as the matrices $\mathbf{A}^{\snu}$, $\widetilde{\mathbf{A}}^{\snu}
-\widetilde{\mathbf{B}}^{\snu}(\widetilde{\mathbf{J}}^{\snu})^{-1}\widetilde{\mathbf{C}}^{\snu}$ and
$\mathbf{A}^{\snu}-\mathbf{B}^{\snu}(\widetilde{\mathbf{J}}^{\snu})^{-1}\mathbf{C}^{\snu}$, as per Definition \eqref{def:matricesmixed} below.
There exists a neighbourhood $D$ of $\e=0$ and a solution
 \begin{equation}
 \vectt^{\snu}: D\to\C^{2n+3} : \e\to \vectt^{\snu}(\e) = (t_1(\e),t_2(\e),t_3(\e), u_1^{\snu}(\e), \dots ,u_{2n}^{\snu}(\e)) \;,
 \end{equation}
 of the system \eqref{eq:BLZt}, with the following Puiseux expansion
 \begin{align}  
  t_k(\e) &= \varepsilon^{\frac13}t_k^{\scriptscriptstyle{(0)}} +
  \sum_{l\geq 1} \varepsilon^{\frac{2l+1}{3}}t_k^{\scriptscriptstyle{(l)}} \;,\quad k=1,2,3 \;, \notag \\
  u_a^{\snu}(\e) &= u_{a}^{\snu} + \sum_{l\geq 1}\e^l u_{a}^{\scriptscriptstyle{(l)}} \;,\quad a=1,\dots,2n \;.
\label{eq:ansatzvecmixed}
 \end{align}
 The above solution is unique up to permutation of the indices $1,2,3$, and 
\begin{equation}
 \label{eq:ansatzt0mixed}
  t_k^{\scriptscriptstyle{(0)}} =\omega^{k-1} b \;,\quad k=1,2,3 \;,
 \end{equation}
 with $\omega=e^{\frac{2\pi i}{3}}$ and for some $b\in\C^*$.
 
 Consequently, the corresponding solution of the BLZ system \eqref{eq:algblz} $z_k^{\snu}(L)=\varphi(t_k^{\snu}(\e),\e)$, $k=1,\dots, 2n+3$,
 where $\varphi$ is as per \eqref{eq:varphi} and $\e=L^{-\frac{1}{4}}$, has expansion 
  \begin{align}
  \label{eq:zkmixedd2}
   z_k(L)&= \frac{L}{\alpha}+
  \frac{ \omega^{k-1} b\, (2\alpha+2)^{\frac34}}{\alpha} L^{\frac23} + \mathcal{O}(L^{\frac12}) \;,\quad k=1,2,3 \;, \\
 z_{a+3}^{\snu}(L) &= \frac{L}{\alpha}+
 \frac{(2\alpha+2)^{\frac34}}{\alpha} u_{a}^{\snu}  L^{\frac34} + \mathcal{O}(L^{\frac12}) \;,\quad a=1,\dots, 2n \;,
   \end{align}
  Moreover, letting $P^{\snu}(z;L)=\prod_{k=1}^{2n+3} \big(z-z_k^{\snu}(L)\big)$,  we have that
  \begin{equation*} 
P^{\snu}(z;e^{2\pi i }L) = P^{\snus}(z;L) ,
\end{equation*}
where $P^{\snu}(z;e^{2\pi i }L)$ denotes the analytic continuation of $P^{\snu}(z;L)$ along a small loop about $L=\infty$.
\end{proposition}

The long and technical proof of the Proposition is in the Appendix.

The matrices $\mathbf{A}^{\snu}$, $\widetilde{\mathbf{A}}^{\snu}$,
$\mathbf{B}^{\snu}$,$\widetilde{\mathbf{B}}^{\snu}$, $\mathbf{C}^{\snu}$, $\widetilde{\mathbf{C}}^{\snu}$, $\mathbf{C}^{\snu}$,
and $\mathbf{D}^{\snu}$, which are referred to in the hypothesis
of the Proposition \ref{prop:d2mixed} are as follows
\begin{definition}
\label{def:matricesmixed}
$\widetilde{\mathbf{A}}^{\snu}\in\C^{2,2}$, $\widetilde{\mathbf{B}}^{\snu}\in\C^{2,2n}$, $\widetilde{\mathbf{C}}^{\snu}\in\C^{2n,2}$ are
\begin{equation}
\widetilde{\mathbf{A}}^{\snu} = \frac{1}{6}\left(\begin{array}{cc}
\displaystyle{\sum_{a=1}^{2n}\frac{6}{(u_a^{\snu})^6}} & \displaystyle{1+\sum_{a=1}^{2n}\frac{6}{(u_a^{\snu})^4}} \\
\displaystyle{\frac{63}{50}\sum_{a=1}^{2n}\frac{1}{(u_a^{\snu})^8}+\frac{9}{200}\left(1+\sum_{a=1}^{2n}\frac{6}{(u_a^{\snu})^4}\right)^2} & \displaystyle{\sum_{a=1}^{2n}\frac{6}{(u_a^{\snu})^6}}
\end{array}\right) \notag \;,
\end{equation}
\begin{equation}
\widetilde{\mathbf{B}}^{\snu} = 12\left(\begin{array}{ccc}
(u_1^{\snu})^{-4} & \dots & (u_{2n}^{\snu})^{-4}\\
(u_1^{\snu})^{-6} & \dots & (u_{2n}^{\snu})^{-6}
\end{array}\right) \notag \;,\quad
\widetilde{\mathbf{C}}^{\snu} = 12\left(\begin{array}{cc}
(u_1^{\snu})^{-6} & (u_1^{\snu})^{-4} \\
\vdots & \vdots \\
(u_{2n}^{\snu})^{-6} & (u_{2n}^{\snu})^{-4}
\end{array}\right) .
\end{equation}
$\mathbf{A}^{\snu},\mathbf{D}^{\snu}\in\C^{3,3}$, $\mathbf{B}^{\snu}\in\C^{3,2n}$, $\mathbf{C}^{\snu}\in\C^{2n,3}$ are
\begin{equation}
\mathbf{A}^{\snu} = 6\left(\begin{array}{ccc}
\displaystyle{1+\sum_{a=1}^{2n}\frac{6}{(u_a^{\snu})^4}} & 0 & -8 \\
\displaystyle{\sum_{a=1}^{2n}\frac{6}{(u_a^{\snu})^6}} & \displaystyle{1+\sum_{a=1}^{2n}\frac{6}{(u_a^{\snu})^4}} & 0 \\
\displaystyle{\frac{63}{50}\sum_{a=1}^{2n}\frac{1}{(u_a^{\snu})^8}-\frac{2}{25}\left(1+\sum_{a=1}^{2n}\frac{6}{(u_a^{\snu})^4}\right)^2} & \displaystyle{\sum_{a=1}^{2n}\frac{6}{(u_a^{\snu})^6}} & \displaystyle{1+\sum_{a=1}^{2n}\frac{6}{(u_a^{\snu})^4} }
\end{array}\right) \notag \;,
\end{equation}
\begin{equation}
\mathbf{D}^{\snu} = \frac{9}{10}\left(\begin{array}{ccc}
0 & 0 & 0 \\
0 & 0 & 0 \\
\displaystyle{1+\sum_{a=1}^{2n}\frac{6}{(u_a^{\snu})^4}} & 0 & -8
\end{array}\right) \notag \;,
\end{equation}
\begin{equation}
\mathbf{B}^{\snu} = -36\left(\begin{array}{ccc}
0 & \dots & 0 \\
(u_1^{\snu})^{-4} & \dots & (u_{2n}^{\snu})^{-4}\\
(u_1^{\snu})^{-6} & \dots & (u_{2n}^{\snu})^{-6}
\end{array}\right) \notag \;,\quad
\mathbf{C}^{\snu} = -36\left(\begin{array}{ccc}
(u_1^{\snu})^{-6} & (u_1^{\snu})^{-4} & 0 \\
\vdots & \vdots & \vdots \\
(u_{2n}^{\snu})^{-6} & (u_{2n}^{\snu})^{-4} & 0
\end{array}\right) .
\end{equation}
\end{definition}

\begin{remark}
Numerical computations 
performed for various values of $n$ suggest that the coefficient $b$ in \eqref{eq:ansatzt0mixed} is $b=\kappa^{\frac13}$,
 with $\kappa=\frac{5-2\alpha}{3(2\alpha+2)^{\frac14}}$, as in the case of complete degeneracy of three roots, see equation \eqref{eq:ansatzt0}.
However, we were not able to prove it in general.
It is however clear from the proof that in the case $\alpha=\frac52$, the coefficient $b$ vanishes and the analysis of
the perturbation series must be amended. For sake of brevity, we omit the discussion of the case $\alpha=\frac52$ in the proof.
\end{remark}

\begin{remark}
The hypothesis that $\widetilde{\mathbf{J}}^{\snu}$, as well as the matrices $\mathbf{A}^{\snu}$, $\widetilde{\mathbf{A}}^{\snu} -\widetilde{\mathbf{B}}^{\snu}(\widetilde{\mathbf{J}}^{\snu})^{-1}\widetilde{\mathbf{C}}^{\snu}$ and $\mathbf{A}^{\snu}-\mathbf{B}^{\snu}(\widetilde{\mathbf{J}}^{\snu})^{-1}\mathbf{C}^{\snu}$ are invertible holds for a partition $\nu$ if and only if it holds for the partition $\nu^*$, since the roots of a partition and its conjugate satisfy the relation $v^{\snus}_k=i v_k^{\snu}$. Moreover, numerical computations performed for various values of $n$ suggest that the hypothesis is always satisfied.
\end{remark}

\section{Integer 2 alpha}\label{section:integer}
The functional form of the monster potentials \eqref{eq:Vgen} when $\alpha$ is irrational may be slightly puzzling for the reader.
Sometimes, in order to develop some `physical' intuition, it is better to think of the case $2\alpha \in \bb{N}$.

Let us thus consider a potential of the form
\begin{equation}\label{eq:Minteger}
V_P(x)=\frac{L}{ x^2}+x^{M-2}-2 \frac{d^2}{d x^2}\log  P(x) \;,\quad \bb{N}\ni M \geq 3 \;,
\end{equation}
where $P$ is a monic polynomial.
\begin{definition}\label{defi:particles}
 $\pjm \subset \C^{J+1}$ consists of the ordered pairs $(P,L)$
 where $L$ is a complex number, and $P$ is a monic polynomial of degree $J$ with distinct and non-zero roots, such that
 the potential \eqref{eq:Minteger} has trivial monodromy at every root of $P$.
$\Pjm$ is the closure of $\pjm$ in $\C^{J+1}$.
\end{definition}
After Theorem \ref{thm:dui}, we have that $(P,L) \in \pjm$ if and only if
the roots $x_1,\dots, x_J$ of $P$ satisfy the system
\begin{equation}\label{eq:particles}
 -\frac{2L}{x_k^3}+(M-2)x_k^{M-3}-\sum_{j\neq k}\frac{4}{(x_k-x_j)^3} =0 \;,\quad k=1\dots J \;.
\end{equation}
According to our experience, the latter system is much simpler to analyse than the BLZ system \eqref{eq:algblz}.
It moreover coincides with the system describing the the complex equilibria of the following Calogero-Moser-like classical Hamiltonian
\begin{equation}\label{eq:hamiltonian}
 H=\sum_{j}p_j^2+ x^{M-2}+\frac{L}{x^2}+\sum_{j< j'}\frac{2}{(x_j-x_{j'})^2} \;.
\end{equation}
In fact, it is easy to verify that the system \eqref{eq:particles} is the equation for the critical points
of the potential of the above Hamiltonian, which represents a system of $J$ 
particles with an `inverse-square interaction-potential', subject to the external field $x^{M-2}+\frac{L}{x^2}$
\footnote{A deeper interpretation of the monster potentials with $2\alpha$ integer
in terms of classical integrable systems was obtained by Fioravanti in \cite{fioravanti05}.}.

It is straightforward to show that the higher state (and monster) potentials
belong to $\Pjm$, provided $2\alpha+2=M$ and $J=M \times N$.
\begin{lemma}\label{lem:integerpot}
Assume $2\alpha+2=M\geq 3$ is an integer.
The locus $\bna$ (resp. $\Bna$) coincides with the subset of
the locus of $\pjm$ (resp. $\Pjm$) of all those potentials $P$ such that
\begin{itemize} 
 \item $J=M \times N$;
 \item The polynomial has the symmetry $P(\gamma_M x)= P(x)$, where
 \begin{equation}\label{eq:gammaM}
  \gamma_M=e^{\frac{2\pi i}{M}}\;.
 \end{equation}

\end{itemize}
In particular, if $(P,L) \in \pjm$ satisfies the conditions above, and $x_1$, $\dots$, $x_J$ are
the roots of $P$, then one can construct exactly $N$ distinct quantities of the form $z_k:=x_k^M$ for some $k=1,\dots,J$,
and these solve the BLZ system \eqref{eq:algblz}.
\begin{proof}
 This follows from a simple fact. If $P(\gamma_M x)=P(x)$ then 
 there exists a polynomial $Q$ of degree $N=\frac{J}{M}$ such that
 $P(x)=Q(x^M)$. In particular $x$ is a root of $P$ if and only if $x^M$ is a root of $Q$.
\end{proof}
\end{lemma}

In the rest of this section, we study the locus $\Pjm$ in the large momentum limit, extending the analysis that we have developed for solutions
of the BLZ system. In particular, we state the analogous of Proposition \ref{pro:largez} and Theorem \ref{thm:toharmonic}
for the system \eqref{eq:particles}.
We skip most details of the proofs since these are just a simplified version of the proofs of
Proposition \ref{pro:largez} and Theorem \ref{thm:toharmonic}.
Therefore we keep the discussion more informal than in the rest of the paper.

\subsection{Weakly interacting particle subsystems}\label{sub:weakly}
We begin with the analogous of Proposition \ref{pro:largez}
\begin{proposition}\label{pro:Mlargex}
 Fix $\bb{N}\ni M\geq 3$ and $J \in \mathbb{N}$.

\begin{enumerate}
 \item For every $L \in \mathbb{C}$, There exists a $C_{L}>0$ such that if $x_1$, $\dots$, $x_J$ is a solution of the the BLZ system \eqref{eq:particles},
 then $|z_k|<C_{L}$, for all $k=1\dots,J$.
\item  If $L^{\sn}$ is a diverging sequence of complex numbers  $L^{\sn} \to \infty$ and
$x_1^{\sn}$, $\dots$, $x_J^{\sn}$ a sequence of solutions of the BLZ system with $L=L^{\sn}$,
then \begin{equation}\label{eq:largeLparticles}
      \lim_{n \to \infty} \frac{(M-2) (x_k^{\sn})^M}{2 L^{\sn}} \to 1 \;,\quad \forall  k  \in \lbrace 1,\dots,J \rbrace \;.
     \end{equation}
\end{enumerate}
\begin{proof}
We can prove the thesis following the same steps of the proof of Proposition \ref{pro:largez}, but with a simplification.

We notice that \eqref{eq:particles} can be written as
\begin{equation}\label{eq:particleproof}
 V'_G(x_k)+\sum_{j\neq k} I(x_k,I_j)=0 \;,\quad I(x,y)=-\frac{4}{(x-y)^3} \;,
\end{equation}
and $V_G$ is as per \eqref{eq:ground} with $M=2\alpha+2$.
This is the analogous of the decomposition \eqref{eq:BLZlarge}.
The analogous of Lemma \ref{lem:boundedinteratcion} and Lemma \ref{lem:boundedinteratcionsymm} holds:
Let
$x^{\sn},y^{\sn}$ be a sequence of a pair of complex numbers such that their ratio converges to a number $w \in \overline{\C}$,
$\frac{x^{\sn}}{y^{\sn}} \to w \in \overline{\C}$, then
\begin{enumerate}
 \item If $w \neq 1$, there exist a $C>0$ such that $|I(x^{\sn},y^{\sn}) |\leq C | x^{\sn} |^{-3}$.
 \item If $w = 1$, there exists a $C>0$ such that $|I(x^{\sn},y^{\sn}) + I (y^{\sn},x^{\sn})|\leq C | x^{\sn} |^{-3}$.
\end{enumerate}
The first inequality is easily seen to hold, the second is trivially satisfied because $I(x,y)+I(y,x)=0$ since
$I(x,y)$ is anti-symmetric.

The reader can now easily check that the very same proof of Proposition \ref{pro:largez} leads to
thesis.

\end{proof}

\end{proposition}

After Proposition \ref{pro:Mlargex}, the roots of system \eqref{eq:particles} condensate about the points
\begin{equation}\label{eq:minimaparticles}
 m_l=\gamma_M \left(\frac{2 L}{M-2}\right)^{\frac{1}{M}} \;,\quad l=0,\dots,M-1 \;,
\end{equation}
where $\gamma_M$ is as per \eqref{eq:gammaM}. These are the zeros of the derivative
of the external potential $V_G(x)$.

We can therefore study all equilibria configurations in the large $L$ limit, by localising the system \eqref{eq:particles}
about these points. In fact, we show that the system \eqref{eq:particles} splits into $M$ weakly interacting sub-systems,
and within each subsystem the particles lie approximately at the poles of a rational extension of the harmonic oscillator.

To start with, we denote by $N_l$ the number of particles which lay around $m_l$, $l=0,\dots,M-1$, and,
for all $N_l\neq0$, we denote by $x_{l,k}\;,\;1\leq k \leq N_l$, the location of the corresponding particles.
Finally we make the change of variables 
\begin{equation}\label{eq:xlktot}
 x_{l,k}= m_l \big(1+ M^{-\frac14} \, \e \, t_{l,k} \big) \;.
\end{equation}
This coincides with the change of variable \eqref{eq:xtot}, introduced in the study of the monster potentials.
Plugging \eqref{eq:xlktot} in the system \eqref{eq:particles} we obtain
\begin{align} \nonumber
& 2\e^{-4} \frac{-1 + (1 + M^{-\frac14}\e\,t_{l,k})^M}{(1 +  M^{-\frac14}\e\,t_{l,k})^3}-\e^{-3}\sum_{j\neq k}
\frac{ 4 M^{\frac{3}{4}}}{(t_{l,k}-t_{l,j})^3}+\\ \label{eq:weaklyinteracting}
& -\sum_{l'\neq l}\sum_{j=1}^{N_{l'}}
\frac{  4 M^{\frac{3}{4}}}
{\left((1 +  M^{-\frac14} \e\,t_{l,k})-\gamma^{l'-l} (1 +  M^{-\frac14} \e\,t_{l',j})\right)^3} =0 \;,
\end{align}
for all $l=0\dots M-1$, $k=1 \dots N_l$ .
 In the above equation,
we have separated 3 contributions: The interaction with the external field,
the interaction with particles in the same subsystem, the interaction with particles in a different subsystem.

Expanding the equation \eqref{eq:weaklyinteracting} in power of $\e$ and collecting a common factor, we obtain
\begin{equation}\label{eq:weaklysmalle}
 2t_{l,k}- \sum_{j=1,j\neq k}^{N_l}\frac{4}{(t_{l,k}-t_{l,j})^3} + \e \frac{(7-M) }{M^{\frac14}} t_{l,k}^2  +O(\e^2)  =0 \;,
\end{equation}
for all $l=0\dots M-1$, $k=1 \dots N_l$ .
We notice two things. First, for $L$ large, the particle sub-systems asymptotically decouple, as in fact the first
interaction between different sub-systems in \eqref{eq:weaklysmalle} is of order $\mathcal{O}(\e^{3})$.
Second, the above system coincides up to order $\mathcal{O}(\e^2)$ with the system for the rational extensions
of $M$ uncoupled perturbed harmonic oscillators \eqref{eq:rationale} with  $\kappa=\frac{(7-M) }{3M^{\frac14}}$
\footnote{Unless when $M=7$. In which case, we need to consider the potential
$U(t)=t^2+ \e^2  \mu t^4$ where $\mu=-7^{-\frac12}$.}.

\begin{definition}
A $M$-partition of $J\geq 1$ consists of a sequence of $M$, possibly empty, partitions
$\nu^{\scriptscriptstyle{(l)}}$ of non-negative numbers $N_l$, $l=0,\dots,M-1$, which sum up to $J$,
$\sum_{l=0}^{M-1}N_l=J$.

We denote by $p_M(N)$ the number of $M$-partitions of
$N$.
\end{definition}
We have the following Theorem, whose proof we omit
since it follows from the very same arguments which led us to the proof of Theorem \ref{thm:toharmonic}.
\begin{theorem}\label{thm:Mztot}
Fix $\bb{N} \ni M \geq 3$ and $J \in \mathbb{N}$.
Let $(P^{\sn},L^{\sn}) \in \Pjm$ be a sequence such that $L^{\sn} \to \infty$.
Furthermore assume that
$L^{\sn}$ belongs, for $n$ large, to the complex plane minus a cut connecting $0$ with $\infty$, so that
$(L^{\sn})^{\frac14}$ can be unambiguously defined.

The sequence $P^{\sn}$ can be split into $K$ subsequences, with $1\leq K\leq p_M(N) $, each subsequence associated to a unique
$M$-partition $\lbrace \nu^{\scriptscriptstyle{(0)}},\dots,n^{\scriptscriptstyle{(M-1)}}\rbrace$ of $J$ in such a way that the,
appropriately ordered, roots $x_{l,j}^{\sn}$ of $P^{\sn}$ satisfy the following asymptotics
\begin{equation}\label{eq:Mtkaslimit}
x_{l,j}=m_l\left( 1 + M^{-\frac14} v_j^{\scriptscriptstyle{[\nu^{(l)}]}} \big(L^{\sn}\big)^{-\frac14}+ o\left(|L^{\sn}|^{-\frac14} \right)\right)\;.
\end{equation}
Here $v_j^{\scriptscriptstyle{[\nu^{(l)}]}}$, $j=1,\dots,N_l$, are the roots of the polynomial $P^{\scriptscriptstyle{[\nu^{(l)}]}}$, associated to the partition $\nu^{\scriptscriptstyle{(l)}}$ of $N_l$
according to formula \eqref{eq:Pnu}.
\end{theorem}

It remains to prove that for each $M$-partition of $J$ there exists a unique one-parameter family of equilibria
$P(x;L)$, whose roots satisfy \eqref{eq:Mtkaslimit}. This is the very same task that we addressed in Section 6, 7 and 8
for the BLZ system.
If the multiplicity of the zero root is not greater than $6$ for each polynomial $P^{[\nu_l]}$, we can
actually prove that such a unique family of equilibria exists.
The general case is however an open problem.

\begin{remark}
A particularly interesting case is $M=4$ with the symmetry $\bb{Z}_2$
(namely $x$ belongs to the equilibrium if and only if $-x$ belongs to the
equilibrium). In fact, in this case the monster potentials are explicitly expressed
in term of Wronskians of Laguerre (or Hermite \footnote{In particular
when $L=l(l+1)$ with $l$ a positive integer, these Wronskian of Laguerre polynomials can be expressed 
in term of Wronskians of Hermite polynomials. In fact, 
due to the $\bb{Z}_2$ symmetry, the coefficients of all odd terms in the Laurent expansion at $0$ of the potential
vanish. Hence, according to Theorem \ref{thm:dui}, the monodromy is trivial at $x=0$, hence
these potentials are rational extensions of the harmonic oscillator, such that $0$ has multiplicity
$\frac{l(l+1)}{2}$}) polynomials \cite{dunningpers}.
According to our Theorem \ref{thm:Mztot}, the non-trivial roots of these Wronskians
of Laguerre (or Hermite) polynomials
condensate about the points $L^{\frac14}$, and they obey the asymptotics
\eqref{eq:Mtkaslimit} for a symmetric $4-$partition of $2N$, i.e. for a $2-$partition of $N$.
This explains the numerical findings, and it is in accordance with the theoretical description, of \cite{dunning20,dunningpers}.
\end{remark}

\begin{remark}
Proposition \ref{pro:Mlargex} and Theorem \ref{thm:Mztot} can be easily extended to
study the large $L$ limit of the following more general problem.
Fixed an arbitrary polynomial $\Pi$ of degree $M-2, M \geq 3$,
find the monic polynomials $P$ of degree $J$ such that the potential
$$V(x)=\Pi(x)+\frac{L}{x^2}-2\frac{d^2}{dx^2} \log P(x)$$
has trivial monodromy at all roots of $P$.

One discovers that the roots of $P$ condensate about the points $x$ such that $\Pi'(x)-2\frac{L}{x^3}=0$, and 
that about such points the roots are distributed approximately as the poles of a rational extension of the harmonic oscillator.
\end{remark}

\section{Numerics}

In this section, we test our results on the location of the roots of the BLZ system when $L$ is large, against numerical solutions of the same equation.

Recall that if one defines
\begin{equation*}
 z_k=\frac{L}{\alpha}\left( 1+ (2\alpha+2)^{-\frac14} t_k(L) L^{-\frac{1}{4}} \right)^{2\alpha+2} \;,\quad k=1,\dots, N, \;,
\end{equation*}
then the unknowns $t_1(L)$, $\dots$, $t_N(L)$ satisfy the perturbation series \eqref{eq:BLZt}, with initial conditions
\begin{equation*}
 \lim_{L \to \infty}t_k(L)=v_k^{\snu} \;.
\end{equation*}
In the above formula $\nu$ is a given partition of $N$, and $v_1^{\snu}$, $\dots$, $v_N^{\snu}$ are the roots of the Wronskian of Hermite polynomials
$P^{\snu}(t)$.

We test our formulas for $N=5$, $L=7 \times 10^4$ and $\alpha=\frac{\pi}{3}$.
For every partition $\nu$ of $5$, we compute the first non-trivial term in the perturbations series \eqref{eq:BLZt},
and we compare this against a numerical solution of the BLZ system \eqref{eq:algblz}.
There are $7$ partitions of $5$.
\begin{itemize}
 \item $5$ partitions are non-degenerate, and the roots of $P^{\snu}$ are all distinct.
 According to Proposition \ref{thm:notdegenerate}, the solutions of the BLZ system have the following asymptotics
 \begin{equation}\label{eq:zknotnumerics}
  z_k=\frac{L}{\alpha} + \frac{(2\alpha+2)^{\frac34}}{\alpha} v_k^{\snu} L^{\frac34} + \mathcal{O}(L^{\frac12}) \;.
 \end{equation}
 \item $2$ partitions are partially degenerate: $P^{\snu}$ has two distinct non-zero roots, while
 $0$ is a triple root. The corresponding solution of the perturbation series is considered in Proposition \ref{prop:d2mixed},
 according to which the following formula holds
 \begin{align}
  \label{eq:zkmixedd2num}
   & z_k(L)= \frac{L}{\alpha}+
  \frac{ e^{\frac{2\pi i k}{3}} \, (2\alpha+2)^{\frac34}}{\alpha}
  \kappa^{\frac13} L^{\frac23} + \mathcal{O}(L^{\frac12}) \;,\quad k=1,2,3 \;, \\
  & z_{a+3}^{\snu}(L) = \frac{L}{\alpha}+
 \frac{(2\alpha+2)^{\frac34}}{\alpha} u_{a}^{\snu}  L^{\frac34} + \mathcal{O}(L^{\frac12}) \;,\quad a=1,2 \;,
   \end{align}
   where $\kappa=\frac{5-2\alpha}{3 (2\alpha+2)^{\frac14}}$, and $u_{1}^{\snu}$, $u_{2}^{\snu}$ are the two non-trivial roots of $P^{\snu}$.
\end{itemize}
We also make the same test in the case $N=6$, for the partition
$(3,2,1)$ whose polynomial has a unique root, namely $0$, of multiplicity $6$. 
The solution of the perturbation series corresponding to such a partition is considered in Proposition \ref{prop:d3},
where we obtained the following asymptotics
  \begin{align}\nonumber
   & z_k(L)= \frac{L}{\alpha}+
\frac{ a^{\scriptscriptstyle{-}}e^{\frac{2\pi i k}{3}} \kappa^{\frac13} (2\alpha+2)^{\frac34}}{\alpha} L^{\frac23} + \mathcal{O}(L^{\frac13}) 
\;,\quad  k=1,2,3 \;, \\ \label{eq:zkd3num}
& z_k(L)= \frac{L}{\alpha}+
\frac{a^{\scriptscriptstyle{+}}e^{\frac{2\pi i k}{3}} \kappa^{\frac13} (2\alpha+2)^{\frac34}}{\alpha} L^{\frac23} + \mathcal{O}(L^{\frac13})
\;,\quad  k=4,5,6 \;,
   \end{align}
 with $\kappa=\frac{5-2\alpha}{3 (2\alpha+2)^{\frac14}}$, and
 $a^{\scriptscriptstyle{\pm}}=\left(\frac{5\pm3\sqrt{5}}{2}\right)^{\frac13}$.

\begin{figure}[H]
	\centering
	\begin{tabular}{ c c c }
		\includegraphics[width=6cm]{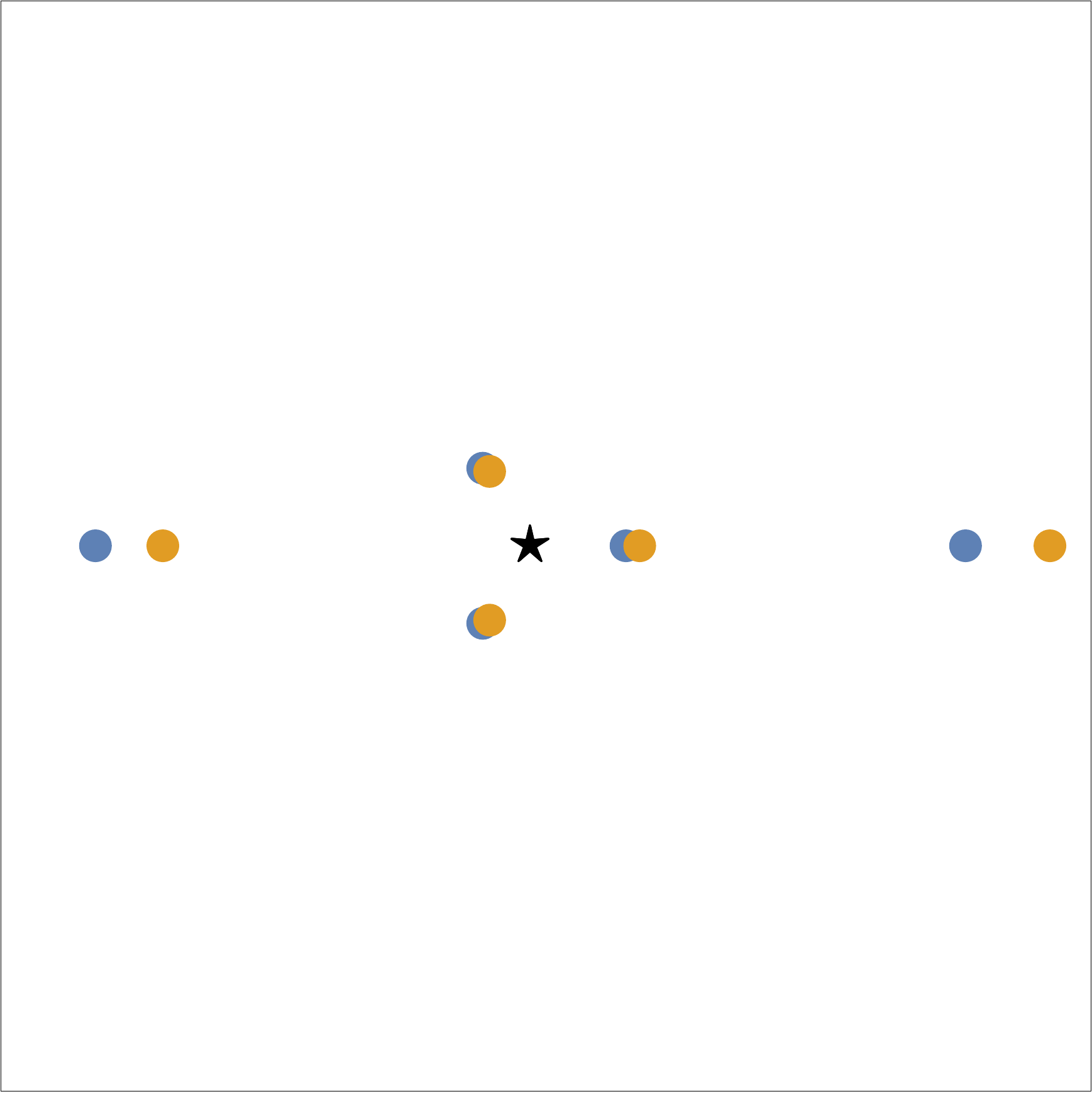} & \includegraphics[width=6cm]{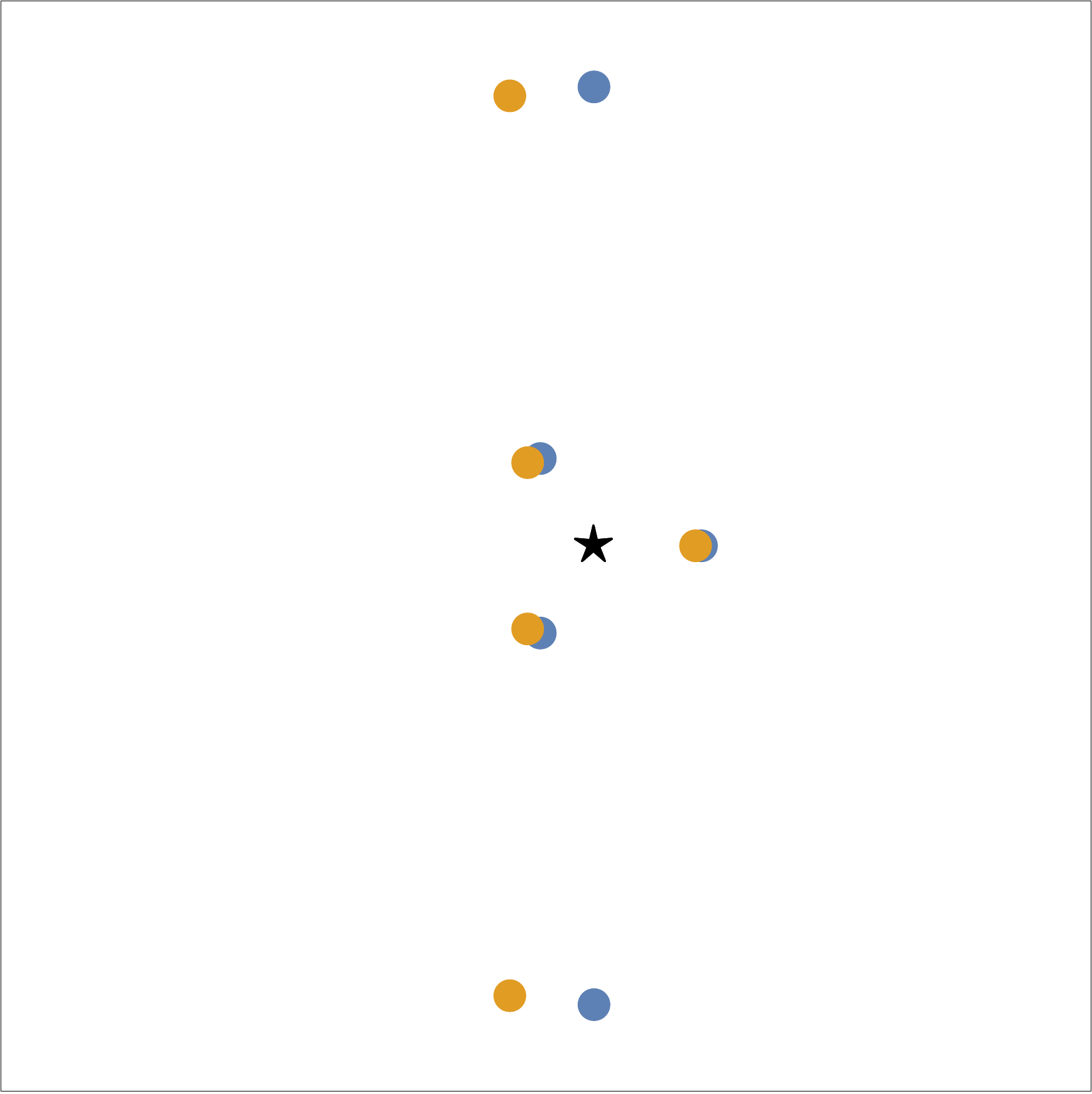} 
	\end{tabular}
	\caption{On the left, the case of the degenerate partition $(4,1)$, and on the right the conjugate case $(2,1,1,1)$.
	In yellow the numerical solution with $L=7 \times 10^4$ and $\alpha=\frac{\pi}{3}$, in blue the perturbative solution according to 
	formula \eqref{eq:zkmixedd2num}.
	The star is the point $z=\frac{L}{\alpha}=
	\frac{ 7 \pi 10^4}{3}$. }\label{fig:num1}
\end{figure}

\begin{figure}[H]
	\centering
	\begin{tabular}{ c c c }
		\includegraphics[width=6cm]{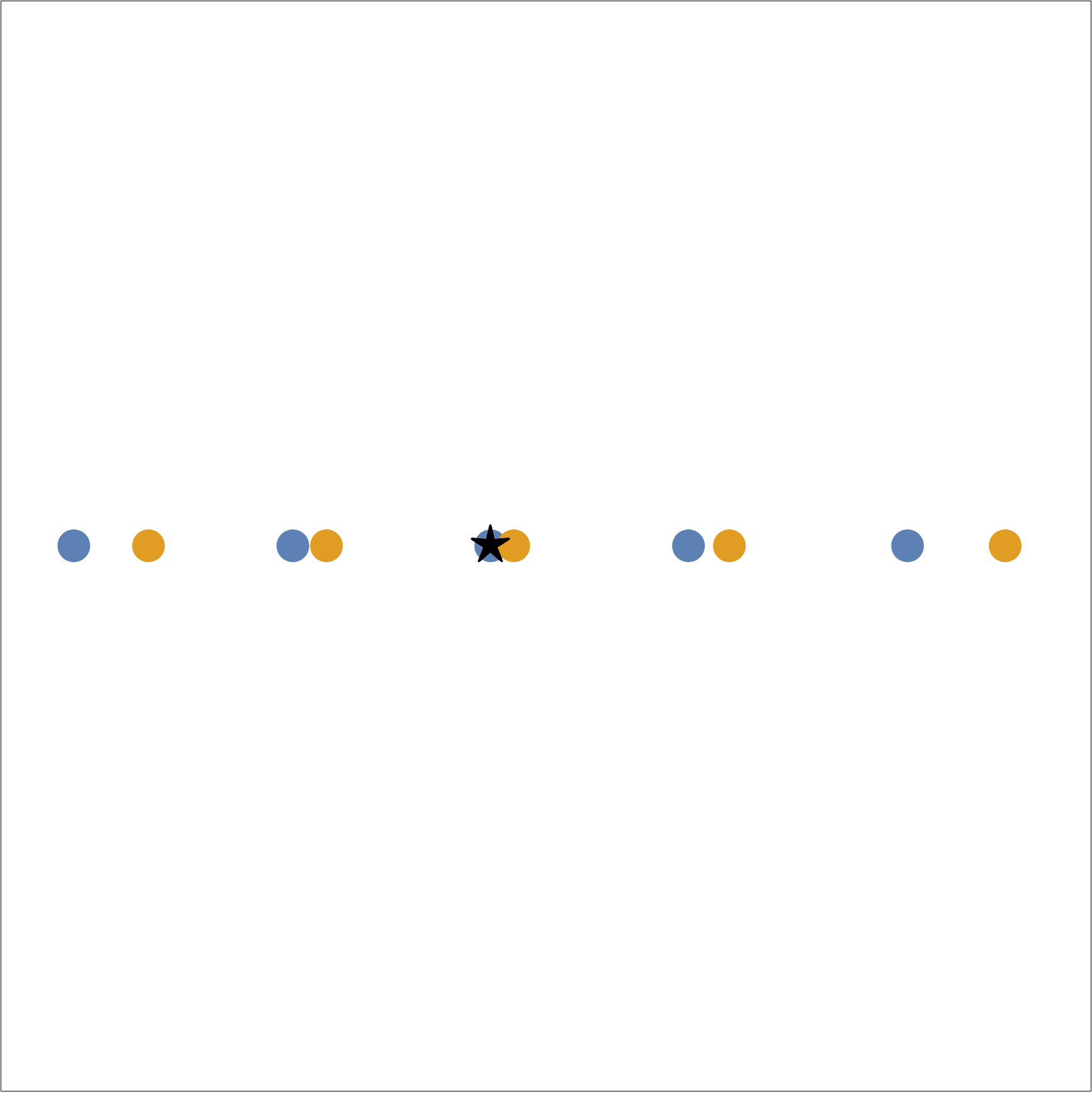} & \includegraphics[width=6cm]{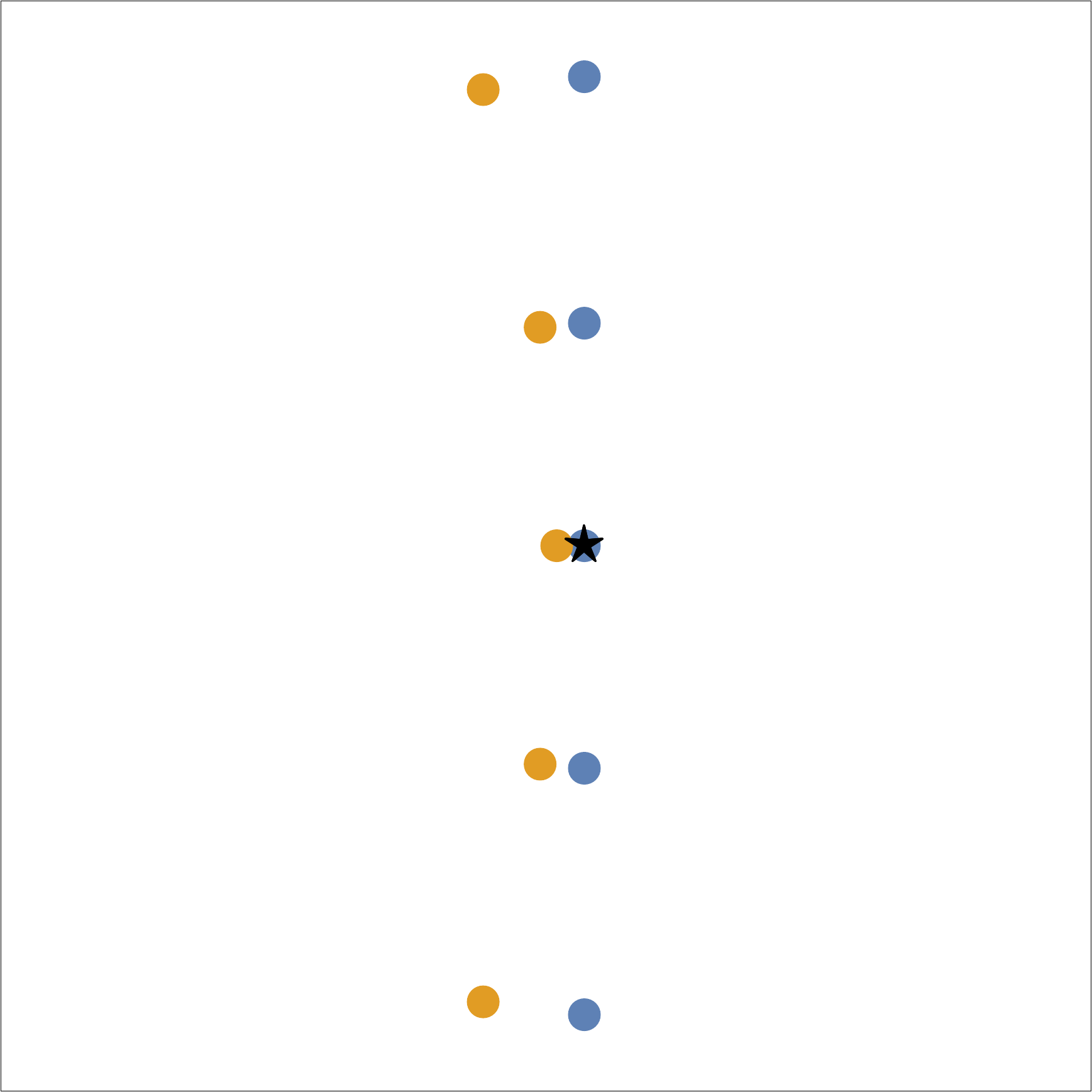} 
	\end{tabular}
	\caption{On the left, the case of the non-degenerate partition $(5)$, and on the right the conjugate case $(1,1,1,1,1)$.
	In yellow the numerical solution, in blue the perturbative solution according to 
	formula \eqref{eq:zknotnumerics}.
	The star is the point $z=\frac{L}{\alpha}=
	\frac{ 7 \pi 10^4}{3}$. }\label{fig:num2}
\end{figure}

\begin{figure}[H]
	\centering
	\begin{tabular}{ c c c }
		\includegraphics[width=6cm]{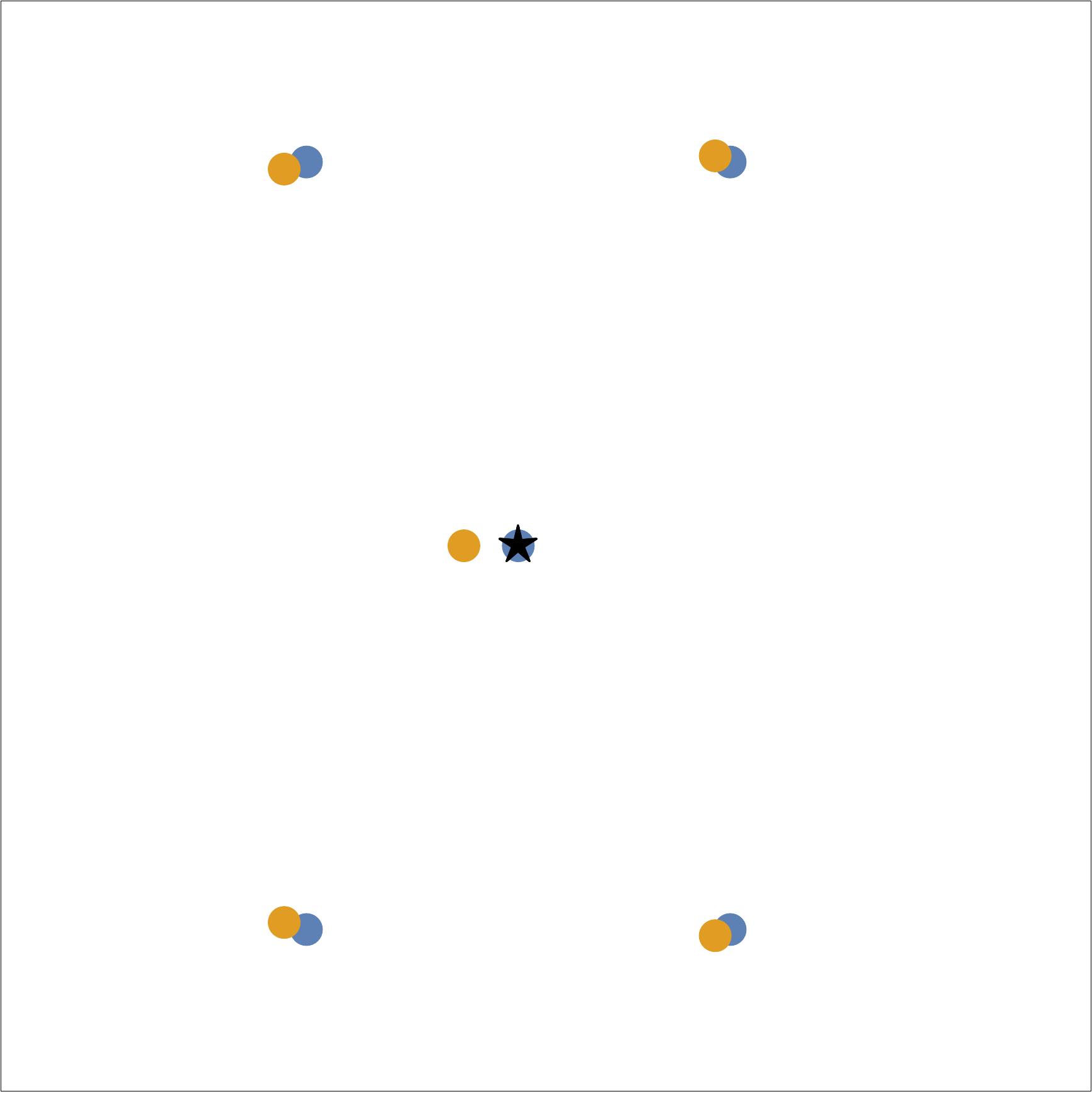} & \includegraphics[width=6cm]{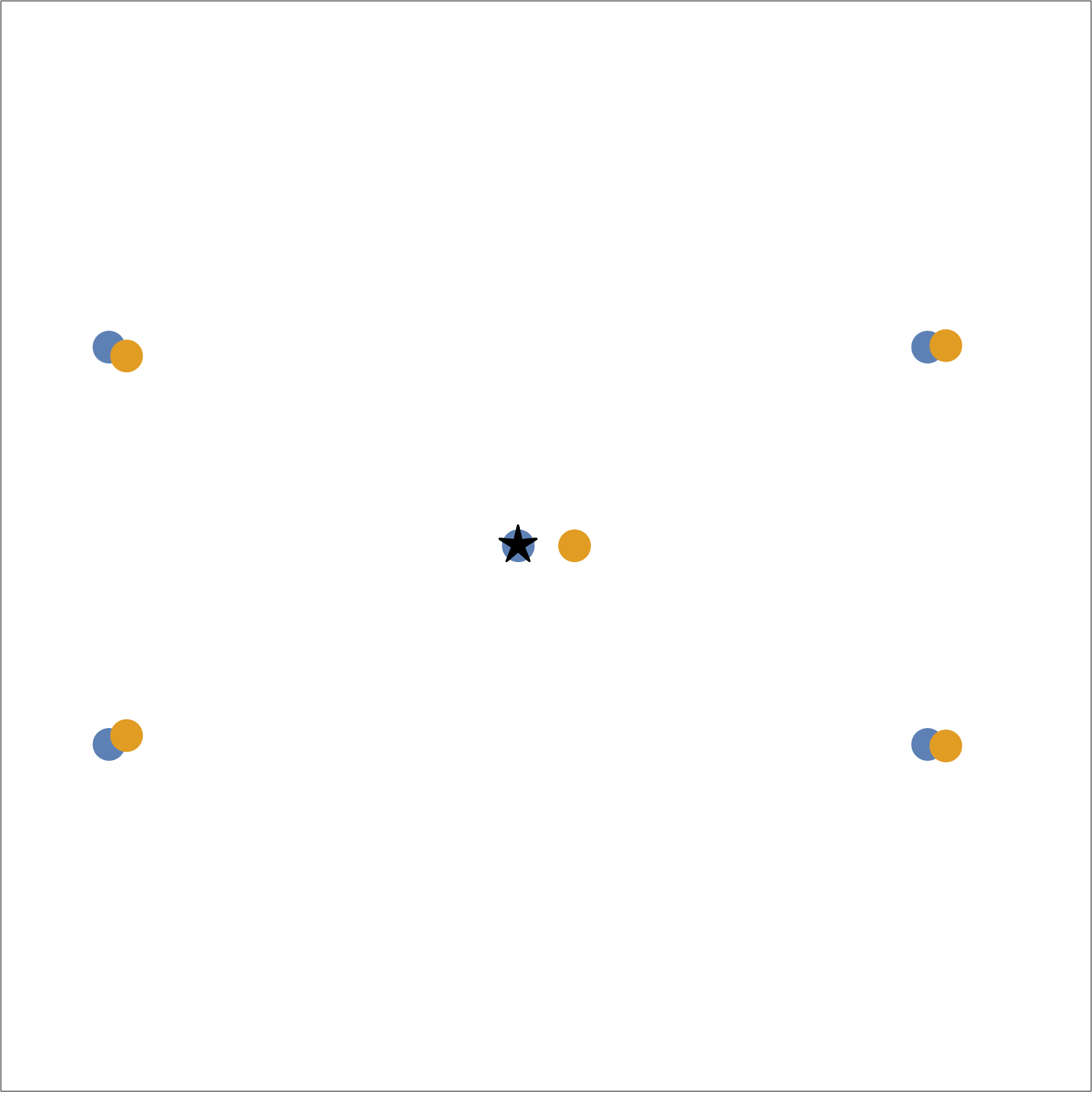} 
	\end{tabular}
	\caption{On the left, the case of the non-degenerate partition $(2,2,1)$, and on the right the conjugate case $(3,2)$.
	In yellow the numerical solution, in blue the perturbative solution according to 
	formula \eqref{eq:zknotnumerics}.
	The star is the point $z=\frac{L}{\alpha}=
	\frac{ 7 \pi 10^4}{3}$. }\label{fig:num3}
\end{figure}

\begin{figure}[H]
	\centering
	\begin{tabular}{ c c c }
		\includegraphics[width=6cm]{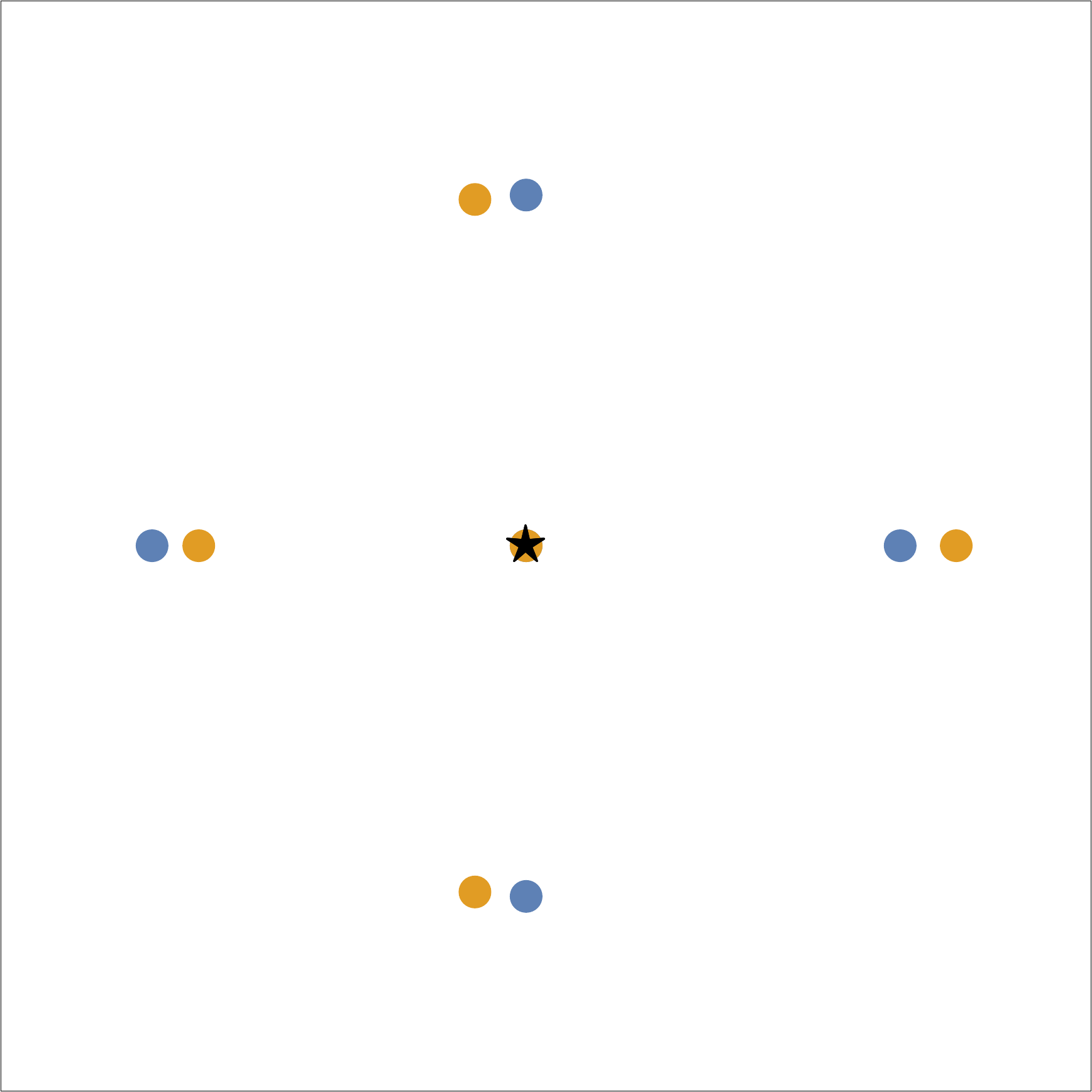} & \includegraphics[width=6cm]{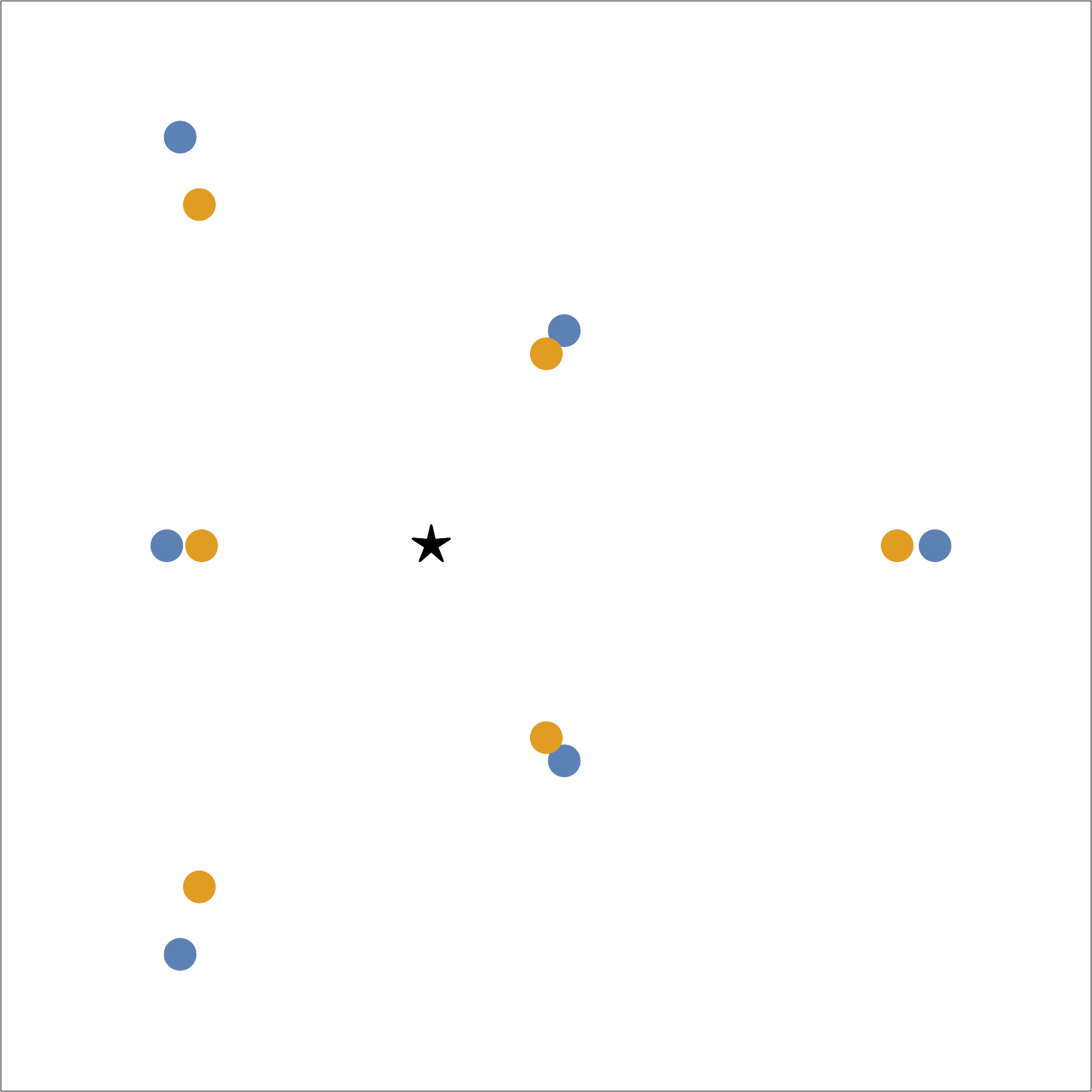} 
	\end{tabular}
	\caption{On the left, the case of the non-degenerate self-conjugate partition $(3,1,1)$ of $N=5$, and on the right the completely degenerate
	partition $(3,2,1)$ of $N=6$.
	In yellow the numerical solution, in blue the perturbative solution according to 
	formula \eqref{eq:zknotnumerics} for the case $(3,1,1)$, and according to formula \eqref{eq:zkd3num} for the case $(3,2,1)$.
	The star is the point $z=\frac{L}{\alpha}=
	\frac{ 7 \pi 10^4}{3}$. }\label{fig:num4}
\end{figure}

\section{Concluding Remarks}

The Bazhanov-Lukyanov-Zamolodchikov paper \cite{BLZ04}
has inspired many important developments in the theory of the ODE/IM correspondence, see e.g.
\cite{FF11,fioravanti05,bazhanov14,bazhanov19}. However, little we knew of the solutions of the BLZ system
\eqref{eq:algblz} prior to the present paper. This is possibly because the BLZ system is
a formidable system of algebraic equations whose solution is difficult even
numerically.

The large momentum limit gave us an entry to the study of the solutions of the BLZ system and revealed us many of their mysteries. 
In fact, we were able -- after thousands of pages of computations which ended up in the trash bin of our offices
and computers -- to reduce the study of the monster potentials to a perturbation of a much easier and better understood
problem, the rational extensions of the harmonic oscillator.
This fact alone provides us with so many information about the monster potentials that the BLZ conjecture seems
finally within reach.

Let us comment about this. According to our notation, the Weak BLZ conjecture is the fact that the number
of monster potentials with $N$ roots is $p(N)$, for generic $\alpha,L$.  As we showed, the Weak BLZ conjecture
follow directly from a much simpler and merely technical conjecture, namely our Conjecture \ref{conj:Pte}, which states that the perturbation
series \eqref{eq:BLZt} has a unique solution for every partition $\nu$ of $N$.

Assuming that Conjecture \ref{conj:Pte} holds (of which there is little doubt), one should be able to prove the full
BLZ conjecture by comparing the large momentum expansion
of the spectrum of the radial eigenvalue problem for the monster potentials, with the
large momentum expansion of solutions
of the Bethe Ansatz equations for the excited states of the Quantum KdV model. This again is a
technical, if beautiful and difficult, task. Our asymptotic formula of the radial spectrum, see equation \eqref{eq:spectrumeps},
is a good starting point in this direction, and further study is underway.

Another interesting problem which our paper opens is the study of the $p(N)$-sheeted Riemann surface $\Bna$, whose points
$(P(z),L)$ are the monster (more precisely, higher-state) potentials for fixed $N,\alpha$. We only know such a surface locally
in a neighbourhood of $L=\infty$. In fact, according to our results,
if $P^{\snu}(z,L)$ is the branch of this surface
which, for $L$ large, is associated
to the partition $\nu$ via the asymptotic formula \eqref{eq:zkexpthm}, then $P^{\snu}(z,e^{2\pi i}L)=P^{\snus}(z,L)$
where $\nu^*$ is the transpose of $\nu$ and $P^{\snu}(z,e^{2\pi i}L)$ the result of the analytic continuation along a small loop about $L=\infty$.
If one's only tool is the BLZ system \eqref{eq:algblz}, the global structure of Riemann surface $\Bna$ seems out of reach.
However, it is very possible,
and indeed desirable, that the IM side of the correspondence can give us more insight in this problem.

Finally, let us mention that the ODE/IM correspondence exist also for the
(generalised) Quantum $\mathfrak{g}$-KdV models, where $\mathfrak{g}$ is an untwisted Kac-Moody algebra
(the Quantum KdV model coincides with the case $\mathfrak{g}=\mathfrak{sl}_2^{(1)}$), and massive deformations of these models
\cite{lukyanov10,carr19}.
In the massless case, the analogous of the monster potentials are called Quantum KdV opers, they were
introduced by Feigin and Frenkel in \cite{FF11} and explicitly constructed in \cite{mara18,mara19}.
Of these opers very little is known and the analogous of the BLZ system
exists, but it is too intricate to be manipulated by hand. In the massive case,
hardly anything is known, even when $\mathfrak{g}=\mathfrak{sl}_2^{(1)}$. The large momentum limit
is a very promising tool to study both the massless and the massive case, and it will possibly uncover beautiful and unknown mathematical structures.

\section{Appendix}

\paragraph*{\it Proof of Proposition \ref{prop:d2mixed}}
Reasoning as in the beginning of the proof of Proposition \ref{prop:d2},
we deduce that we only need to prove the existence and uniqueness of a formal solution satisfying the Ansatz
\eqref{eq:ansatzvecmixed}. To prove the latter statement we will adopt a
strategy and a notation similar to that used in the proof of Proposition \ref{prop:d2}.
Hence the reader should recall the definition of the basis-vectors $\vecX_1$, $\vecX_2$,
$\vecX_3$ of $\C^3$, the $n$-linear maps $\vecM_n$ and the Hermitian product $\langle,\rangle$ as
per Definitions \ref{def:basis}, \ref{def:multimaps} and \ref{def:hadamard}, respectively.

The solvability of the perturbation series involves the invertibility of the following block matrices
\begin{equation}
\label{eq:PQmatrices}
\mathbf{P}_k^{\snu} = \left(\begin{array}{cc}
\mathbf{A}^{\snu}+k\mathbf{D}^{\snu} & \mathbf{B}^{\snu} \\
\mathbf{C}^{\snu} & \widetilde{\mathbf{J}}^{\snu}
\end{array}\right) \;,\quad
\mathbf{Q}_k^{\snu} = \left(\begin{array}{cc}
\mathbf{A}^{\snu}+k\mathbf{D}^{\snu} & \mathbf{0}_{3,2n} \\
\mathbf{0}_{2n,3} & \widetilde{\mathbf{J}}^{\snu}
\end{array}\right) \;,\quad k\in\C \;,
\end{equation}
where $\mathbf{0}_{p,q}$ denotes a $p\times q$ null matrix while $\mathbf{A}^{\snu}$, $\widetilde{\mathbf{A}}^{\snu}$, $\mathbf{B}^{\snu}$, $\widetilde{\mathbf{B}}^{\snu}$, $\mathbf{C}^{\snu}$, $\widetilde{\mathbf{C}}^{\snu}$, $\mathbf{D}^{\snu}$ and $\widetilde{\mathbf{J}}^{\snu}$ are as per Definition \ref{def:matricesmixed} and equation \eqref{eq:jtildesnu}, respectively.

From \eqref{eq:PQmatrices}, it is straightforward to prove the following Lemma.
\begin{lemma}
\label{prop:lemmamatricesmixed}
Assume that $\mathbf{A}^{\snu}$ is invertible. Then, for any $k\in\C$, one has
\begin{enumerate}
\item $\det{(\mathbf{A}^{\snu}+k\mathbf{D}^{\snu})}= \det{(\mathbf{A}^{\snu})} \;,$
\item $(\mathbf{A}^{\snu}+k\mathbf{D}^{\snu})^{-1}\mathbf{B}^{\snu}=(\mathbf{A}^{\snu})^{-1}\mathbf{B}^{\snu} \;.$
\end{enumerate}
\end{lemma}
An immediate consequence is this second Lemma.
\begin{lemma}
\label{prop:lemmablockmatricesmixed}
Assume that $\mathbf{A}^{\snu}$ is invertible. Then, for any $k\in\C$, one has
\begin{enumerate}
\item $\det{(\mathbf{P}_k^{\snu})}= \det{\big(\widetilde{\mathbf{J}}^{\snu}\big)}\det{\big(\mathbf{A}^{\snu}-\mathbf{B}^{\snu}(\widetilde{\mathbf{J}}^{\snu})^{-1}\mathbf{C}^{\snu}\big)} \;,$
\item $\det{(\mathbf{Q}_k^{\snu})}= \det{\big(\widetilde{\mathbf{J}}^{\snu}\big)}\det{(\mathbf{A}^{\snu})} \;.$
\end{enumerate}
\end{lemma}

Similarly to Proposition \ref{prop:d2}, we shall construct iteratively a solution $\vectt^{\snu}(\e)$ of the form \eqref{eq:ansatzvecmixed} to the system $F_k(\vectt,\e)=0 \;,\; k=1,\dots,2n+3$, with $F_k(\vectt,\e)$ defined as per \eqref{eq:Fk}, and show that such solution is unique.
Plugging the Ansatz \eqref{eq:ansatzvecmixed} in $F_k(\vectt,\e)$, the first perturbative contribution to the expansion of
$F_k(\vectt^{\snu}(\e),\e)$ about $\e=0$ is $\mathcal{O}(\e^{-1})$ and its cancellation corresponds to the set of equations
\eqref{eq:locust0} for $\vectt^{\scriptscriptstyle{(0)}}$, from which
\begin{equation}
 \label{eq:t0mixed}
 \vectt^{\scriptscriptstyle{(0)}} = a\vecX_1 + b \vecX_2 \;,\quad a,b\in\C \;,\quad b\neq 0 \;.
 \end{equation}
To fix the coefficients $a$ and $b$ in \eqref{eq:t0mixed}, we need to take into account higher perturbative contributions in the expansion of $F_k(\vectt^{\snu}(\e),\e)$. Setting $ \vectt^{\scriptscriptstyle{(0)}}$ as per \eqref{eq:t0mixed} one finds
\begin{align}
\label{eq:Fexpan}
 F_k(\vectt^{\snu}(\e),\e) &= -\frac{1}{4M^\frac14}\Biggl(A_k^{\scriptscriptstyle{(-1)}}\e^{-\frac13} + \sum_{l\geq 0}A_k^{\scriptscriptstyle{(l)}}\e^{\frac{l}{3}}\Biggr) \;,\quad k=1,2,3 \;, \notag \\
 F_{a+3}(\vectt^{\snu}(\e),\e) &=  -\frac{1}{4M^\frac14}\sum_{l\geq 0}B_{a}^{\scriptscriptstyle{(l)}}\e^{\frac{l}{3}} \;,\quad a=1,\dots,2n \;,
 \end{align}
 with $M=2\alpha+2$. The system $F_k(\vectt^{\snu}(\e),\e)=0 \;,\; k=1,\dots,2n+3$ can thus be expressed perturbatively as
\begin{equation}
\label{eq:systemvecmixed}
S:\begin{cases}
\vecA^{\scriptscriptstyle{(m)}} = \underline{0} \;,\quad m\geq -1 \\
\vecB^{\scriptscriptstyle{(m)}} = \underline{0}  \;,\quad m\geq 0
\end{cases} \;,
\end{equation} 
where $\vecA^{\scriptscriptstyle{(m)}} = (A_1^{\scriptscriptstyle{(m)}},A_2^{\scriptscriptstyle{(m)}},A_3^{\scriptscriptstyle{(m)}})$ , $\vecB^{\scriptscriptstyle{(m)}} = (B_1^{\scriptscriptstyle{(m)}},\dots,B_{2n}^{\scriptscriptstyle{(m)}})$ and the first few orders are given by
 \begin{align}
&A_k^{\scriptscriptstyle{(-1)}} = 12M_{1,k}(\vectt^{\scriptscriptstyle{(1)}}) \;,\quad A_k^{\scriptscriptstyle{(0)}} = \sum_{a=1}^{2n} \frac{4}{(u_a^{\snu})^3} \;,\quad \vecA^{\scriptscriptstyle{(2)}} = (t_k^{\scriptscriptstyle{(0)}})^{2}\sum_{a=1}^{2n}\frac{24}{(u_a^{\snu})^5}\;,\notag \\
&A_k^{\scriptscriptstyle{(1)}} = 12M_{1,k}(\vectt^{\scriptscriptstyle{(2)}}) - 24M_{2,k}(\vectt^{\scriptscriptstyle{(1)}},\vectt^{\scriptscriptstyle{(1)}}) + 2t_k^{\scriptscriptstyle{(0)}}\left(1+\sum_{a=1}^{2n}\frac{6}{(u_a^{\snu})^4}\right) \;, \notag \\
&A_k^{\scriptscriptstyle{(3)}} = 12M_{1,k}(\vectt^{\scriptscriptstyle{(3)}}) - 48M_{2,k}(\vectt^{\scriptscriptstyle{(1)}},\vectt^{\scriptscriptstyle{(2)}}) + 2t_k^{\scriptscriptstyle{(1)}}\left(1+\sum_{a=1}^{2n} \frac{6}{(u_a^{\snu})^4}\right) +\frac{36\,t_k^{\scriptscriptstyle{(0)}}}{M^{\frac14}}M_{1,k}(\vectt^{\scriptscriptstyle{(1)}}) \notag \\
&+40M_{3,k}(\vectt^{\scriptscriptstyle{(1)}},\vectt^{\scriptscriptstyle{(1)}},\vectt^{\scriptscriptstyle{(1)}})+(t_k^{\scriptscriptstyle{(0)}})^{3}\sum_{a=1}^{2n}\frac{40}{(u_a^{\snu})^6} -\sum_{a=1}^{2n}\frac{12\,u_a^{\scriptscriptstyle{(1)}}}{(u_a^{\snu})^4}\;, \notag \\
&\vdots \notag 
% &\vecA^{\scriptscriptstyle{(-1)}} = 12\vecM_1(\vectt^{\scriptscriptstyle{(1)}}) \;,\quad \vecA^{\scriptscriptstyle{(0)}} = 4\vecX_1\sum_{a=1}^{2n} \frac{1}{(u_a^{\snu})^3} \;,\quad \vecA^{\scriptscriptstyle{(2)}} = (\vectt^{\scriptscriptstyle{(0)}})^{\circ 2}\sum_{a=1}^{2n}\frac{24}{(u_a^{\snu})^5}\;,\notag \\
% &\vecA^{\scriptscriptstyle{(1)}} = 12\vecM_1(\vectt^{\scriptscriptstyle{(2)}}) - 24\vecM_2(\vectt^{\scriptscriptstyle{(1)}},\vectt^{\scriptscriptstyle{(1)}}) + 2\vectt^{\scriptscriptstyle{(0)}}\left(1+\sum_{a=1}^{2n}\frac{6}{(u_a^{\snu})^4}\right) \;, \notag \\
%&\vecA^{\scriptscriptstyle{(3)}} = 12\vecM_1(\vectt^{\scriptscriptstyle{(3)}}) - 48\vecM_2(\vectt^{\scriptscriptstyle{(1)}},\vectt^{\scriptscriptstyle{(2)}}) + 2\vectt^{\scriptscriptstyle{(1)}}\left(1+\sum_{a=1}^{2n} \frac{6}{(u_a^{\snu})^4}\right) +40 \vecM_3(\vectt^{\scriptscriptstyle{(1)}},\vectt^{\scriptscriptstyle{(1)}},\vectt^{\scriptscriptstyle{(1)}})\notag \\ &+\frac{36}{M^{\frac14}}\vectt^{\scriptscriptstyle{(0)}}\circ\vecM_1(\vectt^{\scriptscriptstyle{(1)}}) +(\vectt^{\scriptscriptstyle{(0)}})^{\circ 3}\sum_{a=1}^{2n}\frac{40}{(u_a^{\snu})^6} -12\vecX_1 \sum_{a=1}^{2n}\frac{u_a^{\scriptscriptstyle{(1)}}}{(u_a^{\snu})^4} \;, \notag \\
%&\vdots \notag 
\end{align}
\begin{align}
B_a^{\scriptscriptstyle{(0)}} &= \widetilde{F}_a^{\snu} \;,\quad B_a^{\scriptscriptstyle{(1)}} = -\frac{12}{(u_a^{\snu})^4}\sum_{k=1}^3 t_k^{\scriptscriptstyle{(0)}} \;,\quad
B_a^{\scriptscriptstyle{(2)}} = -\frac{24}{(u_a^{\snu})^5}\sum_{k=1}^3 \left(t_k^{\scriptscriptstyle{(0)}}\right)^2 \;, \notag \\
B_a^{\scriptscriptstyle{(3)}} &= \sum_{b=1}^{2n} \left(u_b^{\scriptscriptstyle{(1)}}\widetilde{\mathbf{J}}_{ab}^{\snu}\right) - \frac{12}{(u_a^{\snu})^4}\sum_{k=1}^3 t_k^{\scriptscriptstyle{(1)}} -\frac{40}{(u_a^{\snu})^6}\sum_{k=1}^3 (t_k^{\scriptscriptstyle{(0)}})^3 + \frac{M-7}{M^{\frac14}}(u_a^{\snu})^{2} \;. \notag \\
&\vdots 
%\vecB^{\scriptscriptstyle{(0)}} &= \underline{\widetilde{F}}^{\snu} \;,\quad \vecB^{\scriptscriptstyle{(1)}} = -12 \left(\vecu^{\snu}\right)^{\circ -4}\sum_{k=1}^3 t_k^{\scriptscriptstyle{(0)}} \;,\quad
%\vecB^{\scriptscriptstyle{(2)}} = -24 \left(\vecu^{\snu}\right)^{\circ -5}\sum_{k=1}^3 \left(t_k^{\scriptscriptstyle{(0)}}\right)^2 \;, \notag \\
%\vecB^{\scriptscriptstyle{(3)}} &= \vecu^{\scriptscriptstyle{(1)}}\widetilde{\mathbf{J}}^{\snu} - 12(\vecu^{\snu})^{\circ -4}\sum_{k=1}^3 t_k^{\scriptscriptstyle{(1)}} -40(\vecu^{\snu})^{\circ -6}\sum_{k=1}^3 (t_k^{\scriptscriptstyle{(0)}})^3 + (M-7)M^{-\frac14}(\vecu^{\snu})^{\circ 2} \;. \notag \\
%&\vdots 
%\notag \\
%\vecB^{\scriptscriptstyle{(3m)}} &= \vecu^{\scriptscriptstyle{(m)}}\tilde{\mathbf{J}}(\vecu^{\scriptscriptstyle{(0)}}) +\underline{g}_B^{\scriptscriptstyle{(3m)}}\left( \vectt^{\scriptscriptstyle{(0)}},\vectt^{\scriptscriptstyle{(1)}},\vecu^{\scriptscriptstyle{(0)}},\vecu^{\scriptscriptstyle{(1)}},\vectt^{\scriptscriptstyle{(2m-1-\lfloor m/2 \rfloor)}}\right) \notag \\
%&+\underline{f}_B^{\scriptscriptstyle{(3m)}}\left( \vectt^{\scriptscriptstyle{(0)}},\dots,\vectt^{\scriptscriptstyle{(2m-2-\lfloor m/2 \rfloor)}},\vecu^{\scriptscriptstyle{(0)}},\dots,\vecu^{\scriptscriptstyle{(m-1)}} \right)\;,\quad m\geq 3 \;.
 \end{align}
%-12(\vecu^{\scriptscriptstyle{(0)}})^{\circ -4}\sum_{k=1}^3 t_k^{\scriptscriptstyle{(m-1)}} -120(\vecu^{\scriptscriptstyle{(0)}})^{\circ -6}\sum_{k=1}^3 (t_k^{\scriptscriptstyle{(0)}})^2t_k^{\scriptscriptstyle{(m-2)}}
Notice that $\vecA^{\scriptscriptstyle{(0)}},\vecA^{\scriptscriptstyle{(2)}}=\underline{0}$ together with $\vecB^{\scriptscriptstyle{(0)}}=\underline{0}$, correspond to the system \eqref{eq:Fta} for $\vecu^{\snu}$, whose solution has the following structure
\begin{equation}
\label{eq:usnu}
\vecu^{\snu} = \sum_{i=1}^{n}d_i^{\snu}\vecY_i^{\scriptscriptstyle{-}}\;,
\end{equation}
where
\begin{equation}\label{eq:basisC}
 \vecY_i^{\scriptscriptstyle{\pm}} = \underline{e}_{2i-1} \pm \underline{e}_{2i} \;,
\end{equation}
and $\underline{e}_1,\dots,\underline{e}_{2n}$ are the standard basis-vectors of $\C^{2n}$. Furthermore, equations $\vecB^{\scriptscriptstyle{(1)}},\vecB^{\scriptscriptstyle{(2)}}=\underline{0}$ imply $a=0$, whence
 \begin{equation}
 \label{eq:t0mixed2}
 \vectt^{\scriptscriptstyle{(0)}} =b\vecX_2 \;.
 \end{equation}
Let us consider the system 
\begin{equation}
\widetilde{S}:\begin{cases}
\vecA^{\scriptscriptstyle{(-1)}},\vecA^{\scriptscriptstyle{(1)}} = \underline{0} & \\
\vecA^{\scriptscriptstyle{(m)}} = \underline{0} &,\quad m\geq 3 \\
\vecB^{\scriptscriptstyle{(m)}} = \underline{0}  &,\quad m\geq 3
\end{cases} \;,
\end{equation}
that is the system $S$ deprived of the equations $\vecA^{\scriptscriptstyle{(0)}},\vecA^{\scriptscriptstyle{(2)}} = \underline{0}$ and $\vecB^{\scriptscriptstyle{(0)}},\vecB^{\scriptscriptstyle{(1)}},\vecB^{\scriptscriptstyle{(2)}} = \underline{0}$ that we already exploited to fix $\vecu^{\snu}$ and $\vectt^{\scriptscriptstyle{(0)}}$ as per \eqref{eq:usnu} and \eqref{eq:t0mixed2}, respectively. Moreover we split $\widetilde{S}$ into $\widetilde{S}=\widetilde{S}_1\cup \widetilde{S}_2$, where the subsystems $\widetilde{S}_1$ and $\widetilde{S}_2$ are defined as
\begin{equation}
\label{eq:subsystemvecmixed}
\widetilde{S}_1:\begin{cases}
\vecA^{\scriptscriptstyle{(2m-1)}} = \underline{0} &,\quad m\geq 0 \\
\vecB^{\scriptscriptstyle{(3m)}} = \underline{0} &,\quad m\geq 1
\end{cases} \;,\quad 
\widetilde{S}_2:\begin{cases}
\vecA^{\scriptscriptstyle{(2m)}} = \underline{0} &,\quad m\geq 2 \\
\vecB^{\scriptscriptstyle{(2m-1-\lfloor m/2 \rfloor)}} = \underline{0} &,\quad m\geq 3
\end{cases} \;.
\end{equation} 
We divide the proof of the Proposition in two parts. The first step is to show that the subsystem $\widetilde{S}_1$ admits a unique solution $\vectt^{\snu}(\e)$ of the form \eqref{eq:ansatzvecmixed} for each $\vecu^{\snu}$ and depending on the parameter $\alpha\neq\frac52$. To this aim, we exploit the fact that the dependence of $\vecA^{\scriptscriptstyle{(2m-1)}}$ and $\vecB^{\scriptscriptstyle{(3m)}}$ on the higher order terms converges to the following expressions for $m$ sufficiently large:
\begin{align}
&A_k^{\scriptscriptstyle{(2m-1)}} = 12M_{1,k}(\vectt^{\scriptscriptstyle{(m+1)}}) - 48M_{k,2}(\vectt^{\scriptscriptstyle{(1)}},\vectt^{\scriptscriptstyle{(m)}}) + 2t_k^{\scriptscriptstyle{(m-1)}}\left(1+\sum_{a=1}^{2n}\frac{6}{(u_a^{\snu})^4}\right) \notag \\
&+\frac{36\,t_k^{\scriptscriptstyle{(0)}}}{M^{\frac14}}M_{1,k}(\vectt^{\scriptscriptstyle{(m-1)}}) - 48M_{2,k}(\vectt^{\scriptscriptstyle{(2)}},\vectt^{\scriptscriptstyle{(m-1)}}) +120M_{3,k}(\vectt^{\scriptscriptstyle{(1)}},\vectt^{\scriptscriptstyle{(1)}},\vectt^{\scriptscriptstyle{(m-1)}})\notag \\
&+12\,t_k^{\scriptscriptstyle{(m-2)}}\left(\frac{3}{M^{\frac14}}M_{1,k}(\vectt^{\scriptscriptstyle{(1)}})+(t_k^{\scriptscriptstyle{(0)}})^{2}\sum_{a=1}^{2n}\frac{10}{(u_a^{\snu})^6}\right) - 48M_{2,k}(\vectt^{\scriptscriptstyle{(3)}},\vectt^{\scriptscriptstyle{(m-2)}}) \notag \\
&+\frac{36}{M^{\frac14}}\left(t_k^{\scriptscriptstyle{(1)}}M_{1,k}(\vectt^{\scriptscriptstyle{(m-2)}}) -4\,t_k^{\scriptscriptstyle{(0)}}M_{2,k}(\vectt^{\scriptscriptstyle{(1)}},\vectt^{\scriptscriptstyle{(m-2)}})\right) -240M_{4,k}(\vectt^{\scriptscriptstyle{(1)}},\vectt^{\scriptscriptstyle{(1)}},\vectt^{\scriptscriptstyle{(1)}},\vectt^{\scriptscriptstyle{(m-2)}}) \notag \\
&+ 240 M_{3,k}(\vectt^{\scriptscriptstyle{(1)}},\vectt^{\scriptscriptstyle{(2)}},\vectt^{\scriptscriptstyle{(m-2)}})+\frac{2\,t_k^{\scriptscriptstyle{(m-3)}}}{M^{\frac14}}\Biggl( (M-1)\,t_k^{\scriptscriptstyle{(0)}} + 18M_{1,k}(\vectt^{\scriptscriptstyle{(2)}}) +\sum_{a=1}^{2n}\frac{36\,t_k^{\scriptscriptstyle{(0)}}}{(u_a^{\snu})^4} \notag \\
&- 36M_{2,k}(\vectt^{\scriptscriptstyle{(1)}},\vectt^{\scriptscriptstyle{(1)}}) + \sum_{a=1}^{2n}\frac{210M^{\frac14}(t_k^{\scriptscriptstyle{(0)}})^4}{(u_a^{\snu})^8} + \sum_{a=1}^{2n}\frac{120M^{\frac14}t_k^{\scriptscriptstyle{(0)}}t_k^{\scriptscriptstyle{(1)}}}{(u_a^{\snu})^6} -\sum_{a=1}^{2n}\frac{120M^{\frac14}t_k^{\scriptscriptstyle{(0)}}u_a^{\scriptscriptstyle{(1)}}}{(u_a^{\snu})^6} \Biggr) \notag \\
&+\frac{36}{M^{\frac12}}\Bigl((t_k^{\scriptscriptstyle{(0)}})^2+M^{\frac14}t_k^{\scriptscriptstyle{(2)}}\Bigr)M_{1,k}(\vectt^{\scriptscriptstyle{(m-3)}}) - 48M_{2,k}(\vectt^{\scriptscriptstyle{(4)}},\vectt^{\scriptscriptstyle{(m-3)}}) + 120M_{3,k}(\vectt^{\scriptscriptstyle{(2)}},\vectt^{\scriptscriptstyle{(2)}},\vectt^{\scriptscriptstyle{(m-3)}}) \notag \\
&-\frac{144}{M^{\frac14}}\left(t_k^{\scriptscriptstyle{(0)}}M_{2,k}(\vectt^{\scriptscriptstyle{(2)}},\vectt^{\scriptscriptstyle{(m-3)}})+t_k^{\scriptscriptstyle{(1)}}M_{2,k}(\vectt^{\scriptscriptstyle{(1)}},\vectt^{\scriptscriptstyle{(m-3)}})\right) + \frac{360\,t_k^{\scriptscriptstyle{(0)}}}{M^{\frac14}}M_{3,k}(\vectt^{\scriptscriptstyle{(1)}},\vectt^{\scriptscriptstyle{(1)}},\vectt^{\scriptscriptstyle{(m-3)}})  \notag \\
& - 720M_{4,k}(\vectt^{\scriptscriptstyle{(1)}},\vectt^{\scriptscriptstyle{(1)}},\vectt^{\scriptscriptstyle{(2)}},\vectt^{\scriptscriptstyle{(m-3)}}) + 420M_{5,k}(\vectt^{\scriptscriptstyle{(1)}},\vectt^{\scriptscriptstyle{(1)}},\vectt^{\scriptscriptstyle{(1)}},\vectt^{\scriptscriptstyle{(1)}},\vectt^{\scriptscriptstyle{(m-3)}}) \notag \\
&+ 240M_{3,k}(\vectt^{\scriptscriptstyle{(1)}},\vectt^{\scriptscriptstyle{(3)}},\vectt^{\scriptscriptstyle{(m-3)}}) + g_{A,k}^{\scriptscriptstyle{(2m-1)}}\left(\vecu^{\scriptscriptstyle{(\lfloor (2m-1)/3\rfloor-1)}},\vecu^{\scriptscriptstyle{(\lfloor (2m-1)/3\rfloor)}}\right) \notag \\
&+ f_{A,k}^{\scriptscriptstyle{(2m-1)}}\left( \vectt^{\scriptscriptstyle{(0)}},\dots,\vectt^{\scriptscriptstyle{(m-4)}},\vecu^{\snu},\dots,\vecu^{\scriptscriptstyle{(\lfloor (2m-1)/3\rfloor-2)}} \right) \;,\quad m\geq 8 \;,
\label{eq:Ageneral}
\end{align}
and
\begin{align}
&B_a^{\scriptscriptstyle{(3m)}} = \sum_{b=1}^{2n}\left(u_b^{\scriptscriptstyle{(m)}}\widetilde{\mathbf{J}}_{ab}^{\snu}\right) + g_{B,a}^{\scriptscriptstyle{(3m)}}\left( \vectt^{\scriptscriptstyle{(2m-2-\lfloor m/2 \rfloor)}},\vectt^{\scriptscriptstyle{(2m-1-\lfloor m/2 \rfloor)}}\right) \notag \\
&+f_{B,a}^{\scriptscriptstyle{(3m)}}\left( \vectt^{\scriptscriptstyle{(0)}},\dots,\vectt^{\scriptscriptstyle{(2m-3-\lfloor m/2 \rfloor)}},\vecu^{\snu},\dots,\vecu^{\scriptscriptstyle{(m-1)}} \right) \;,\quad m\geq 3 \;,
%&\vecB^{\scriptscriptstyle{(3m)}} = \vecu^{\scriptscriptstyle{(m)}}\widetilde{\mathbf{J}}^{\snu} +\underline{g}_B^{\scriptscriptstyle{(3m)}}\left( \vectt^{\scriptscriptstyle{(2m-2-\lfloor m/2 \rfloor)}},\vectt^{\scriptscriptstyle{(2m-1-\lfloor m/2 \rfloor)}}\right) \notag \\
%&+\underline{f}_B^{\scriptscriptstyle{(3m)}}\left( \vectt^{\scriptscriptstyle{(0)}},\dots,\vectt^{\scriptscriptstyle{(2m-3-\lfloor m/2 \rfloor)}},\vecu^{\snu},\dots,\vecu^{\scriptscriptstyle{(m-1)}} \right) \;,
\label{eq:Bgeneral}
\end{align}
where the functions $\underline{g}_A^{\scriptscriptstyle{(2m-1)}} = \bigl(g_{A,1}^{\scriptscriptstyle{(2m-1)}},g_{A,2}^{\scriptscriptstyle{(2m-1)}},g_{A,3}^{\scriptscriptstyle{(2m-1)}}\bigr)$ and $\underline{g}_B^{\scriptscriptstyle{(2m-1)}} = \bigl(g_{B,1}^{\scriptscriptstyle{(2m-1)}}$, $\dots$, $g_{B,2n}^{\scriptscriptstyle{(2m-1)}}\bigr)$ are 
\begin{align}
&g_{A,k}^{\scriptscriptstyle{(2m-1)}} = -720\,(t_k^{\scriptscriptstyle{(0)}})^2\Biggl( \sum_{a=1}^{2n} \frac{u_a^{\scriptscriptstyle{(\lfloor (2m-1)/3\rfloor)}}}{6\,(u_a^{\snu})^6}+ t_k^{\scriptscriptstyle{(1)}} \sum_{a=1}^{2n} \frac{u_a^{\scriptscriptstyle{(\lfloor (2m-1)/3\rfloor-1)}}}{(u_a^{\snu})^7} + \sum_{a=1}^{2n}\frac{u_a^{\scriptscriptstyle{(\lfloor (2m-1)/3\rfloor-1)}}}{5M^{\frac14}(u_a^{\snu})^5} \notag \\
&- \sum_{a=1}^{2n} \frac{u_a^{\scriptscriptstyle{(\lfloor (2m-1)/3\rfloor-1)}}u_a^{\scriptscriptstyle{(1)}}}{(u_a^{\snu})^7} \Biggr) -672\,(t_k^{\scriptscriptstyle{(0)}})^{5} \sum_{a=1}^{2n} \frac{u_a^{\scriptscriptstyle{(\lfloor (2m-1)/3\rfloor-1)}}}{(u_a^{\snu})^9} -48\,t_k^{\scriptscriptstyle{(2)}} \sum_{a=1}^{2n} \frac{u_a^{\scriptscriptstyle{(\lfloor (2m-1)/3\rfloor-1)}}}{(u_a^{\snu})^5} \notag \\
&,\quad m \cong 0 \mod 3 \;, \notag 
%&\underline{g}_A^{\scriptscriptstyle{(2m-1)}} =-120\,(\vectt^{\scriptscriptstyle{(0)}})^{\circ 2} \sum_{a=1}^{2n} \frac{u_a^{\scriptscriptstyle{(\lfloor (2m-1)/3\rfloor)}}}{(u_a^{\snu})^6} -48\,\vectt^{\scriptscriptstyle{(2)}} \sum_{a=1}^{2n} \frac{u_a^{\scriptscriptstyle{(\lfloor (2m-1)/3\rfloor-1)}}}{(u_a^{\snu})^5} \notag \\
%&-720\,\vectt^{\scriptscriptstyle{(1)}}\circ(\vectt^{\scriptscriptstyle{(0)}})^{\circ 2} \sum_{a=1}^{2n} \frac{u_a^{\scriptscriptstyle{(\lfloor (2m-1)/3\rfloor-1)}}}{(u_a^{\snu})^7}  +720\,(\vectt^{\scriptscriptstyle{(0)}})^{\circ 2} \sum_{a=1}^{2n} \frac{u_a^{\scriptscriptstyle{(\lfloor (2m-1)/3\rfloor-1)}}u_a^{\scriptscriptstyle{(1)}}}{(u_a^{\snu})^7} \notag \\
%& -672\,(\vectt^{\scriptscriptstyle{(0)}})^{\circ 5} \sum_{a=1}^{2n} \frac{u_a^{\scriptscriptstyle{(\lfloor (2m-1)/3\rfloor-1)}}}{(u_a^{\snu})^9} -144\,(\vectt^{\scriptscriptstyle{(0)}})^{\circ 2}\sum_{a=1}^{2n}\frac{u_a^{\scriptscriptstyle{(\lfloor (2m-1)/3\rfloor-1)}}}{M^{\frac14}(u_a^{\snu})^5} \;,\quad m \cong 0 \mod 3 \;, \notag 
\end{align}
\begin{align}
&g_{A,k}^{\scriptscriptstyle{(2m-1)}} =-240\,t_k^{\scriptscriptstyle{(0)}}\Biggl( \sum_{a=1}^{2n} \frac{u_a^{\scriptscriptstyle{(\lfloor (2m-1)/3\rfloor)}}}{5\,(u_a^{\snu})^5} - \sum_{a=1}^{2n}\frac{u_a^{\scriptscriptstyle{(\lfloor (2m-1)/3\rfloor-1)}}u_a^{\scriptscriptstyle{(1)}}}{(u_a^{\snu})^6} + \sum_{a=1}^{2n}\frac{3\,u_a^{\scriptscriptstyle{(\lfloor (2m-1)/3\rfloor-1)}}}{20M^{\frac14}(u_a^{\snu})^4} \notag \\
&+t_k^{\scriptscriptstyle{(1)}}\sum_{a=1}^{2n}\frac{u_a^{\scriptscriptstyle{(\lfloor (2m-1)/3\rfloor-1)}}}{(u_a^{\snu})^6} \Biggr) -420\,(t_k^{\scriptscriptstyle{(0)}})^{4}\sum_{a=1}^{2n}\frac{u_a^{\scriptscriptstyle{(\lfloor (2m-1)/3\rfloor-1)}}}{(u_a^{\snu})^8} \;,\quad m \cong 1 \mod 3 \;, \notag
%&\underline{g}_A^{\scriptscriptstyle{(2m-1)}} =-48\,\vectt^{\scriptscriptstyle{(0)}}\sum_{a=1}^{2n} \frac{u_a^{\scriptscriptstyle{(\lfloor (2m-1)/3\rfloor)}}}{(u_a^{\snu})^5} +240\,\vectt^{\scriptscriptstyle{(0)}}\sum_{a=1}^{2n}\frac{u_a^{\scriptscriptstyle{(\lfloor (2m-1)/3\rfloor-1)}}u_a^{\scriptscriptstyle{(1)}}}{(u_a^{\snu})^6} \notag \\
%&-420\,(\vectt^{\scriptscriptstyle{(0)}})^{\circ 4}\sum_{a=1}^{2n}\frac{u_a^{\scriptscriptstyle{(\lfloor (2m-1)/3\rfloor-1)}}}{(u_a^{\snu})^8}  -240\,\vectt^{\scriptscriptstyle{(0)}}\circ\vectt^{\scriptscriptstyle{(1)}}\sum_{a=1}^{2n}\frac{u_a^{\scriptscriptstyle{(\lfloor (2m-1)/3\rfloor-1)}}}{(u_a^{\snu})^6} \;,\quad m \cong 1 \mod 3 \;,
%- 240(\vectt^{\scriptscriptstyle{(0)}})^{\circ 3}\sum_{a=1}^{2n}u_a^{\scriptscriptstyle{(\lfloor (2m-1)/3\rfloor-1)}}(u_a^{\scriptscriptstyle{(0)}})^{-7} - 48\,\vectt^{\scriptscriptstyle{(1)}}\sum_{a=1}^{2n}u_a^{\scriptscriptstyle{(\lfloor (2m-1)/3\rfloor-1)}}(u_a^{\scriptscriptstyle{(0)}})^{-5} +48\vecX_1\sum_{a=1}^{2n}u_a^{\scriptscriptstyle{(\lfloor (2m-1)/3\rfloor-1)}}u_a^{\scriptscriptstyle{(1)}}(u_a^{\scriptscriptstyle{(0)}})^{-5}
\end{align}
\begin{align}
&g_{A,k}^{\scriptscriptstyle{(2m-1)}} =-12\left(\sum_{a=1}^{2n}\frac{u_a^{\scriptscriptstyle{(\lfloor (2m-1)/3\rfloor)}}}{(u_a^{\snu})^4}-4\sum_{a=1}^{2n}\frac{u_a^{\scriptscriptstyle{(\lfloor (2m-1)/3\rfloor-1)}}u_a^{\scriptscriptstyle{(1)}}}{(u_a^{\snu})^5} +4\,t_k^{\scriptscriptstyle{(1)}}\sum_{a=1}^{2n} \frac{u_a^{\scriptscriptstyle{(\lfloor (2m-1)/3\rfloor-1)}}}{(u_a^{\snu})^5}\right) \notag \\
&- 240\,(t_k^{\scriptscriptstyle{(0)}})^{3}\sum_{a=1}^{2n}\frac{u_a^{\scriptscriptstyle{(\lfloor (2m-1)/3\rfloor-1)}}}{(u_a^{\snu})^7} \;,\quad m \cong 2 \mod 3 \;,
%&\underline{g}_A^{\scriptscriptstyle{(2m-1)}} =
%-12\vecX_1\sum_{a=1}^{2n}\frac{u_a^{\scriptscriptstyle{(\lfloor (2m-1)/3\rfloor)}}}{(u_a^{\snu})^4} - 240\,(\vectt^{\scriptscriptstyle{(0)}})^{\circ 3}\sum_{a=1}^{2n}\frac{u_a^{\scriptscriptstyle{(\lfloor (2m-1)/3\rfloor-1)}}}{(u_a^{\snu})^7}   \notag \\
%&- 48\,\vectt^{\scriptscriptstyle{(1)}}\sum_{a=1}^{2n} \frac{u_a^{\scriptscriptstyle{(\lfloor (2m-1)/3\rfloor-1)}}}{(u_a^{\snu})^5} +48\vecX_1\sum_{a=1}^{2n}\frac{u_a^{\scriptscriptstyle{(\lfloor (2m-1)/3\rfloor-1)}}u_a^{\scriptscriptstyle{(1)}}}{(u_a^{\snu})^5} \;,\quad m \cong 2 \mod 3 \;,\notag 
\end{align}
and
\begin{align}
&g_{B,a}^{\scriptscriptstyle{(2m-1)}} = -\frac{48}{(u_a^{\snu})^5}\Biggl( \sum_{k=1}^3  t_k^{\scriptscriptstyle{(0)}}t_k^{\scriptscriptstyle{(2m-1-\lfloor m/2 \rfloor)}} + \sum_{k=1}^3 t_k^{\scriptscriptstyle{(1)}} t_k^{\scriptscriptstyle{(2m-2-\lfloor m/2 \rfloor)}} -u_a^{\scriptscriptstyle{(1)}}\sum_{k=1}^3 t_k^{\scriptscriptstyle{(2m-2-\lfloor m/2 \rfloor)}} \notag \\ 
&+\frac{5}{(u_a^{\snu})^2} \sum_{k=1}^3 ( t_k^{\scriptscriptstyle{(0)}})^3t_k^{\scriptscriptstyle{(2m-2-\lfloor m/2 \rfloor)}}\Biggr) -\frac{36}{M^{\frac14}(u_a^{\snu})^3}\sum_{k=1}^3 t_k^{\scriptscriptstyle{(2m-2-\lfloor m/2 \rfloor)}} \;,\quad m \cong 0 \mod 2 \;,\notag
%&\underline{g}_B^{\scriptscriptstyle{(3m)}} = -48\,(\vecu^{\snu})^{\circ -5}\sum_{k=1}^3  t_k^{\scriptscriptstyle{(0)}}t_k^{\scriptscriptstyle{(2m-1-\lfloor m/2 \rfloor)}}  -240\,(\vecu^{\snu})^{\circ -7}\sum_{k=1}^3 ( t_k^{\scriptscriptstyle{(0)}})^3t_k^{\scriptscriptstyle{(2m-2-\lfloor m/2 \rfloor)}} \notag \\
%&-48\,(\vecu^{\snu})^{\circ -5}\sum_{k=1}^3 t_k^{\scriptscriptstyle{(1)}} t_k^{\scriptscriptstyle{(2m-2-\lfloor m/2 \rfloor)}}  +48\,\vecu^{\scriptscriptstyle{(1)}}\circ(\vecu^{\snu})^{\circ -5}\sum_{k=1}^3 t_k^{\scriptscriptstyle{(2m-2-\lfloor m/2 \rfloor)}} \notag \\
%&-\frac{36}{M^{\frac14}}(\vecu^{\snu})^{\circ -3}\sum_{k=1}^3 t_k^{\scriptscriptstyle{(2m-2-\lfloor m/2 \rfloor)}} \;,\quad m \cong 0 \mod 2 \;,\notag
\end{align}
\begin{equation}
g_{B,a}^{\scriptscriptstyle{(2m-1)}} = -\frac{12}{(u_a^{\snu})^4}\sum_{k=1}^3 t_k^{\scriptscriptstyle{(2m-1-\lfloor m/2 \rfloor)}} -\frac{120}{(u_a^{\snu})^6}\sum_{k=1}^3 (t_k^{\scriptscriptstyle{(0)}})^2t_k^{\scriptscriptstyle{(2m-2-\lfloor m/2 \rfloor)}}  \;,\quad m \cong 1 \mod 2 \;,
%\underline{g}_B^{\scriptscriptstyle{(3m)}} = -12\,(\vecu^{\snu})^{\circ -4}\sum_{k=1}^3 t_k^{\scriptscriptstyle{(2m-1-\lfloor m/2 \rfloor)}} -120 \,(\vecu^{\snu})^{\circ -6} \sum_{k=1}^3 (t_k^{\scriptscriptstyle{(0)}})^2t_k^{\scriptscriptstyle{(2m-2-\lfloor m/2 \rfloor)}} \\
%,\quad m \cong 1 \mod 2 \;, 
\end{equation}
%&- 240(\vecu^{\scriptscriptstyle{(0)}})^{\circ -6} \sum_{k=1}^3 t_k^{\scriptscriptstyle{(0)}}t_k^{\scriptscriptstyle{(1)}}t_k^{\scriptscriptstyle{(2m-2-\lfloor m/2 \rfloor)}} +240\,\vecu^{\scriptscriptstyle{(1)}}\circ(\vecu^{\scriptscriptstyle{(0)}})^{\circ -6} \sum_{k=1}^3 t_k^{\scriptscriptstyle{(0)}}t_k^{\scriptscriptstyle{(2m-2-\lfloor m/2 \rfloor)}} \notag \\
%&- 420(\vecu^{\scriptscriptstyle{(0)}})^{\circ -8} \sum_{k=1}^3 (t_k^{\scriptscriptstyle{(0)}})^4 t_k^{\scriptscriptstyle{(2m-2-\lfloor m/2 \rfloor)}} \;,\quad m\in 2\mathbbm{N}-1\;, \notag \\
while the functions $\underline{f}_A^{\scriptscriptstyle{(2m-1)}} = \bigl(f_{A,1}^{\scriptscriptstyle{(2m-1)}},f_{A,2}^{\scriptscriptstyle{(2m-1)}},f_{A,3}^{\scriptscriptstyle{(2m-1)}}\bigr)$ and $\underline{f}_B^{\scriptscriptstyle{(2m-1)}} = \bigl(f_{B,1}^{\scriptscriptstyle{(2m-1)}}$, $\dots$, $f_{B,2n}^{\scriptscriptstyle{(2m-1)}}\bigr)$ collect all the lower order terms. 

The second step is instead to show that the coefficients $\lbrace \vectt^{\scriptscriptstyle{(m)}}\rbrace_{m\geq 0}$ and $\lbrace \vecu^{\scriptscriptstyle{(m)}}\rbrace_{m\geq 1}$ of the solution $\vectt^{\snu}(\e)$ admits a cyclic structure of the form
%Moreover, one can show by induction that the unique solution admits the following cyclic structure
\begin{equation}
 \vectt^{\scriptscriptstyle{(m)}} = \begin{cases}
 c_2^{\scriptscriptstyle{(m)}}\vecX_2 &,\quad m \cong 0 \mod 3 \\
 c_1^{\scriptscriptstyle{(m)}}\vecX_1 &,\quad m \cong 1 \mod 3 \\
 c_3^{\scriptscriptstyle{(m)}}\vecX_3 &,\quad m \cong 2 \mod 3
 \end{cases} \;,\quad \vecu^{\scriptscriptstyle{(m)}} = \begin{cases}
\displaystyle{\sum_{i=1}^n d_i^{\scriptscriptstyle{(m)}}Y_i^{\scriptscriptstyle{-}}} &,\quad m \cong 0 \mod 2 \\
\displaystyle{\sum_{i=1}^n \widetilde{d}_i^{\scriptscriptstyle{(m)}}Y_i^{\scriptscriptstyle{+}}} &,\quad m \cong 1 \mod 2
\end{cases} \;,
\label{eq:cyclicity}
 \end{equation}
and that the subsystem $\widetilde{S}_2$ is trivially fulfilled by such a solution. In this way we prove the existence of a unique solution to $F_k(\vectt,\e)=0$, $k=1,\dots,2n+3$ of the form \eqref{eq:ansatzvecmixed} which possesses the cyclic structure \eqref{eq:cyclicity}.

Let us focus on the first part. Notice that before attacking the general case, we must first determine the coefficients $\vectt^{\scriptscriptstyle{(0)}},\dots,\vectt^{\scriptscriptstyle{(4)}}$ which appear explicitly in the expressions of $\vecA^{\scriptscriptstyle{(2m-1)}}$ and $\vecB^{\scriptscriptstyle{(3m)}}$. The strategy that we shall follow is similar to that of Proposition \ref{prop:d2}, although the computation is much more involved. First of all we express $\vectt^{\scriptscriptstyle{(l)}} \;,\; l\geq 1$ in the basis $\mathcal{B}$ with coefficients $c_1^{\scriptscriptstyle{(l)}},c_2^{\scriptscriptstyle{(l)}},c_3^{\scriptscriptstyle{(l)}}$ as per \eqref{eq:tldecomp}. Moreover, we introduce the row vectors $\vecx^{\scriptscriptstyle{(m)}}\in\C^{2n+3}$ and $\underline{y}^{\scriptscriptstyle{(m)}}\in\C^3$ defined as
\begin{align}
\vecx^{\scriptscriptstyle{(m)}}&=\left(10b^2\,c_2^{\scriptscriptstyle{(m-1)}},c_1^{\scriptscriptstyle{(m)}},\frac{c_3^{\scriptscriptstyle{(m+1)}}}{2b^2},u_1^{\scriptscriptstyle{(\lfloor (2m+2)/3 \rfloor)}},\dots,u_{2n}^{\scriptscriptstyle{(\lfloor (2m+2)/3 \rfloor)}}\right) \;, \notag \\
\underline{y}^{\scriptscriptstyle{(m)}}&=\left(10b^2\,c_2^{\scriptscriptstyle{(m-1)}},c_1^{\scriptscriptstyle{(m)}},\frac{c_3^{\scriptscriptstyle{(m+1)}}}{2b^2}\right) \;.
\label{eq:vecxydef}
\end{align}
The equation $\vecA^{\scriptscriptstyle{(-1)}}=\underline{0}$ implies that $\vectt^{\scriptscriptstyle{(1)}}\in\ker(\vecM_1)$, whence
\begin{equation}
\label{eq:c31mixed}
c_3^{\scriptscriptstyle{(1)}}=0 \;.
\end{equation}
In order to fix $b$, hence $\vectt^{\scriptscriptstyle{(0)}}$, we combine the projection of the vector equations $\vecA^{\scriptscriptstyle{(1)}},\vecA^{\scriptscriptstyle{(3)}},\vecA^{\scriptscriptstyle{(5)}}=\underline{0}$ along the basis $\mathcal{B}$ with $\vecB^{\scriptscriptstyle{(3)}}=\underline{0}$ so as to arrive at the following system of equations
 \begin{align}
 \label{eq:systemt1mixed}
 \widetilde{S}_1^{\scriptscriptstyle{(1)}}:\begin{cases}
 \langle \underline{A}^{\scriptscriptstyle{(1)}},\vecX_2\rangle = 0\\%&= 6b\left(1+6\sum_{a=1}^{2n}(u_a^{\scriptscriptstyle{(0)}})^{-4}\right) - \frac{12}{b^4}\,c_3^{\scriptscriptstyle{(2)}} = 0 \\
 \langle \underline{A}^{\scriptscriptstyle{(3)}},\vecX_1\rangle =0 \\%&= 120\,b^3\sum_{a=1}^{2n}(u_a^{\scriptscriptstyle{(0)}})^{-6} + 6\,c_1^{\scriptscriptstyle{(1)}}\left(1+6\sum_{a=1}^{2n}(u_a^{\scriptscriptstyle{(0)}})^{-4}\right)  \\ &-36\sum_{a=1}^{2n}u_a^{\scriptscriptstyle{(1)}}(u_a^{\scriptscriptstyle{(0)}})^{-4}= 0 \\
 \langle \underline{A}^{\scriptscriptstyle{(5)}},\vecX_3\rangle = 0 \\ %&= 252\,b^5\sum_{a=1}^{2n}(u_a^{\scriptscriptstyle{(0)}})^{-8} + \frac{24}{b^5}\left(c_3^{\scriptscriptstyle{(2)}}\right)^2 + 6\,c_3^{\scriptscriptstyle{(2)}}\left(1+6\sum_{a=1}^{2n}(u_a^{\scriptscriptstyle{(0)}})^{-4}\right)  \\ &+9b^2\left(\kappa - 40\sum_{a=1}^{2n}u_a^{\scriptscriptstyle{(1)}}(u_a^{\scriptscriptstyle{(0)}})^{-6} + 40\,c_1^{\scriptscriptstyle{(1)}}\sum_{a=1}^{2n}(u_a^{\scriptscriptstyle{(0)}})^{-6}\right) = 0 \\
\vecB^{\scriptscriptstyle{(3)}} = \underline{0}%&=\vecu^{\scriptscriptstyle{(1)}}\tilde{\mathbf{J}}(\vecu^{\scriptscriptstyle{(0)}}) - 36(\vecu^{\scriptscriptstyle{(0)}})^{\circ -4}c_1^{\scriptscriptstyle{(1)}} - 120(\vecu^{\scriptscriptstyle{(0)}})^{\circ -6}b^3 = \underline{0}
%3b^2\left( \frac{M-7}{M^{1/4}} -120s_{-6,1}^{\scriptscriptstyle{(0,1)}} + 120 c_1^{\scriptscriptstyle{(1)}}s_{-6}^{\scriptscriptstyle{(0)}}+ 3 b^3\left(28s_{-8}^{\scriptscriptstyle{(0)}}+(1+6s_{-4}^{\scriptscriptstyle{(0)}})^2 \right) \right) = 0
 \end{cases} 
\longrightarrow \quad 
 \begin{array}{c}
 \displaystyle{c_3^{\scriptscriptstyle{(2)}} = \frac{b^5}{2}\left(1+\sum_{a=1}^{2n}\frac{6}{(u_a^{\snu})^4}\right)} \\
\mathbf{M}^{\scriptscriptstyle{(1)}}(\widetilde{\vecx}^{\scriptscriptstyle{(1)}})^{\text{T}} =(\widetilde{\underline{a}}^{\scriptscriptstyle{(1)}})^{\text{T}}
\end{array} \;,
%120\,b^3\sum_{a=1}^{2n}(u_a^{\scriptscriptstyle{(0)}})^{-6} + 6\,c_1^{\scriptscriptstyle{(1)}}\left(1+6\sum_{a=1}^{2n}(u_a^{\scriptscriptstyle{(0)}})^{-4}\right) \\
%- 36 \sum_{a=1}^{2n}u_a^{\scriptscriptstyle{(1)}}(u_a^{\scriptscriptstyle{(0)}})^{-4} = 0 \\
%b^3\left(84\sum_{a=1}^{2n}(u_a^{\scriptscriptstyle{(0)}})^{-8}+3\left(1+6\sum_{a=1}^{2n}(u_a^{\scriptscriptstyle{(0)}})^{-4}\right)^2 \right) \\ + 120\,c_1^{\scriptscriptstyle{(1)}}\sum_{a=1}^{2n}(u_a^{\scriptscriptstyle{(0)}})^{-6} - 120\sum_{a=1}^{2n}u_a^{\scriptscriptstyle{(1)}}(u_a^{\scriptscriptstyle{(0)}})^{-6} =- 3\kappa   \\ \vecu^{\scriptscriptstyle{(1)}}\tilde{\mathbf{J}}(\vecu^{\scriptscriptstyle{(0)}}) - 36(\vecu^{\scriptscriptstyle{(0)}})^{\circ -4}c_1^{\scriptscriptstyle{(1)}} - 120(\vecu^{\scriptscriptstyle{(0)}})^{\circ -6}b^3 = \underline{0}\;,
 \end{align}
where $\star^{\text{T}}$ denotes the transposition, $\widetilde{\vecx}^{\scriptscriptstyle{(1)}}=\big(-20\,b^3,-6\,c_1^{\scriptscriptstyle{(1)}},u_1^{\scriptscriptstyle{(1)}},\dots,u_{2n}^{\scriptscriptstyle{(1)}}\big)\in\C^{2n+2}$ and
\begin{align}
\mathbf{M}^{\scriptscriptstyle{(1)}} &=\left(\begin{array}{cc}
\widetilde{\mathbf{A}}^{\snu} & \widetilde{\mathbf{B}}^{\snu} \\
\widetilde{\mathbf{C}}^{\snu} & \widetilde{\mathbf{J}}^{\snu} 
\end{array}\right)\in\C^{2n+2,2n+2} \;, \notag \\
\widetilde{\underline{a}}^{\scriptscriptstyle{(1)}} &= \widetilde{\underline{a}}^{\scriptscriptstyle{(1)}}(\vecu^{\snu};\alpha) = \frac{M-7}{M^{\frac14}}\left(0,\frac{1}{20},-(u_1^{\snu})^2,\dots,-(u_{2n}^{\snu})^{2}\right)\in\C^{2n+2} \;.
\end{align}
Under the assumptions $\det{\big(\widetilde{\mathbf{J}}^{\snu}\big)},\det{\big(\widetilde{\mathbf{A}}^{\snu}-\widetilde{\mathbf{B}}^{\snu}(\widetilde{\mathbf{J}}^{\snu})^{-1}\widetilde{\mathbf{C}}^{\snu}\big)}\neq 0$, it follows that $\det{(\mathbf{M}^{\scriptscriptstyle{(1)}})}\neq 0$. Therefore, the linear system $\mathbf{M}^{\scriptscriptstyle{(1)}}(\widetilde{\vecx}^{\scriptscriptstyle{(1)}})^{\text{T}} =(\widetilde{\underline{a}}^{\scriptscriptstyle{(1)}})^{\text{T}}$ admits a unique solution $\widetilde{\vecx}^{\scriptscriptstyle{(1)}}= \widetilde{\vecx}^{\scriptscriptstyle{(1)}}(\vecu^{\snu};\alpha)$ which fixes uniquely 
%\footnote{For $M=7$, i.e. $\alpha=\frac52$, the solution to the system $\mathbf{M}^{\scriptscriptstyle{(1)}}(\widetilde{\vecx}^{\scriptscriptstyle{(1)}})^{\text{T}} =(\widetilde{\underline{a}}^{\scriptscriptstyle{(1)}})^{\text{T}}$  is the trivial one, namely $\widetilde{\vecx}^{\scriptscriptstyle{(1)}}=\underline{0}$, which implies $b=c_1^{\scriptscriptstyle{(1)}}=0$ and $\vecu^{\scriptscriptstyle{(1)}}=\underline{0}$. In the following we shall consider the case $\alpha\neq\frac52$.}
%As we noticed in the proof of Proposition \ref{prop:d2}, when the coefficient $b$ vanishes, the matrices $\vecM_n $ become singular.  However, this is just a fictitious singularity of the system: the solution in the case $b=0$ can be obtained by suitably taking the limit $b \to 0$ of the solution constructed below for all $b \neq 0$.}
\begin{equation}
\label{eq:solsystemt1mixed}
b=b\left(\vecu^{\snu};\alpha\right) \;,\quad c_1^{\scriptscriptstyle{(1)}}=c_1^{\scriptscriptstyle{(1)}}(\vecu^{\snu};\alpha) \;,\quad c_3^{\scriptscriptstyle{(2)}}=c_3^{\scriptscriptstyle{(2)}}(\vecu^{\snu};\alpha) \;,\quad \vecu^{\scriptscriptstyle{(1)}}= \vecu^{\scriptscriptstyle{(1)}}(\vecu^{\snu};\alpha) \;.
\end{equation}
Notice that for $\alpha=\frac52$, i.e. $M=7$, the solution to the linear system $\mathbf{M}^{\scriptscriptstyle{(1)}}(\widetilde{\vecx}^{\scriptscriptstyle{(1)}})^{\text{T}} =(\widetilde{\underline{a}}^{\scriptscriptstyle{(1)}})^{\text{T}}$ is the trivial one $\widetilde{\vecx}^{\scriptscriptstyle{(1)}}\left(\vecu^{\snu};\frac52\right)=\underline{0}$, whence $b\left(\vecu^{\snu};\frac52\right)=0$. In the following we shall set $\alpha\neq \frac52$ so that $b\left(\vecu^{\snu};\alpha\right)\neq 0$.

In summary, we found
\begin{equation}
\label{eq:t0u1mixed}
\vectt^{\scriptscriptstyle{(0)}}(\vecu^{\snu};\alpha)= b\left(\vecu^{\snu};\alpha\right)\vecX_2 \;,\quad \vecu^{\scriptscriptstyle{(1)}}= \vecu^{\scriptscriptstyle{(1)}}(\vecu^{\snu};\alpha) \;.
\end{equation}
Decomposing the equations of $\widetilde{S}_1$ in a similar way, we can determine the higher order coefficients iteratively. To completely fix $\vectt^{\scriptscriptstyle{(1)}}$ and $\vecu^{\scriptscriptstyle{(2)}}$, we combine the projection of the vector equations $\vecA^{\scriptscriptstyle{(3)}}$, $\vecA^{\scriptscriptstyle{(5)}}$, $\vecA^{\scriptscriptstyle{(7)}}=\underline{0}$ along the basis $\mathcal{B}$ with $\vecB^{\scriptscriptstyle{(6)}} = \underline{0}$. Using \eqref{eq:c31mixed} and \eqref{eq:solsystemt1mixed} we arrive at the following system of equations
\begin{align}
 \label{eq:systemt2mixed}
 \widetilde{S}_1^{\scriptscriptstyle{(2)}}:\begin{cases}
 \langle \vecA^{\scriptscriptstyle{(3)}},\vecX_2\rangle = 0 \\
 \langle \vecA^{\scriptscriptstyle{(5)}},\vecX_1\rangle = 0 \\
 \langle \vecA^{\scriptscriptstyle{(7)}},\vecX_3 \rangle = 0 \\
\vecB^{\scriptscriptstyle{(6)}} = \underline{0}
%\sum_{a=1}^{N-3} B_a^{\scriptscriptstyle{(2l+3)}} = -360\,b^2c_2^{\scriptscriptstyle{(l)}}s_{-6}^{\scriptscriptstyle{(0)}} - 36\,c_1^{\scriptscriptstyle{(l+1)}}s_{-4}^{\scriptscriptstyle{(0)}} = 0
 \end{cases} \longrightarrow\quad \mathbf{M}^{\scriptscriptstyle{(2)}}(\underline{x}^{\scriptscriptstyle{(2)}})^{\text{T}}=(\underline{a}^{\scriptscriptstyle{(2)}})^{\text{T}} \;,
%30\,c_2^{\scriptscriptstyle{(1)}}\left(1+6\sum_{a=1}^{2n}(u_a^{\scriptscriptstyle{(0)}})^{-4}\right) - \frac{12}{b^4}\,c_3^{\scriptscriptstyle{(3)}} = 0 \\ 360\,b^2c_2^{\scriptscriptstyle{(1)}}\sum_{a=1}^{2n}(u_a^{\scriptscriptstyle{(0)}})^{-6} + 6\,c_1^{\scriptscriptstyle{(2)}}\left(1+6\sum_{a=1}^{2n}(u_a^{\scriptscriptstyle{(0)}})^{-4}\right)= 0 \\ - 9b^2\,c_2^{\scriptscriptstyle{(1)}}\left(28\sum_{a=1}^{2n}(u_a^{\scriptscriptstyle{(0)}})^{-8} -\frac{16}{9}\left(1+6\sum_{a=1}^{2n}(u_a^{\scriptscriptstyle{(0)}})^{-4}\right)^2 \right) \\ +120\,c_1^{\scriptscriptstyle{(2)}}\sum_{a=1}^{2n}(u_a^{\scriptscriptstyle{(0)}})^{-6} + \frac{10}{b^2}\,c_3^{\scriptscriptstyle{(3)}}\left(1+6\sum_{a=1}^{2n}(u_a^{\scriptscriptstyle{(0)}})^{-4}\right) = 0 \\ \vecu^{\scriptscriptstyle{(2)}}\tilde{\mathbf{J}}(\vecu^{\scriptscriptstyle{(0)}}) = \vecf\left(b, c_1^{\scriptscriptstyle{(1)}},c_3^{\scriptscriptstyle{(2)}},\vecu^{\scriptscriptstyle{(1)}}\right)
 \end{align}
where $\mathbf{M}^{\scriptscriptstyle{(2)}} = \mathbf{Q}_k^{\snu}$ with $k=(M^{\frac14}b^3)^{-1}$, $\displaystyle{\vecx^{\scriptscriptstyle{(2)}}=\left(10b^2\,c_2^{\scriptscriptstyle{(1)}},c_1^{\scriptscriptstyle{(2)}},\frac{c_3^{\scriptscriptstyle{(3)}}}{2b^2},u_1^{\scriptscriptstyle{(2)}},\dots,u_{2n}^{\scriptscriptstyle{(2)}}\right)}$ as per \eqref{eq:vecxydef} and $\underline{a}^{\scriptscriptstyle{(2)}} = \underline{a}^{\scriptscriptstyle{(2)}}(\vecu^{\snu};\alpha)\in\C^{2n+3}$ is some vector function with $a_1^{\scriptscriptstyle{(2)}}$, $a_2^{\scriptscriptstyle{(2)}}$, $a_3^{\scriptscriptstyle{(2)}}=0$. Using (2) of Lemma \ref{prop:lemmablockmatricesmixed} and the assumptions $\det{(\mathbf{A}^{\snu})},\det{(\widetilde{\mathbf{J}}^{\snu})}\neq 0$, it follows that $\det{(\mathbf{M}^{\scriptscriptstyle{(2)}})}\neq 0$. Hence, the linear system \eqref{eq:systemt2mixed} admits a unique solution $\vecx^{\scriptscriptstyle{(2)}}=\underline{x}^{\scriptscriptstyle{(2)}}(\vecu^{\snu};\alpha)$ with $x_1^{\scriptscriptstyle{(2)}},x_2^{\scriptscriptstyle{(2)}},x_3^{\scriptscriptstyle{(2)}}= 0$ which fixes uniquely
\begin{equation}
\label{eq:solsystemt2mixed}
c_2^{\scriptscriptstyle{(1)}} = c_1^{\scriptscriptstyle{(2)}} = c_3^{\scriptscriptstyle{(3)}} = 0 \;,\quad \vecu^{\scriptscriptstyle{(2)}}=\vecu^{\scriptscriptstyle{(2)}}(\vecu^{\snu};\alpha) \;.
\end{equation}
Combining \eqref{eq:c31mixed}, \eqref{eq:solsystemt1mixed} and \eqref{eq:solsystemt2mixed} together we find
\begin{equation}
\label{eq:t1u2mixed}
\vectt^{\scriptscriptstyle{(1)}}(\vecu^{\snu};\alpha) = c_1^{\scriptscriptstyle{(1)}}(\vecu^{\snu};\alpha)\,\vecX_1 \;,\quad  \vecu^{\scriptscriptstyle{(2)}}=\vecu^{\scriptscriptstyle{(2)}}(\vecu^{\snu};\alpha) \;.
\end{equation}
Going further, we can completely fix $\vectt^{\scriptscriptstyle{(2)}}$ plugging \eqref{eq:c31mixed}, \eqref{eq:solsystemt1mixed} and \eqref{eq:solsystemt2mixed} inside the system
\begin{align}
 \label{eq:systemt3mixed}
 \widetilde{S}_1^{\scriptscriptstyle{(3)}}:\begin{cases}
 \langle \vecA^{\scriptscriptstyle{(5)}},\vecX_2\rangle = 0 \\
 \langle \vecA^{\scriptscriptstyle{(7)}},\vecX_1\rangle = 0 \\
 \langle \vecA^{\scriptscriptstyle{(9)}},\vecX_3 \rangle = 0 
%\sum_{a=1}^{N-3} B_a^{\scriptscriptstyle{(2l+3)}} = -360\,b^2c_2^{\scriptscriptstyle{(l)}}s_{-6}^{\scriptscriptstyle{(0)}} - 36\,c_1^{\scriptscriptstyle{(l+1)}}s_{-4}^{\scriptscriptstyle{(0)}} = 0
 \end{cases} \longrightarrow\quad
\mathbf{N}^{\scriptscriptstyle{(3)}}(\underline{y}^{\scriptscriptstyle{(3)}})^{\text{T}} = (\underline{b}^{\scriptscriptstyle{(3)}})^{\text{T}} \;,
 \end{align}
where $\mathbf{N}^{\scriptscriptstyle{(3)}} = \mathbf{A^{\snu}}+k\mathbf{D}^{\snu}$ with $k=(M^{\frac14}b^3)^{-1}$, $\displaystyle{\underline{y}^{\scriptscriptstyle{(3)}}=\left(10b^2\,c_2^{\scriptscriptstyle{(2)}},c_1^{\scriptscriptstyle{(3)}},\frac{c_3^{\scriptscriptstyle{(4)}}}{2b^2}\right)}$ as per \eqref{eq:vecxydef} and $\C^3\ni\underline{b}^{\scriptscriptstyle{(3)}}=\underline{0}$. Using (1) of Lemma \ref{prop:lemmamatricesmixed} and the assumption $\det{(\mathbf{A^{\snu}})}\neq 0$, it follows that the linear system \eqref{eq:systemt3mixed} admits the trivial solution $\underline{y}^{\scriptscriptstyle{(3)}}=\underline{0}$ which fixes
\begin{equation}
\label{eq:solsystemt3mixed}
c_2^{\scriptscriptstyle{(2)}}= c_1^{\scriptscriptstyle{(3)}}=c_3^{\scriptscriptstyle{(4)}}=0 \;.
\end{equation}
Combining \eqref{eq:solsystemt1mixed}, \eqref{eq:solsystemt2mixed} and \eqref{eq:solsystemt3mixed} together we find
\begin{equation}
\label{eq:t2mixed}
\vectt^{\scriptscriptstyle{(2)}}(\vecu^{\snu};\alpha) = c_3^{\scriptscriptstyle{(2)}}(\vecu^{\snu};\alpha)\,\vecX_3 \;.
\end{equation}
The construction displayed above suggests an iterative procedure to construct the coefficients $\vectt^{\scriptscriptstyle{(m)}}$ and $\vecu^{\scriptscriptstyle{(m)}}$
%$\vectt^{\scriptscriptstyle{(m+1)}}$ and $\vecu^{\scriptscriptstyle{(\lfloor (2m+4)/3\rfloor)}}$
 from the previous ones. The idea is indeed to straighten the system $\widetilde{S}_1$ defined as per \eqref{eq:subsystemvecmixed} by taking into account suitable combinations of its constituent vector equations coming together to form linear systems. %that allow to recover $\vectt^{\scriptscriptstyle{(m+1)}}$ and $\vecu^{\scriptscriptstyle{(\lfloor (2m+4)/3\rfloor)}}$ from the previous order coefficients for any $m$. 
The structure of the system suggests that the combinations to be considered are 
\begin{equation}
 \label{eq:generalsystem1mixed}
 \widetilde{S}_1^{\scriptscriptstyle{(m)}}:\begin{cases}
 \langle \vecA^{\scriptscriptstyle{(2m-1)}},\vecX_2\rangle = 0 \\
 \langle \vecA^{\scriptscriptstyle{(2m+1)}},\vecX_1\rangle = 0 \\
 \langle \vecA^{\scriptscriptstyle{(2m+3)}},\vecX_3\rangle = 0 \\
\vecB^{\scriptscriptstyle{(3m)}}=\underline{0}
 \end{cases} \;,\quad (m\cong 1\mod 3\;\;\vee\;\;m\cong 2\mod 3) \;,\quad m\geq 1 \;,
 \end{equation}
and
\begin{equation}
 \label{eq:generalsystem2mixed}
 \widetilde{S}_1^{\scriptscriptstyle{(m)}}:\begin{cases}
 \langle \vecA^{\scriptscriptstyle{(2m-1)}},\vecX_2\rangle = 0 \\
 \langle \vecA^{\scriptscriptstyle{(2m+1)}},\vecX_1\rangle = 0 \\
 \langle \vecA^{\scriptscriptstyle{(2m+3)}},\vecX_3\rangle = 0 
 \end{cases} \;,\quad m\cong 0\mod 3 \;,\quad m\geq 3 \;.
 \end{equation}
which are such that 
$$\widetilde{S}_1=\bigcup_{m\geq 1}\widetilde{S}_1^{\scriptscriptstyle{(m)}}\;,$$
since $\langle\vecA^{\scriptscriptstyle{(1)}},\vecX_1\rangle$, $\langle\vecA^{\scriptscriptstyle{(1)}},\vecX_3\rangle$ and $\vecA^{\scriptscriptstyle{(-1)}}$ vanish identically upon setting $c_3^{\scriptscriptstyle{(1)}}=0$ as per \eqref{eq:c31mixed}.

We still need to compute $\vectt^{\scriptscriptstyle{(3)}}$ and $\vectt^{\scriptscriptstyle{(4)}}$ by iterating once again the procedure described above. To fix $\vectt^{\scriptscriptstyle{(3)}}$ and $\vecu^{\scriptscriptstyle{(3)}}$ we use \eqref{eq:t0u1mixed}, \eqref{eq:solsystemt2mixed}, \eqref{eq:t1u2mixed}, \eqref{eq:solsystemt3mixed} and \eqref{eq:t2mixed} inside the system
\begin{align}
 \label{eq:systemt4mixed}
 \widetilde{S}_1^{\scriptscriptstyle{(4)}}:\begin{cases}
 \langle \vecA^{\scriptscriptstyle{(7)}},\vecX_2\rangle = 0 \\
 \langle \vecA^{\scriptscriptstyle{(9)}},\vecX_1\rangle = 0 \\
 \langle \vecA^{\scriptscriptstyle{(11)}},\vecX_3 \rangle = 0 \\
\vecB^{\scriptscriptstyle{(9)}} = \underline{0}
%\sum_{a=1}^{N-3} B_a^{\scriptscriptstyle{(2l+3)}} = -360\,b^2c_2^{\scriptscriptstyle{(l)}}s_{-6}^{\scriptscriptstyle{(0)}} - 36\,c_1^{\scriptscriptstyle{(l+1)}}s_{-4}^{\scriptscriptstyle{(0)}} = 0
 \end{cases} \longrightarrow\quad
\mathbf{M}^{\scriptscriptstyle{(4)}}(\vecx^{\scriptscriptstyle{(4)}})^{\text{T}} = (\underline{a}^{\scriptscriptstyle{(4)}})^{\text{T}} \;,
 \end{align}
where $\mathbf{M}^{\scriptscriptstyle{(4)}}=\mathbf{P}_k^{\snu}$ with $k=(M^{\frac14}b^3)^{-1}$, $\displaystyle{\vecx^{\scriptscriptstyle{(4)}}=\left(10b^2\,c_2^{\scriptscriptstyle{(3)}},c_1^{\scriptscriptstyle{(4)}},\frac{c_3^{\scriptscriptstyle{(5)}}}{2b^2},u_1^{\scriptscriptstyle{(3)}},\dots,u_{2n}^{\scriptscriptstyle{(3)}}\right)}$ as per \eqref{eq:vecxydef} and $\underline{a}^{\scriptscriptstyle{(4)}}=\underline{a}^{\scriptscriptstyle{(4)}}\left(\vecu^{\snu};\alpha\right)\in\C^{2n+3}$ is some vector function. Using (1) of Lemma \ref{prop:lemmablockmatricesmixed} and the assumptions $\det{\big(\widetilde{\mathbf{J}}^{\snu}\big)}$, $\det{(\mathbf{A}^{\snu})}$, $\det{(\mathbf{A}^{\snu}-\mathbf{B}^{\snu}(\widetilde{\mathbf{J}}^{\snu})^{-1}\mathbf{C}^{\snu})}\neq 0$, it follows that the linear system \eqref{eq:systemt4mixed} admits a unique solution $\underline{x}^{\scriptscriptstyle{(4)}}= \underline{x}^{\scriptscriptstyle{(4)}}\left(\vecu^{\snu};\alpha\right)$ which fixes uniquely
\begin{equation}
\label{eq:solsystemt4mixed}
c_2^{\scriptscriptstyle{(3)}}=c_2^{\scriptscriptstyle{(3)}}(\vecu^{\snu};\alpha) \;,\quad c_1^{\scriptscriptstyle{(4)}}=c_1^{\scriptscriptstyle{(4)}}(\vecu^{\snu};\alpha) \;,\quad c_3^{\scriptscriptstyle{(5)}}=c_3^{\scriptscriptstyle{(5)}}(\vecu^{\snu};\alpha) \;,\quad \vecu^{\scriptscriptstyle{(3)}} = \vecu^{\scriptscriptstyle{(3)}}(\vecu^{\snu};\alpha) \;.
\end{equation}
Combining \eqref{eq:solsystemt2mixed}, \eqref{eq:solsystemt3mixed} and \eqref{eq:solsystemt4mixed} together gives
\begin{equation}
\label{eq:t3u3mixed}
\vectt^{\scriptscriptstyle{(3)}}(\vecu^{\snu};\alpha) = c_2^{\scriptscriptstyle{(3)}}(\vecu^{\snu};\alpha)\,\vecX_2 \;,\quad \vecu^{\scriptscriptstyle{(3)}} = \vecu^{\scriptscriptstyle{(3)}}(\vecu^{\snu};\alpha) \;.
\end{equation}
 Finally, to fix $\vectt^{\scriptscriptstyle{(4)}}$ and $\vecu^{\scriptscriptstyle{(4)}}$ we use \eqref{eq:t0u1mixed}, \eqref{eq:t1u2mixed}, \eqref{eq:solsystemt3mixed}, \eqref{eq:t2mixed}, \eqref{eq:solsystemt4mixed} and \eqref{eq:t3u3mixed} inside the system
\begin{align}
 \label{eq:systemt5mixed}
 \widetilde{S}_1^{\scriptscriptstyle{(5)}}:\begin{cases}
 \langle \vecA^{\scriptscriptstyle{(9)}},\vecX_2\rangle = 0 \\
 \langle \vecA^{\scriptscriptstyle{(11)}},\vecX_1\rangle = 0 \\
 \langle \vecA^{\scriptscriptstyle{(13)}},\vecX_3 \rangle = 0 \\
\vecB^{\scriptscriptstyle{(12)}} = \underline{0}
%\sum_{a=1}^{N-3} B_a^{\scriptscriptstyle{(2l+3)}} = -360\,b^2c_2^{\scriptscriptstyle{(l)}}s_{-6}^{\scriptscriptstyle{(0)}} - 36\,c_1^{\scriptscriptstyle{(l+1)}}s_{-4}^{\scriptscriptstyle{(0)}} = 0
 \end{cases} \longrightarrow\quad
\mathbf{M}^{\scriptscriptstyle{(5)}}(\vecx^{\scriptscriptstyle{(5)}})^{\text{T}} = (\underline{a}^{\scriptscriptstyle{(5)}})^{\text{T}} \;,
 \end{align}
where $\mathbf{M}^{\scriptscriptstyle{(5)}}=\mathbf{Q}_k^{\snu}$ with $k=(M^{\frac14}b^3)^{-1}$, $\displaystyle{\vecx^{\scriptscriptstyle{(5)}}=\left(10b^2\,c_2^{\scriptscriptstyle{(4)}},c_1^{\scriptscriptstyle{(5)}},\frac{c_3^{\scriptscriptstyle{(6)}}}{2b^2},u_1^{\scriptscriptstyle{(4)}},\dots,u_{2n}^{\scriptscriptstyle{(4)}}\right)}$ as per \eqref{eq:vecxydef} and $\underline{a}^{\scriptscriptstyle{(5)}}=\underline{a}^{\scriptscriptstyle{(5)}}\left(\vecu^{\snu};\alpha\right)\in\C^{2n+3}$ is some vector function with $a_1^{\scriptscriptstyle{(5)}}$, $a_2^{\scriptscriptstyle{(5)}}$, $a_3^{\scriptscriptstyle{(5)}}=0$. Therefore, under the assumptions $\det{(\mathbf{A}^{\snu})},\det{(\widetilde{\mathbf{J}}^{\snu})}\neq 0$, the linear system \eqref{eq:systemt5mixed} admits a unique solution $\underline{x}^{\scriptscriptstyle{(5)}}= \underline{x}^{\scriptscriptstyle{(5)}}\left(\vecu^{\snu};\alpha\right)$ with $x_1^{\scriptscriptstyle{(5)}},x_2^{\scriptscriptstyle{(5)}},x_3^{\scriptscriptstyle{(5)}}=0$ which fixes uniquely
\begin{equation}
\label{eq:solsystemt5mixed}
c_2^{\scriptscriptstyle{(4)}}=c_1^{\scriptscriptstyle{(5)}}=c_3^{\scriptscriptstyle{(6)}}=0 \;,\quad \vecu^{\scriptscriptstyle{(4)}}=\vecu^{\scriptscriptstyle{(4)}}(\vecu^{\snu};\alpha) \;.
\end{equation}
Combining \eqref{eq:solsystemt3mixed}, \eqref{eq:solsystemt4mixed} and \eqref{eq:solsystemt5mixed} together gives
\begin{equation}
\vectt^{\scriptscriptstyle{(4)}}(\vecu^{\snu};\alpha) = c_1^{\scriptscriptstyle{(4)}}(\vecu^{\snu};\alpha)\,\vecX_1 \;,\quad \vecu^{\scriptscriptstyle{(4)}}=\vecu^{\scriptscriptstyle{(4)}}(\vecu^{\snu};\alpha) \;.
\end{equation}
We are now ready to show by induction that the system $\widetilde{S}_1$ admits a unique solution. In the following, we shall refer to the general expressions \eqref{eq:Ageneral} and \eqref{eq:Bgeneral}. Let us fix an arbitrary $m \cong 1 \mod 3$, with $m\geq 10$, and determine $\vectt^{\scriptscriptstyle{(m+1)}}$ and $\vecu^{\scriptscriptstyle{((2m+4)/3)}}$ in terms of $\vectt^{\scriptscriptstyle{(0)}}$, $\dots$, $\vectt^{\scriptscriptstyle{(m)}}$, $\vecu^{\snu}$, $\dots$, $\vecu^{\scriptscriptstyle{((2m+1)/3)}}$. According to \eqref{eq:generalsystem1mixed}, since $m \cong 1 \mod 3$ the first system to consider is
\begin{align}
 \label{eq:systemtmmixed}
 \widetilde{S}_1^{\scriptscriptstyle{(m)}}:\begin{cases}
 \langle \vecA^{\scriptscriptstyle{(2m-1)}},\vecX_2\rangle = 0 \\
 \langle \vecA^{\scriptscriptstyle{(2m+1)}},\vecX_1\rangle = 0 \\
 \langle \vecA^{\scriptscriptstyle{(2m+3)}},\vecX_3 \rangle = 0 \\
\vecB^{\scriptscriptstyle{(2m+1)}} = \underline{0}
 \end{cases} \longrightarrow\quad
\mathbf{M}^{\scriptscriptstyle{(m)}}(\vecx^{\scriptscriptstyle{(m)}})^{\text{T}} = (\underline{a}^{\scriptscriptstyle{(m)}})^{\text{T}} \;,
 \end{align}
where $\mathbf{M}^{\scriptscriptstyle{(m)}}=\mathbf{P}_k^{\snu}$ with $k=(M^{\frac14}b^3)^{-1}$, $\vecx^{\scriptscriptstyle{(m)}}$ is as per \eqref{eq:vecxydef}, namely
$$\vecx^{\scriptscriptstyle{(m)}}=\left(10b^2\,c_2^{\scriptscriptstyle{(m-1)}},c_1^{\scriptscriptstyle{(m)}},\frac{c_3^{\scriptscriptstyle{(m+1)}}}{2b^2},u_1^{\scriptscriptstyle{((2m+1)/3)}},\dots,u_{2n}^{\scriptscriptstyle{((2m+1)/3)}}\right) \;,$$
and $\underline{a}^{\scriptscriptstyle{(m)}}\in\C^{2n+3}$ is some vector function that depends on $\vectt^{\scriptscriptstyle{(0)}}$, $\dots$, $\vectt^{\scriptscriptstyle{(m-2)}}$, $\vecu^{\snu}$, $\dots$, $\vecu^{\scriptscriptstyle{((2m-2)/3)}}$. The invertibility of $\mathbf{P}_k^{\snu}$ ensures the existence of a unique solution $\vecx^{\scriptscriptstyle{(m)}}$ to \eqref{eq:systemtmmixed} expressed as a function of $\vectt^{\scriptscriptstyle{(0)}}$, $\dots$, $\vectt^{\scriptscriptstyle{(m-2)}}$, $\vecu^{\snu}$, $\dots$, $\vecu^{\scriptscriptstyle{((2m-2)/3)}}$ which fixes uniquely
\begin{align}
&c_2^{\scriptscriptstyle{(m-1)}}=c_2^{\scriptscriptstyle{(m-1)}}(\vectt^{\scriptscriptstyle{(0)}},\dots,\vectt^{\scriptscriptstyle{(m-2)}},\vecu^{\snu},\dots,\vecu^{\scriptscriptstyle{((2m-2)/3)}}) \;, \notag \\
&c_1^{\scriptscriptstyle{(m)}}=c_1^{\scriptscriptstyle{(m)}}(\vectt^{\scriptscriptstyle{(0)}},\dots,\vectt^{\scriptscriptstyle{(m-2)}},\vecu^{\snu},\dots,\vecu^{\scriptscriptstyle{((2m-2)/3)}}) \;, \notag \\
&c_3^{\scriptscriptstyle{(m+1)}}=c_3^{\scriptscriptstyle{(m+1)}}(\vectt^{\scriptscriptstyle{(0)}},\dots,\vectt^{\scriptscriptstyle{(m-2)}},\vecu^{\snu},\dots,\vecu^{\scriptscriptstyle{((2m-2)/3)}}) \;,\notag \\
&\vecu^{\scriptscriptstyle{((2m+1)/3)}}=\vecu^{\scriptscriptstyle{((2m+1)/3)}}(\vectt^{\scriptscriptstyle{(0)}},\dots,\vectt^{\scriptscriptstyle{(m-2)}},\vecu^{\snu},\dots,\vecu^{\scriptscriptstyle{((2m-2)/3)}}) \;.
\label{eq:solsystemtmmixed}
\end{align}
Using \eqref{eq:solsystemtmmixed}, the second system yields
\begin{align}
 \label{eq:systemtm1mixed}
 \widetilde{S}_1^{\scriptscriptstyle{(m+1)}}:\begin{cases}
 \langle \vecA^{\scriptscriptstyle{(2m+1)}},\vecX_2\rangle = 0 \\
 \langle \vecA^{\scriptscriptstyle{(2m+3)}},\vecX_1\rangle = 0 \\
 \langle \vecA^{\scriptscriptstyle{(2m+5)}},\vecX_3 \rangle = 0 \\
\vecB^{\scriptscriptstyle{(2m+4)}} = \underline{0}
 \end{cases} \longrightarrow\quad
\mathbf{M}^{\scriptscriptstyle{(m+1)}}(\vecx^{\scriptscriptstyle{(m+1)}})^{\text{T}} = (\underline{a}^{\scriptscriptstyle{(m+1)}})^{\text{T}} \;,
 \end{align}
where $\mathbf{M}^{\scriptscriptstyle{(m+1)}}=\mathbf{Q}_k^{\snu}$ with $k=(M^{\frac14}b^3)^{-1}$, $\vecx^{\scriptscriptstyle{(m+1)}}$ is as per \eqref{eq:vecxydef}, namely
$$\vecx^{\scriptscriptstyle{(m+1)}}=\left(10b^2\,c_2^{\scriptscriptstyle{(m)}},c_1^{\scriptscriptstyle{(m+1)}},\frac{c_3^{\scriptscriptstyle{(m+2)}}}{2b^2},u_1^{\scriptscriptstyle{((2m+4)/3)}},\dots,u_{2n}^{\scriptscriptstyle{((2m+4)/3)}}\right) \;,$$
and $\underline{a}^{\scriptscriptstyle{(m+1)}}\in\C^{2n+3}$ is some vector function that depends on $\vectt^{\scriptscriptstyle{(0)}}$, $\dots$, $\vectt^{\scriptscriptstyle{(m-1)}}$, $\vecu^{\snu}$, $\dots$, $\vecu^{\scriptscriptstyle{((2m-2)/3)}}$. Again, the invertibility of $\mathbf{Q}_k^{\snu}$ ensures that there exists a unique solution $\vecx^{\scriptscriptstyle{(m+1)}}$ to \eqref{eq:systemtm1mixed} expressed as a function of $\vectt^{\scriptscriptstyle{(0)}}$, $\dots$, $\vectt^{\scriptscriptstyle{(m-1)}}$, $\vecu^{\snu}$, $\dots$, $\vecu^{\scriptscriptstyle{((2m-2)/3)}}$ which fixes uniquely
\begin{align}
&c_2^{\scriptscriptstyle{(m)}}=c_2^{\scriptscriptstyle{(m)}}(\vectt^{\scriptscriptstyle{(0)}},\dots,\vectt^{\scriptscriptstyle{(m-1)}},\vecu^{\snu},\dots,\vecu^{\scriptscriptstyle{((2m-2)/3)}}) \;, \notag \\
&c_1^{\scriptscriptstyle{(m+1)}}=c_1^{\scriptscriptstyle{(m+1)}}(\vectt^{\scriptscriptstyle{(0)}},\dots,\vectt^{\scriptscriptstyle{(m-1)}},\vecu^{\snu},\dots,\vecu^{\scriptscriptstyle{((2m-2)/3)}}) \;, \notag \\
&c_3^{\scriptscriptstyle{(m+2)}}=c_3^{\scriptscriptstyle{(m+2)}}(\vectt^{\scriptscriptstyle{(0)}},\dots,\vectt^{\scriptscriptstyle{(m-1)}},\vecu^{\snu},\dots,\vecu^{\scriptscriptstyle{((2m-2)/3)}}) \;,\notag \\
&\vecu^{\scriptscriptstyle{((2m+4)/3)}}=\vecu^{\scriptscriptstyle{((2m+4)/3)}}(\vectt^{\scriptscriptstyle{(0)}},\dots,\vectt^{\scriptscriptstyle{(m-1)}},\vecu^{\snu},\dots,\vecu^{\scriptscriptstyle{((2m-2)/3)}}) \;.
\label{eq:solsystemtm1mixed}
\end{align}
 Finally, using \eqref{eq:systemtmmixed} and \eqref{eq:systemtm1mixed} the last system is
\begin{align}
 \label{eq:systemtm2mixed}
 \widetilde{S}_1^{\scriptscriptstyle{(m+2)}}:\begin{cases}
 \langle \vecA^{\scriptscriptstyle{(2m+3)}},\vecX_2\rangle = 0 \\
 \langle \vecA^{\scriptscriptstyle{(2m+5)}},\vecX_1\rangle = 0 \\
 \langle \vecA^{\scriptscriptstyle{(2m+7)}},\vecX_3 \rangle = 0 
%\sum_{a=1}^{N-3} B_a^{\scriptscriptstyle{(2l+3)}} = -360\,b^2c_2^{\scriptscriptstyle{(l)}}s_{-6}^{\scriptscriptstyle{(0)}} - 36\,c_1^{\scriptscriptstyle{(l+1)}}s_{-4}^{\scriptscriptstyle{(0)}} = 0
 \end{cases} \longrightarrow\quad
\mathbf{N}^{\scriptscriptstyle{(m+2)}}(\underline{y}^{\scriptscriptstyle{(m+2)}})^{\text{T}} = (\underline{b}^{\scriptscriptstyle{(m+2)}})^{\text{T}} \;,
 \end{align}
where $\mathbf{N}^{\scriptscriptstyle{(m+2)}} = \mathbf{A}^{\snu}+ k\mathbf{D}^{\snu}$ with $k=(M^{\frac14}b^3)^{-1}$, $\underline{y}^{\scriptscriptstyle{(m+2)}}$ is as per \eqref{eq:vecxydef}, namely
$$\underline{y}^{\scriptscriptstyle{(m)}}=\left(10b^2\,c_2^{\scriptscriptstyle{(m+1)}},c_1^{\scriptscriptstyle{(m+2)}},\frac{c_3^{\scriptscriptstyle{(m+3)}}}{2b^2}\right) \;,$$
and $\underline{b}^{\scriptscriptstyle{(m+2)}}\in\C^3$ is some vector function that depends on $\vectt^{\scriptscriptstyle{(0)}}$, $\dots$, $\vectt^{\scriptscriptstyle{(m)}}$, $\vecu^{\snu}$, $\dots$, $\vecu^{\scriptscriptstyle{((2m-2)/3)}}$.
As before, the invertibility of $\mathbf{A}^{\snu}$ provides a unique solution $\underline{y}^{\scriptscriptstyle{(m+2)}}$ to  \eqref{eq:systemtm2mixed} expressed as a function of $\vectt^{\scriptscriptstyle{(0)}}$, $\dots$, $\vectt^{\scriptscriptstyle{(m)}}$, $\vecu^{\snu}$, $\dots$, $\vecu^{\scriptscriptstyle{((2m-2)/3)}}$ which fixes uniquely
\begin{align}
&c_2^{\scriptscriptstyle{(m+1)}}=c_2^{\scriptscriptstyle{(m+1)}}(\vectt^{\scriptscriptstyle{(0)}},\dots,\vectt^{\scriptscriptstyle{(m)}},\vecu^{\snu},\dots,\vecu^{\scriptscriptstyle{((2m-2)/3)}}) \;, \notag \\
&c_1^{\scriptscriptstyle{(m+2)}}=c_1^{\scriptscriptstyle{(m+2)}}(\vectt^{\scriptscriptstyle{(0)}},\dots,\vectt^{\scriptscriptstyle{(m)}},\vecu^{\snu},\dots,\vecu^{\scriptscriptstyle{((2m-2)/3)}}) \;, \notag \\
&c_3^{\scriptscriptstyle{(m+3)}}=c_3^{\scriptscriptstyle{(m+3)}}(\vectt^{\scriptscriptstyle{(0)}},\dots,\vectt^{\scriptscriptstyle{(m)}},\vecu^{\snu},\dots,\vecu^{\scriptscriptstyle{((2m-2)/3)}}) \;,\notag \\
&\vecu^{\scriptscriptstyle{((2m+4)/3)}}=\vecu^{\scriptscriptstyle{((2m+4)/3)}}(\vectt^{\scriptscriptstyle{(0)}},\dots,\vectt^{\scriptscriptstyle{(m-1)}},\vecu^{\snu},\dots,\vecu^{\scriptscriptstyle{((2m-2)/3)}}) \;.
\label{eq:solsystemtm2mixed}
\end{align}
In conclusion, combining \eqref{eq:solsystemtmmixed}, \eqref{eq:solsystemtm1mixed} and \eqref{eq:solsystemtm2mixed} together gives
\begin{align}
&\vectt^{\scriptscriptstyle{(m+1)}}=\vectt^{\scriptscriptstyle{(m+1)}}(\vectt^{\scriptscriptstyle{(0)}},\dots,\vectt^{\scriptscriptstyle{(m)}},\vecu^{\snu},\dots,\vecu^{\scriptscriptstyle{((2m-2)/3)}}) \;, \notag \\
&\vecu^{\scriptscriptstyle{((2m+4)/3)}}=\vecu^{\scriptscriptstyle{((2m+4)/3)}}(\vectt^{\scriptscriptstyle{(0)}},\dots,\vectt^{\scriptscriptstyle{(m)}},\vecu^{\snu},\dots,\vecu^{\scriptscriptstyle{((2m-2)/3)}}) \;.
\end{align}
Going further with the same reasoning, it is easy to fix $\vectt^{\scriptscriptstyle{(m+2)}}$ and $\vectt^{\scriptscriptstyle{(m+3)}}$ in terms of $\vectt^{\scriptscriptstyle{(0)}}$, $\dots$, $\vectt^{\scriptscriptstyle{(m+1)}}$, $\vecu^{\snu}$, $\dots$, $\vecu^{\scriptscriptstyle{((2m-2)/3)}}$ and $\vectt^{\scriptscriptstyle{(0)}}$, $\dots$, $\vectt^{\scriptscriptstyle{(m+2)}}$, $\vecu^{\snu}$, $\dots$, $\vecu^{\scriptscriptstyle{((2m-2)/3)}}$, respectively, thus proving that the closure of the system $\widetilde{S}_1$ holds for any $m$. In conclusion, we proved that the subsystem $\widetilde{S}_1$ admits a unique solution $\vectt^{\snu}(\e)$ of the form \eqref{eq:ansatzvecmixed} for each $\vecu^{\snu}$ and $\alpha\neq\frac52$.

Let us now deal with the second part of the proof. We start by observing that the transformation
\begin{equation}
(t_1,\dots,t_{2n+3};\e) \to (-t_{\sigma(1)},\dots,-t_{\sigma(2n+3)};-\e) \;,
\end{equation}
is a symmetry of the system $F_k(\vectt,\e)=0$, $k=1,\dots,2n+3$, for any $\sigma\in\mathcal{S}_{2n+3}$. In particular, this implies that if $t_1^{\snu}(\e)$, $\dots$, $t_{2n+3}^{\snu}(\e)$ as per \eqref{eq:ansatzvecmixed} solves $F_k(\vectt,\e)=0$, $k=1,\dots,2n+3$ -- and consequently the subsystem $\widetilde{S}_1$ -- so does $-t_{\sigma(1)}^{\snu}(-\e)$, $\dots$, $-t_{\sigma(2n+3)}^{\snu}(-\e)$. Now, given that $\lim_{\e\to 0}\vectt^{\snu}(\e)=\vecv^{\snu}$, with $v_1^{\snu}$, $v_2^{\snu}$, $v_3^{\snu}=0$ and $v_{a+3}^{\snu}=u_{a}^{\snu}$, $a=1,\dots,2n$, is a fixed point of the latter symmetry and that the solution $\vectt^{\snu}(\e)$ to $\widetilde{S}_1$ is unique for each $\vecu^{\snu}$, it follows that 
\begin{equation}
t_k^{\snu}(\e) = t_{\sigma(k)}^{\snu}(-\e) \;,\quad k=1,\dots,2n+3 \;,
\end{equation}
for some $\sigma\in\mathcal{S}_{2n+3}$. In particular, choosing $t_k^{\scriptscriptstyle{(0)}} = \omega^{k-1}\,b$, $k=1,2,3$ as per \eqref{eq:ansatzt0mixed} and $\vecu^{\snu}=\sum_{i=1}^n d_i^{\snu}Y_i^{\scriptscriptstyle{(-)}}$ as per \eqref{eq:usnu} and \eqref{eq:basisC}, it can be easily checked that $\sigma=(123)(45)(67)\dots(2n+2,2n+3)$ and the coefficients $\lbrace \vectt^{\scriptscriptstyle{(m)}}\rbrace_{m\geq 1}$ and $\lbrace \vecu^{\scriptscriptstyle{(m)}}\rbrace_{m\geq 1}$ of $\vectt^{\snu}(\e)$ must possess the cyclic structure \eqref{eq:cyclicity}.

Next, let us show that the subsystem $\widetilde{S}_2$ as per \eqref{eq:subsystemvecmixed} vanish identically provided the coefficients $\lbrace \vectt^{\scriptscriptstyle{(m)}}\rbrace_{m\geq 0}$ and $\lbrace \vecu^{\scriptscriptstyle{(m)}}\rbrace_{m\geq 1}$ of $\vectt^{\snu}(\e)$ fulfil \eqref{eq:cyclicity}. To this aim, we start by observing that the terms $\vecA^{\scriptscriptstyle{(2m)}}$ and $\vecB^{\scriptscriptstyle{(2m-1-\lfloor m/2 \rfloor)}}$, whose cancellations form the subsystem $\widetilde{S}_2$, originate from the expansion of 
%\begin{align}
%&F_k(\vectt^{\snu}(\e);\e) = \frac{\e^{3} (1-4 \alpha^2)}{8 M} - \frac{t_k(\e)+ \e \sigma(t_k(\e);\e)}{2M^{\frac14}} \notag \\
%&+\sum_{j \neq k}\frac{M^{-\frac94}\mathcal{F}(t_k(\e),t_j(\e);\e)}{\big(t_k(\e)-t_j(\e)+ \e (\sigma(t_k(\e);\e)-\sigma(t_j(\e);\e))\big)^3} \notag \\
%&+ \sum_{a=1}^{2n}\frac{M^{-\frac94}\mathcal{F}(t_k(\e),u_a^{\snu}(\e);\e)}{\big(t_k(\e)-u_a^{\snu}(\e)+ \e (\sigma(t_k(\e);\e)-\sigma(u_a^{\snu}(\e);\e))\big)^3} \;,\quad k=1,2,3 \;, \label{eq:Fk3mixed}
%\end{align}
%and
%\begin{align}
%&F_{a+3}(\vectt^{\snu}(\e);\e) = \frac{\e^{3} (1-4 \alpha^2)}{8 M} - \frac{u_a^{\snu}(\e)+ \e \sigma(u_a^{\snu}(\e);\e)}{2M^{\frac14}} \notag \\
%&+\sum_{b \neq a}\frac{M^{-\frac94}\mathcal{F}(u_a^{\snu}(\e),u_b^{\snu}(\e);\e)}{\big(u_a^{\snu}(\e)-u_b^{\snu}(\e)+ \e (\sigma(u_a^{\snu}(\e);\e)-\sigma(u_b^{\snu}(\e);\e))\big)^3} \notag \\
%&+ \sum_{k=1}^{3}\frac{M^{-\frac94}\mathcal{F}(u_a^{\snu}(\e),t_k(\e);\e)}{\big(u_a^{\snu}(\e)-t_k(\e)+ \e (\sigma(u_a^{\snu}(\e);\e)-\sigma(t_k(\e);\e))\big)^3} \;,\quad a=1,\dots,2n \;. \label{eq:Fk2nmixed}
%\end{align}
%Notice that the terms $\vecA^{\scriptscriptstyle{(2m)}}$ and $\vecB^{\scriptscriptstyle{(2m-1-\lfloor m/2 \rfloor)}}$ appearing in the expansion \eqref{eq:Fexpan} of $F_k(\vectt(\e);\e)$ and whose cancellations form the subsystem $S_2$, originate from the expansion about $\e=0$ of the last sum in \eqref{eq:Fk3mixed}, i.e.
\begin{equation}
\label{eq:int3mixed}
\sum_{a=1}^{2n}\frac{M^{-\frac94}\mathcal{F}(t_k(\e),u_a^{\snu}(\e);\e)}{\big(t_k(\e)-u_a^{\snu}(\e)+ \e (\sigma(t_k(\e);\e)-\sigma(u_a^{\snu}(\e);\e))\big)^3} \;,\quad k=1,2,3 \;,
\end{equation}
and
\begin{equation}
\label{eq:int2nmixed}
\sum_{k=1}^{3}\frac{M^{-\frac94}\mathcal{F}(u_a^{\snu}(\e),t_k(\e);\e)}{\big(u_a^{\snu}(\e)-t_k(\e)+ \e (\sigma(u_a^{\snu}(\e);\e)-\sigma(t_k(\e);\e))\big)^3} \;,\quad a=1,\dots,2n \;,
\end{equation}
respectively, in \eqref{eq:Fk}. Both \eqref{eq:int3mixed} and \eqref{eq:int2nmixed} are, by construction, symmetric functions in the variables $u_1(\e)$, $\dots$, $u_{2n}(\e)$ and $t_1(\e)$, $t_2(\e)$, $t_3(\e)$, respectively. Therefore, they can be expressed in terms of the elementary symmetric polynomials in those variables.

Imposing that $\lbrace \vectt^{\scriptscriptstyle{(m)}} \rbrace_{m\geq 0}$ fulfil \eqref{eq:cyclicity},
it is easy to show that the elementary symmetric polynomials in $t_1(\e)$, $t_2(\e)$, $t_3(\e)$ are analytic functions of
$\e$. Then \eqref{eq:int2nmixed} is also analytic in $\e$, whence the coefficients
$\lbrace \vecB^{\scriptscriptstyle{(2m-1-\lfloor m/2 \rfloor)}} \rbrace_{m\geq 3}$ vanish identically.
A similar argument shows that if $\lbrace \vecu^{\scriptscriptstyle{(m)}} \rbrace_{m\geq 0}$ fulfil \eqref{eq:cyclicity}
then \eqref{eq:int3mixed} admits an expansion in odd powers of $\e^{\frac13}$ about $\e=0$,
whence also the coefficients $\lbrace\vecA^{\scriptscriptstyle{(2m)}}\rbrace_{m\geq 2}$ vanish identically.

In conclusion, we proved that there exists a unique solution of the form
\eqref{eq:ansatzvecmixed} to $F_k(\vectt,\e)=0 \;,\; k=1,\dots,2n+3$, which possesses the cyclic structure \eqref{eq:cyclicity}.\\

\def\cprime{$'$} \def\cprime{$'$} \def\cprime{$'$} \def\cprime{$'$}
  \def\cprime{$'$} \def\cprime{$'$} \def\cprime{$'$} \def\cprime{$'$}
  \def\cprime{$'$} \def\cprime{$'$} \def\cydot{\leavevmode\raise.4ex\hbox{.}}
  \def\cprime{$'$} \def\cprime{$'$} \def\cprime{$'$}

\end{document}